\documentclass[%
 mph,%
 amsmath,amssymb,
 reprint,%
superscriptaddress,
longbibliography
]{revtex4-1}

\usepackage{graphicx}
\usepackage{dcolumn}
\usepackage{bm}
\usepackage{xcolor}
\usepackage{ulem}
\usepackage[
pagebackref=false,
colorlinks=true,
linkcolor=blue,
urlcolor=blue,
filecolor=black,
citecolor=red,
pdfstartview=FitV,
pdftitle={},
pdfauthor={},
pdfsubject={},
pdfkeywords={},
pdfpagemode=None,
bookmarksopen=true
]{hyperref}

\begin{document}

\newtheorem{definition}{Definition}[section]
\newcommand{\be}{\begin{equation}}
\newcommand{\ee}{\end{equation}}
\newcommand{\bea}{\begin{eqnarray}}
\newcommand{\eea}{\end{eqnarray}}
\newcommand{\LE}{\left[}
\newcommand{\R}{\right]}
\newcommand{\nn}{\nonumber}
\newcommand{\Tr}{\text{Tr}}
\newcommand{\N}{\mathcal{N}}
\newcommand{\G}{\Gamma}
\newcommand{\vf}{\varphi}
\newcommand{\LL}{\mathcal{L}}
\newcommand{\Op}{\mathcal{O}}
\newcommand{\HH}{\mathcal{H}}
\newcommand{\arctanh}{\text{arctanh}}
\newcommand{\up}{\uparrow}
\newcommand{\down}{\downarrow}
\newcommand{\ket}[1]{\left| #1 \right>}
\newcommand{\bra}[1]{\left< #1 \right|}
\newcommand{\ketbra}[1]{\left|#1\right>\left<#1\right|}
\newcommand{\rd}{\partial}
\newcommand{\de}{\partial}
\newcommand{\ba}{\begin{eqnarray}}
\newcommand{\ea}{\end{eqnarray}}
\newcommand{\db}{\bar{\partial}}
\newcommand{\we}{\wedge}
\newcommand{\ca}{\mathcal}
\newcommand{\lr}{\leftrightarrow}
\newcommand{\f}{\frac}
\newcommand{\s}{\sqrt}
\newcommand{\vp}{\varphi}
\newcommand{\hvp}{\hat{\varphi}}
\newcommand{\tvp}{\tilde{\varphi}}
\newcommand{\tp}{\tilde{\phi}}
\newcommand{\ti}{\tilde}
\newcommand{\pr}{\propto}
\newcommand{\mb}{\mathbf}
\newcommand{\ddd}{\cdot\cdot\cdot}
\newcommand{\no}{\nonumber \\}
\newcommand{\la}{\langle}
\newcommand{\lb}{\rangle}
\newcommand{\ep}{\epsilon}
 \def\we{\wedge}
 \def\lr{\leftrightarrow}
 \def\f {\frac}
 \def\ti{\tilde}
 \def\ap{\alpha}
 \def\pr{\propto}
 \def\mb{\mathbf}
 \def\ddd{\cdot\cdot\cdot}
 \def\no{\nonumber \\}
 \def\la{\langle}
 \def\lb{\rangle}
 \def\ep{\epsilon}
\newcommand{\mcl}{\mathcal}
 \def\g{\gamma}
\def\Tr{\text{tr}}

\preprint{AAPM/123-QED}

\title{Non-Equilibrating a Black Hole 
  with Inhomogeneous Quantum Quench}

\author{Kanato Goto}
\affiliation{ 
RIKEN Interdisciplinary Theoretical and Mathematical Sciences (iTHEMS), \\Wako, Saitama 351-0198, Japan
}

\author{Masahiro Nozaki}%
\affiliation{ 
RIKEN Interdisciplinary Theoretical and Mathematical Sciences (iTHEMS), \\Wako, Saitama 351-0198, Japan
}
\affiliation{ 
Kavli Institute for Theoretical Sciences and CAS Center for Excellence in Topological Quantum Computation, University of Chinese Academy of Sciences, Beijing, 100190, China}

\author{Shinsei Ryu}
\affiliation{%
Department of Physics, Princeton University, Princeton, New Jersey, 08544, USA}

\author{Kotaro Tamaoka}
\affiliation{%
Department of Physics, College of Humanities and Sciences, Nihon University, Sakura-josui, Tokyo 156-8550, Japan}

\author{Mao Tian Tan}
\affiliation{Asia Pacific Center for Theoretical Physics, Pohang, Gyeongbuk, 37673, Korea}

\date{\today}

\begin{abstract}
We study quantum quench processes in (1+1)-dimensional conformal field theory (CFT) 
in which the initial thermal equilibrium (Gibbs) state
is time-evolved by spatially inhomogeneous Hamiltonians,
the so-called M\"obius and sine-square-deformed (SSD) Hamiltonians. 
We found that, when the quench is induced by the SSD Hamiltonian,
almost all the degrees of freedom are asymptotically gathered at a single point,
resulting in a point-like excitation.
This excitation, which we dub black hole-like excitation,
carries as much information as the total thermal entropy.
In contrast,
other parts of the system
approach the low-entropy (low-temperature) state at late times.
For the quench by the M\"obius Hamiltonian,
we instead found an eternal periodic oscillation of physical
quantities such as von Neumann entropy for subsystems.
When the CFT admits a holographic dual description,
the SSD quench induces a time-dependent, inhomogeneous deformation
of the bulk black hole horizon,
which, at late enough times, 
``touches'' the boundary.
Our quench setups can be used as a way to  create low-temperature states,
and, also,
simulate the formation and evaporation processes of black holes.
\end{abstract}

\maketitle


\section{Introduction}

Non-equilibrium phenomena in many-body quantum systems
are cutting-edge research topics in modern physics.
For example, thermalization is an important non-equilibrium process
where a thermal equilibrium state emerges dynamically even when the
dynamics is governed by unitary time evolution.
The celebrated eigenstate thermalization hypothesis (ETH) was put forward,
which claims that when a non-equilibrium process is complex (``chaotic'') enough,
the energy eigenstates will follow the thermal statistical distribution
\cite{PhysRevA.43.2046, Srednicki_1994}.
%
%
The final states in these processes tend to be featureless and their quantum mechanical nature, such as the presence of non-local correlations, is destroyed. The search for non-equilibrium processes that result in more interesting states is an active area of ongoing investigation \cite{2018PhRvB..98o5134T,2018arXiv180609624M,2019PhRvL.122q3401L,2018arXiv180701815H,2018NatPh..14..745T,2021arXiv210803460P,Pakrouski:2020hym,2017PhRvE..96b2153M}. Such states avoid thermalization and have potential applications to quantum computing.

Furthermore, these subjects in non-equilibrium quantum many-body systems
are intimately connected to the evaporation process of a black hole,
arguably one of the most interesting non-equilibrium phenomena
\cite{Hawking:1975vcx,Page_1993}.
Despite recent progress towards resolving the information paradox
\cite{Almheiri_2019, Penington:2019npb,
Almheiri_2020,Almheiri_2020_2,Penington:2019kki,Goto:2020wnk}, 
obtaining a full understanding of black hole evaporation remains a
far-reaching goal in quantum gravity.
While experimental simulations of black holes will improve our understanding of them, none of the proposals of experimental simulations to date \cite{PhysRevLett.46.1351,PhysRevLett.105.240401,2014NatPh..10..864S,PhysRevD.92.024043,2016NatPh..12..959S,2019Natur.569..688M,2021NatPh..17..362K,Lapierre_2020} have sufficiently simulated black hole evaporation which remains an important outstanding experimental problem.

Recently, the authors in \cite{PhysRevLett.46.1351,PhysRevLett.105.240401,2014NatPh..10..864S,PhysRevD.92.024043,2016NatPh..12..959S,2019Natur.569..688M,2021NatPh..17..362K,Lapierre_2020,
Wen_2018,
wen2018floquet,Fan_2020,Han_2020,wen2021periodically,Fan_2021, Lapierre_2020_1, Lapierre_2021, Moosavi_2021} found that in the non-equilibrium processes induced by the two-dimensional inhomogeneous CFT Hamiltonians called the M\"obius and sine-square deformed (SSD) Hamiltonians, the system can avoid evolving to the featureless state. These works provide rare examples where
 the dynamics of interacting many-body quantum systems can be solved
analytically, circumventing finite-size effects that plague the numerical studies that are the norm in this field. 

In this paper, we consider a quantum quench process \cite{Calabrese_2006,
  Calabrese_2007,
  Calabrese_2007b,
  Calabrese_2016,
  Bernard_2012,
  Bernard_2014,
  Bernard_2016,
  2005JSMTE..04..010C,
  Alcaraz_2011, 
  Nozaki_2014,
  2015arXiv150901160I,
  2019Sci...364..256L,
  2019Sci...364..260B} in which the Hamiltonian abruptly changes from a spatially homogeneous to a spatially inhomogeneous one
\footnote{
For previous studies 
on inhomogeneous quenches in CFTs, 
and in AdS/CFT, see, for example,
\cite{Sotiriadis:2008ila,Calabrese_2016,Wen:2016bzx,Dubail:2016tsc,Alba:2021eni,Horvath:2021vlx,
Ugajin:2013xxa,Balasubramanian:2013oga,Balasubramanian:2013rva,PhysRevD.96.026012,Arefeva:2017pho,DeJonckheere:2018pbi,Kudler-Flam:2020url}.}.
In particular, we take the post-quench Hamiltonian to be the
 M\"obius/SSD Hamiltonians in $2$d CFTs
\cite{Gendiar01102009,Gendiar01022010,Hikihara_2011,2011PhRvA..83e2118G,Okunishi_2016,2016PhRvB..93w5119W}
\footnote{
  In \cite{2016PhRvB..93w5119W},
  the Mobius deformation is called the regularized sine-square deformation. 
}.
To be concrete, we consider a (1+1)d CFT defined on a spatial circle of length $L$.
Its (undeformed) Hamiltonian $H_0$ is given
in terms of the energy density $h(x)$
as
$
  H_0 = \int^L_0 dx\, h(x), 
$
where $x$ is the coordinate of the spatial direction.
The Hamiltonian $H_0$ can be
deformed by introducing an envelope function $f(x)$, 
$H_0
=
\int^L_0 dx\, h(x)
\to
\int^L_0 dx\, f(x)h(x)
$
\cite{Allegra_2016,
  Dubail:2016tsc,
  SciPostPhys.3.3.019,
  2018JSP...172..353G,
  PhysRevLett.122.020201,
MacCormack_2019}.
The M\"obius Hamiltonian $H_{\theta}$ corresponds to the choice 
$f(x) =  1 - \tanh(2\theta) \cos (2\pi x/L)$, and is given by 
\begin{align}
  H_{\theta}=H_{0}
  -\frac{\tanh{(2\theta)}}{2}\left(H_{+}+H_-\right),
\end{align}
where $H_{\pm}$ are given by 
$
  H_{\pm}
  =
  \int^L_0 dx\,
  \, 
  e^{ \pm 2\pi  xi / L}\,
  h(x).
  $
The limit $\theta \rightarrow +\infty$ defines the sine-square deformation of $H_0$,
\begin{align}\label{SSDDefinition}
  H_{\theta\to +\infty}
  = \int^{L}_{0}dx ~2\sin^{2}{\left(\frac{\pi x}{L}\right)} h(x) 
  \equiv 
  H_{\text{SSD}}.
\end{align}
The envelope function of SSD has a minimum and a maximum at 
$x=X^1_f\equiv 0$ and $x=X^2_f\equiv L/2$, respectively, which we call fixed points. 
In particular, the envelope function vanishes at $x=X^1_f$.

Previous works discussed 
the quantum quenches or Floquet dynamics starting from pure initial states.
In contrast, we study the time-evolution by the M\"obius/SSD Hamiltonian starting from an initial thermal state,  
\begin{align}
\rho(0)=\frac{e^{-2\epsilon H_0}}{Z},
\quad
Z=\Tr\, e^{-2\epsilon H_0}.
\end{align}
Since the evolution is unitary,
the thermal entropy of the total system $S_{\text{thermal}}$ is conserved,
\begin{align}
\label{tot thermal ent}
S_{\text{thermal}}
=
\frac{c\pi L}{6\epsilon},
\end{align}
with the time-independent temperature $T=1/(2\epsilon)$.
Here, $c$ is the central charge of the CFT.
While nothing much seems to happen at least globally,
looking at local portions of the total system
reveals interesting dynamics induced by the inhomogeneous quantum quench.

\begin{figure}
  \begin{tabular}{cc}
    \begin{minipage}[t]{0.33\hsize}
      \centering
      \includegraphics[keepaspectratio, scale=0.024]{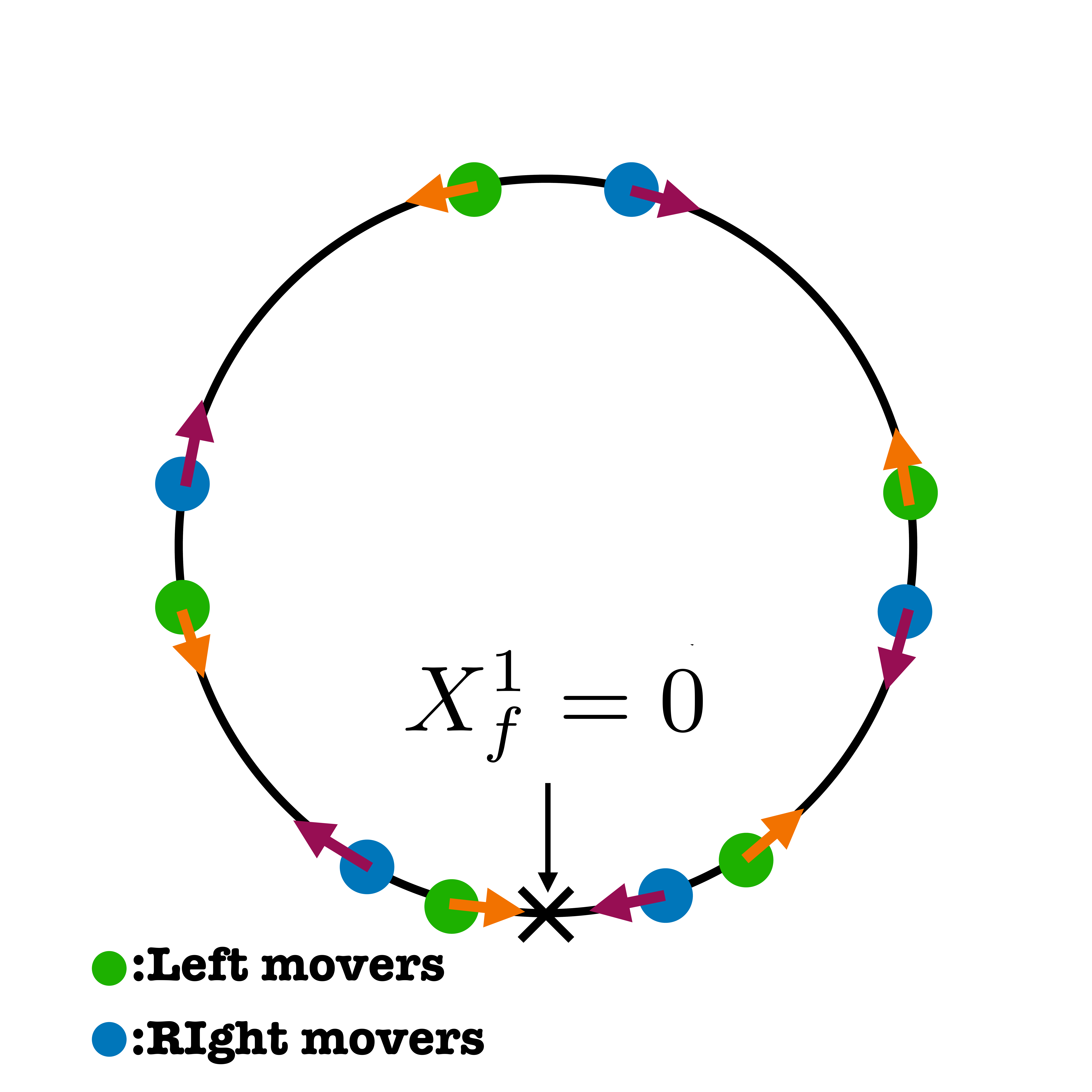}
      \\
      (a) For $0\le t < t_{*,1}$.
    \end{minipage}   
    \begin{minipage}[t]{0.33\hsize}
      \centering
      \includegraphics[keepaspectratio, scale=0.024]{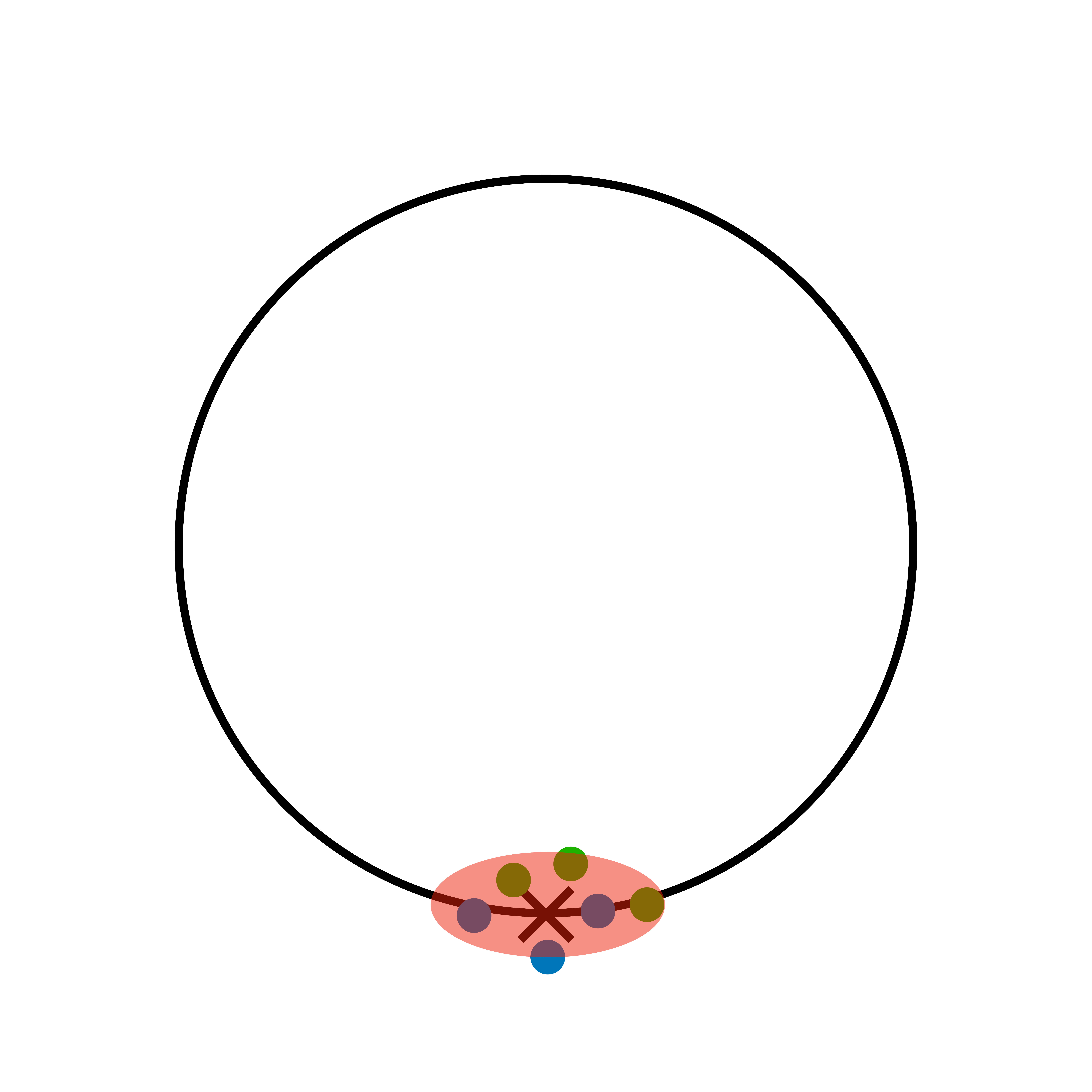}
      \\
      (b) For $t_{*,1} < t\le t_{*,2}$.
    \end{minipage} 
    \begin{minipage}[t]{0.33\hsize}
      \centering
      \includegraphics[keepaspectratio, scale=0.024]{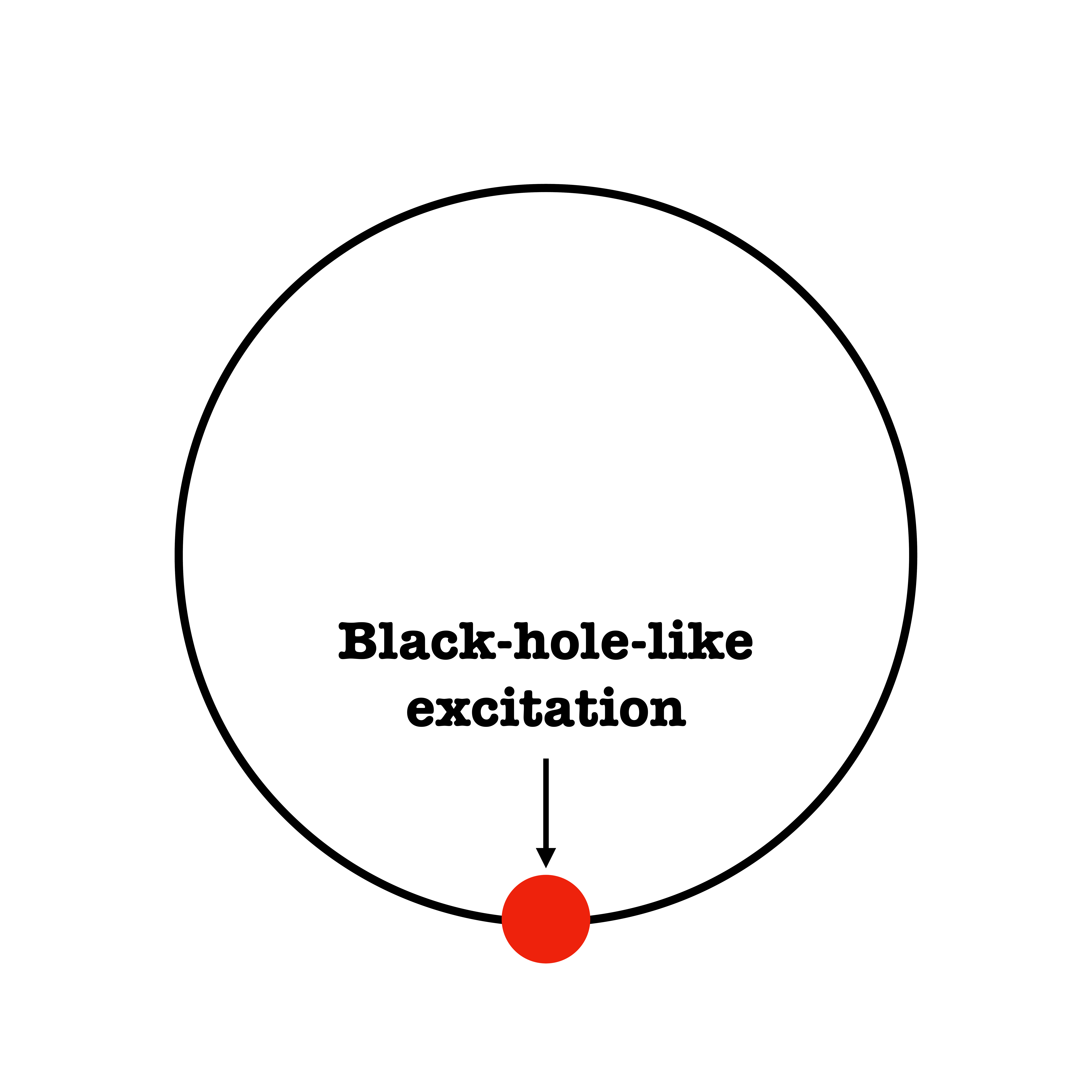}
      \\
      (c) For $t_{*,2} < t$.
    \end{minipage} 
  \end{tabular}
  \caption{A sketch of a black hole-like excitation
    created by the inhomogeneous quench in (1+1)d CFT on a ring, 
    localized around the fixed point $X^1_f=0$. 
    Panel (a) illustrates a quasiparticle picture that describes the time-dependence of the entanglement entropy in $0\le t < t_{*,1}$.
    In this picture, the left- and right-moving quasiparticles propagate to $x=X^1_f$.
    In (b), a black hole-like excitation emerges and can be thought of as a non-local object that corresponds to the red-shaded region. 
    In (c), the system is in a state where the black hole-like excitation is localized at $x\approx X^1_f$.
    }
  \label{evaporation_and_creation_of_BH_main}
\end{figure}

In this paper, we consider the entanglement entropy in holographic CFTs and a free fermion CFT. 
Holographic CFTs are known to be maximally chaotic \cite{2014JHEP...03..067S,2016JHEP...08..106M}. In \cite{2023arXiv230501019G}, even when the quantum chaotic spin chain is SSD/M\"obius deformed, level statistics, a diagnostic of quantum chaos, exhibits chaotic behavior. Thus, the SSD/M\"obius holographic Hamiltonians may have maximal chaoticity in some sense. 
On the other hand, free fermion CFT is integrable and the dynamics of entanglement entropy in this theory is well-described by a quasiparticle picture \footnote{This is true to leading order in $\frac{1}{\epsilon}$ where $\epsilon$ functions as a regulator}.
Nevertheless, the entanglement entropy for small subsystems for both theories are similar \footnote{Due to the difficulty in taking the analytic continuation $n\rightarrow1$ (the replica limit) in the free fermion CFT, we will instead consider the second R\'{e}nyi entropy which serves as a good proxy for the entanglement entropy.}. Therefore, we present mainly the holographic CFT results, mentioning the free fermion results only when it differs from the holographic CFT result. 
As we will show, 
under a M\"{o}bius quench, the entanglement entropy exhibits eternally oscillations with a period of $L\cosh{2\theta}$, breaking ergodicity. 
This is analogous to the quantum revivals studied in holographic systems in
\cite{Freivogel_2012}. This oscillation disappears in the SSD case which corresponds to the $\theta\rightarrow\infty$ limit. During the SSD time evolution, the entanglement entropies of subsystems not including $x=X^1_f$ evolve in time to the entanglement entropy of the vacuum state. On the other hand, when the subsystem includes the fixed point $x=X^1_f$, the entanglement entropy increases to the thermal entropy of the total system. This can be explained by the emergence of an excitation that resembles black holes around $x=X^1_f$.

A more refined understanding of the nature of the quantum correlations can be gleaned from the mutual information. Unlike the entanglement entropy, we find that the mutual information between two subsystems evolves to the mutual information of the vacuum state for both holographic CFTs and free fermion CFTs, even if one of the subsystems includes the fixed point $x=X^1_f$. This suggests the SSD time evolution endows the featureless state with vacuum non-local correlations. 
This may serve as a new quantum quench that cools the subsystems and endows them with non-local correlations even if the undeformed Hamiltonian is maximally chaotic.\footnote{Earlier works on the cooling by using the quantum quenches are \cite{2016arXiv161104591Z,2018PhRvL.120u0604A,2019PhRvB..99j4308M}. In these works, the authors studied the dynamics by using the universal quantities.}.
Furthermore, two time scales, $t_{*,i=1,2}$, characterize the time-dependence of entanglement entropy for a subsystem that includes $x=X^{1}_f$ for holographic CFTs. 
These time scales may depend on the sizes of the system and the subsystem as well as $\epsilon$. We will describe them later 
(see Secs.\ \ref{Section:vneandbkle} and \ref{Section:q-p}).
In the early time regime $0\le t \le t_{*,1}$, the evolution of entanglement entropy is explained by the propagation of quasiparticles to $x=X^1_f$ (see Sec.\ \ref{Section:q-p}). This is followed by an intermediate time regime $t_{*,1} < t < t_{*,2}$ where a non-local excitation with as much information
as the total thermal entropy emerges in a sub-region that includes $x=X^{1}_f$ (Fig.\ \ref{evaporation_and_creation_of_BH_main}). The total information of the $1+1$d system appears to be holographically encoded in a $0+1$d point which is reminiscent of black holes \cite{PhysRevD.7.2333,Bekenstein:1972tm,Bardeen:1973gs,Hawking:1975vcx}, so we call this excitation 
a black hole-like excitation.
At late enough times, when $t_{*,2} \le t$, 
the black hole-like excitation is localized at $x\approx X^{1}_f$. In this late time regime, 
the density matrix can be approximated as
\begin{align}
  \label{frho}
  \rho \approx \rho_{\mathcal{V}} \otimes
  \mathrm{Tr}_{\mathcal{V}}\left(\ket{0} \bra{0}\right),
\end{align}
where $\mathcal{V}$ is a subsystem that includes the origin.
Here, the von Neumann entropy of $\rho_{\mathcal{V}}$
is the total thermal entropy and $\mathrm{Tr}_{\mathcal{V}}\left(\ket{0} \bra{0}\right)$ is the reduced density matrix of the vacuum state (ground state).

\begin{figure*}
  \centering
  \includegraphics[scale=0.5]{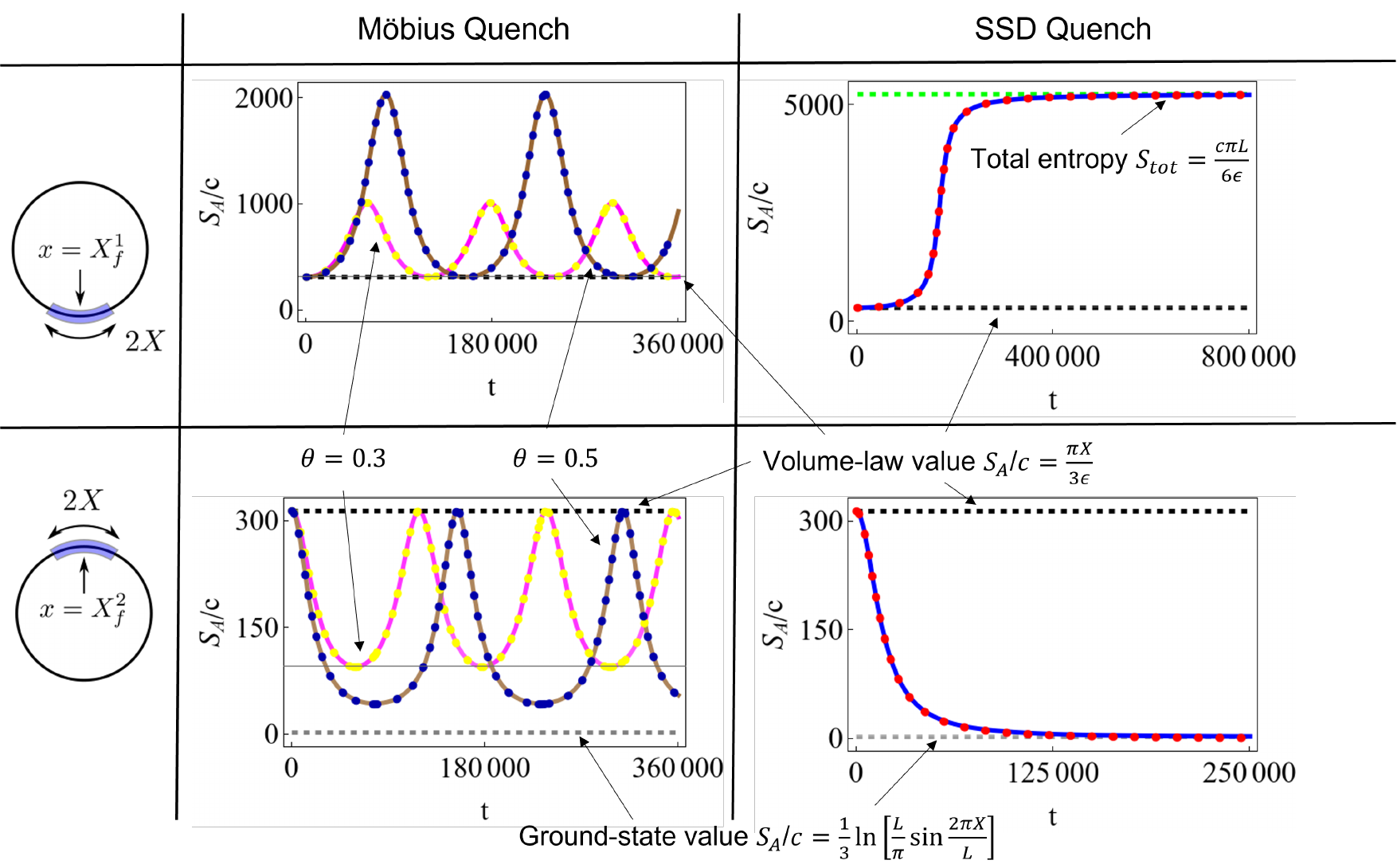}
  \caption{
  The time evolution of von Neumann entropy after 
  the M\"{o}bius (second column) and SSD (third column) quench
  for the subsystems centered around 
  $x=X^1_f$ (second row),
  $x=X^{2}_f$ (third row). The total system size is $L=100000$ and the subsystem size is $2X=6000$ while the regulator has been set to $\epsilon=10$.
  The continuous curves correspond to the holographic entanglement entropy while the dotted lines are the entanglement entropy prediction from the quasiparticle picture.
  \label{time_ee_main}}
\end{figure*}

  
When the CFT admits holographic dual descriptions,
we found 
gravity duals of these systems evolved with $H_{\theta}$ and $H_{\text{SSD}}$. 
From the behavior of these gravity duals,
we found that the periodic behavior of 
the system under the evolution by $H_{\theta}$ is due to the periodic deformation of the black hole horizon.
In contrast, under the evolution by $H_{{\text{SSD}}}$,
the black hole horizon does not oscillate,
but instead has two spikes appearing and touching the asymptotic boundary as $t\to \infty$.
In this sense,  the black hole-like excitation
is in fact an avatar of the bulk black hole.


%
%

\section{Time-dependence of von Neumann entropy and black hole-like excitation \label{Section:vneandbkle}}


The quantum dynamics can be studied by two different pictures  -- the Schr\"odinger and Heisenberg pictures.
Adopting the Schr\"odinger picture, let us begin by computing the time-dependent density matrix explicitly.
In CFT, the regular and M\"obius Hamiltonians 
form an ${\it sl}(2, \mathbb{R})$ algebra. 
(Some details are presented in
Appendix \ref{Time-evolution by Mobius and SSD quench in 2d CFT}.)
By making use of this algebraic structure, 
when $\theta < +\infty$,
the time-dependence of the density matrix can be computed explicitly as
$
\rho(t) = Z^{-1}e^{- 2\epsilon H_0(t)},
$
where
\begin{align}
\label{transformed X}
&
H_0(t) +\frac{2\pi}{L} \frac{c}{12}
 \nonumber \\
&=
\left[ 
\cosh^2(2\theta)
-
\sinh^2(2\theta)
\cos (\Omega t) 
\right]
\left(H_0 +\frac{2\pi}{L}\frac{c}{12}
\right)
\nonumber \\
&\quad
-
\cosh (2\theta)
\sinh (2\theta)
[1- \cos(\Omega t)] 
\frac{1}{2}(H_++ H_-)
\nonumber \\   
&\quad
+ \sinh(2\theta) \sin(\Omega t)  
\frac{i}{2}(H_+ - H_-).
\end{align}
From here, we immediately observe that the system exhibits eternal oscillation.
The periodicity of the oscillation is 
\begin{align}
  \label{frequency}
    \frac{2\pi}{\Omega}
    =
    L\cosh{2\theta}.
\end{align}
The oscillatory behavior after the M\"obius quench
can be understood from  the discrete energy spectrum of the M\"obius Hamiltonian
with the level spacing given by $\sim \Omega$ 
\cite{Okunishi_2016, 2016PhRvB..93w5119W}.
One may then wish to take the SSD limit $\theta \to \infty$,
but it turns out this is a bit subtle:
At the fixed point $x\sim X_f^1$,
the limits $t\to \infty$ and $\theta \to \infty$
do not commute.
We will come back to the Schr\"odinger picture analysis later 
when we analyze the holographic dual description. 
For now, we switch to the Heisenberg picture, which turns out to be
more convenient 
to study the dynamics for generic $\theta$.

Instead of following the time-dependence of the density matrix $\rho(t)$,
we can follow the time-dependence of correlation functions
$\mathrm{Tr}\,
\left[
\mathcal{O}_1(X_1) \mathcal{O}_2(X_2) \cdots  \rho(t)
\right]$
adopting the Heisenberg picture.
In our problem, the time evolution in the Heisenberg picture can be tracked by using
a conformal map (maps).
This allows us to study the time-dependence of various observables,
including von Neumann entropy (mutual information)
(The details of computation are reported in Supplementary Material \ref{Time dependence of operators}. ).
This formalism applies to CFT of any kind.
For presentational simplicity,
in the following, we will focus on
a CFT with a gravity dual (holographic CFT).
We also studied free fermion CFT where the R\'{e}nyi entropy can be computed via bosonization \cite{Herzog2013}.
We will comment on the theory-dependence (i.e., holographic v.s. free fermion CFTs) when necessary.

Let us first look at the von Neumann entropy for subregions. 
Since there is no translation symmetry in our inhomogeneous quenches,
the von Neumann entropy $S_A$ depends both on the size of subregion $A$ and its location.
In the following, we will work with the following two choices of subsystem $A$:
\begin{align}
\begin{split}
A =\begin{cases}
\left\{x\big{|}0 \le x \le X, L-X \le x \le L \right\} & \text{Case (a)} \\
\left\{x\big{|}\f{L}{2}-X \le x \le \f{L}{2}+X\right\} & \text{Case (b)} \\
\end{cases}.
\end{split}
\end{align}
In Case (a), the center of subsystem $A$ is $X^1_f$,
one of the fixed points, and in Case (b) the center is the other fixed point $X^2_f$.
(We also studied other cases, e.g., when the center of subsystem $A$ is the midpoint between $X^1_f$ and $X^2_f$.
$A=\left\{x\big{|}\f{L}{4}-X \le x \le \f{L}{4}+X\right\}$.
Mostly, this case is similar to Case (b) --  see Supplementary Material \ref{Entanglement entropy for single intervals}.)

Let us first study the M\"obius quench with $\theta<\infty$ (Fig.\ \ref{time_ee_main}).
We find that, in all cases, the von Neumann entropy oscillates in time with the periodicity
${2\pi}/{\Omega} = L \cosh{(2\theta)}$,
starting from the volume-law value
\begin{align}
  S_{A}(t=0) \approx {c\pi X}/{3\epsilon}
  \equiv
  S_{{\rm vol}},
\end{align}
in agreement with the analysis in the Schr\"odinger picture.
(Here $2X$ is the size of the subsystem.)
When $\theta$ is sufficiently large, in Case (a), the von Neumann entropy oscillates between the initial value
$S_{{\rm vol}}
$
and the total thermal entropy
$S_{\text{thermal}}
$.
On the other hand, in Case (b) where the subsystem is centered around $X^2_f$
(and once again when $\theta$ is sufficiently large),
the von Neumann entropy oscillates between the initial value
$S_{\text{thermal}}
$
and the ground state value
\begin{align}
S_A = (c/3) \log [(L/\pi)\sin (2\pi X/L)] \equiv  S_{{\rm area}}.
\end{align}

Let us now move on to the SSD limit.
The main difference from the M\"obius quench is the absence of oscillations in the SSD quench.
Plotted in Fig.\ \ref{time_ee_main} is the time evolution of $S_A$ for the setup of Case (a) in the SSD limit.
Once again, $S_A$ is given initially by the von Neumann entropy of the thermal state,
$S_{A}(t\approx 0) \approx S_{{\rm vol}}$.
As time goes by, $S_A$ grows with time.
In the time interval $t_{*,2}>t \gg t_{*,1}$,
$S_A$ can be approximated by the thermal entropy of the total system,
$S_{A}(t_{*,2}>t \gg t_{*,1}) \approx 
{c\pi L}/{6\epsilon}= S_{\text{thermal}},$
which is independent of the subsystem size.
At a sufficiently late time $t > t_{*,2}$, the sub-leading term of $S_A$ is approximately equal to $S_{\text{area}}$. This suggests that the system may evolve to the asymptotic state in (\ref{frho}) according to the equation of motion given by the SSD Hamiltonian. 
The characteristic time $t_{*,1}$ can be estimated
by using the quasiparticle picture or by
directly inspecting the operator evolution while $t_{*,2}$ can be estimated by
directly inspecting the holographic result. 

\begin{figure}
  \includegraphics[keepaspectratio, scale=0.09]{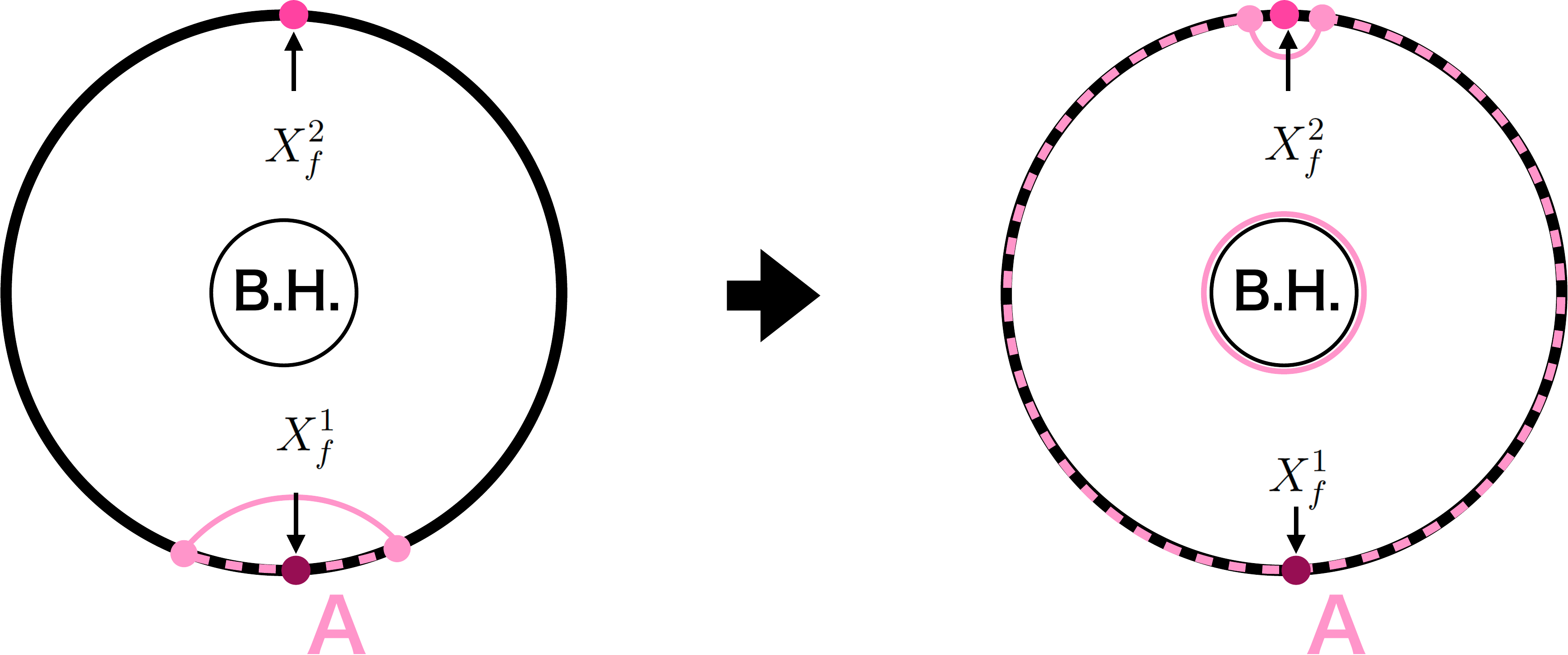}
  \\
  \includegraphics[keepaspectratio, scale=0.09]{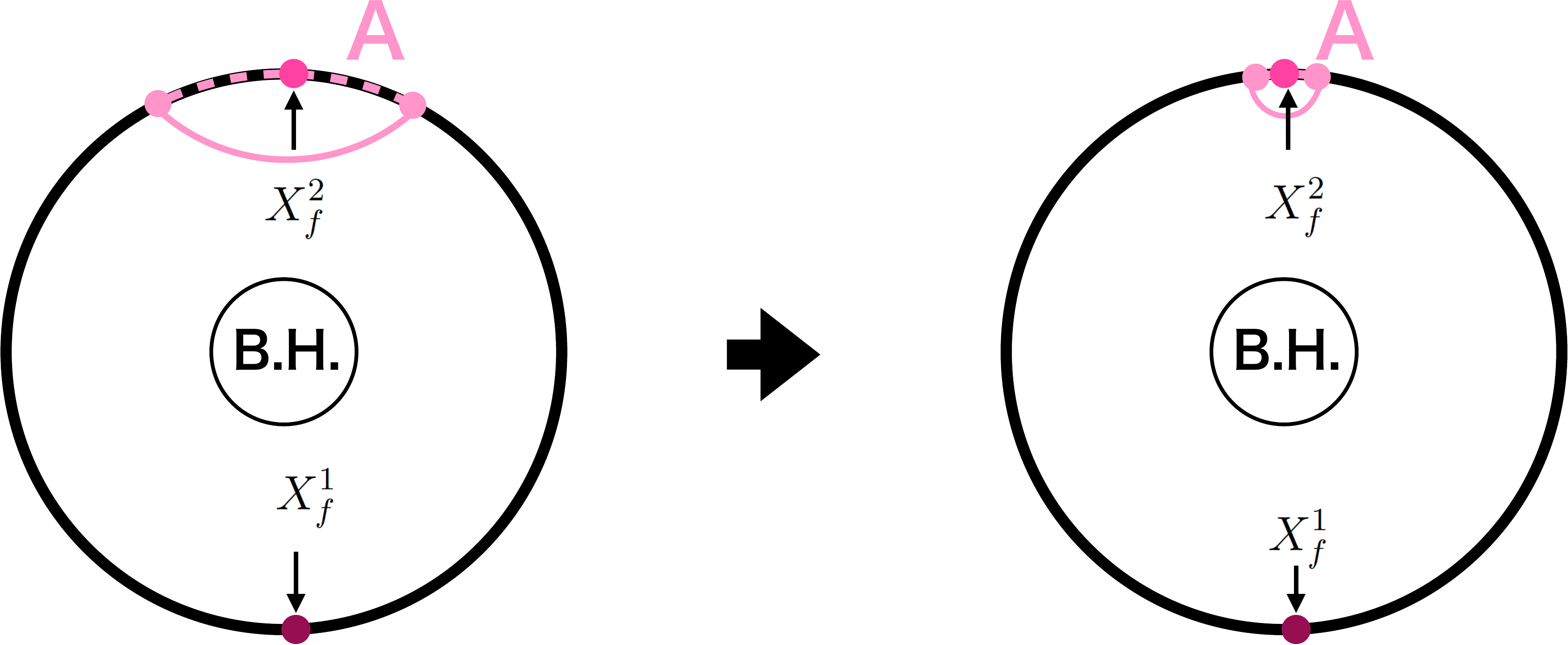}
  \caption{
    The time evolution of the geodesic in the Heisenberg picture 
    for Case (a) (Top) and Case (b) (Bottom).
    In Case (a), we note that, due to the homology condition 
    of the holographic entanglement entropy, the geodesic encircles the black hole
    at late times.
    \label{EE_Heisenberg}}
\end{figure}

The details of the quasiparticle picture and the estimation of $t_{*,1}$ will be discussed in the next section. 
Let us define $t_{*,1}$ as the time for $S_A$ to become half of the thermal entropy of the whole system.
In the Heisenberg picture, subsystem $A$ follows the evolution of the twist and anti-twist operators, expanding or shrinking depending on the location of these operators. 
The time $t_{*,1}$ is approximately equal to the time for the size of $A$ to become half of the whole system.
Either way,  if the size of the subsystem is sufficiently small, $\epsilon \ll 2X \ll L$,
$t_{*,1}$ is inversely proportional to the subsystem size $2X$,
and given by
$t_{*,1}\approx \frac{L^2}{2 \pi ^2 X}$.
In the gravitational bulk, let us define two types of geodesics $\mathcal{L}_{A,1}$ and $\mathcal{L}_{A,2}$ as the surfaces that enclose and do not enclose the black hole respectively. 
Let us define $t_{*,2}$ as the time for the length of $\mathcal{L}_{A,2}$ to be equal to the length of $\mathcal{L}_{A,1}$.
In the late-time region $t>t_{*,2}$, the minimal surface is given by $\mathcal{L}_{A,1}$. 
The time-dependence of $S_A$ can be understood from the 
evolution of the minimal surface (geodesic)
in the Heisenberg picture
(Fig.\ \ref{EE_Heisenberg}(a)).
The asymptotic behavior of $S_A$ for $t>t_{*,2}$ can be understood in terms of these geodesics.
The leading order contribution to the entanglement entropy 
$S_{{\rm thermal}}$ is given by the length of the geodesic enclosing the black hole while a sub-leading contribution to the entanglement entropy $S_{{\rm area}}$ is by a geodesic that connects the edges of the subsystem.

%

On the other hand, in Case (b), the von Neumann entropy decreases monotonically
(Fig.\ \ref{time_ee_main}),
since the geodesic becomes smaller with time
(Fig.\ \ref{EE_Heisenberg}).
The von Neumann entropy asymptotically approaches
the vacuum entanglement entropy \cite{Calabrese_2004,Calabrese_2009}
after a sufficient time has passed,
$S_{A}(t\rightarrow \infty) \approx S_{{\rm area}}$.
The von Neumann entropy, for the cases when the subsystem does not contain $x=X^1_f$,
thus undergoes a crossover from the volume-law to area-law entanglement entropy.

To summarize, when $A$ is centered around the fixed point $x=X^1_f$,
 for large $t$, the leading term of $S_A(t)$ saturates to $S_{\text{thermal}}$ 
independent of the subsystem size and the sub-leading term is asymptotically equal to $S_{\rm area}$, 
while when subsystem $A$ does not include the fixed point $x=X^1_f$,
$S_A(t)$ is well approximated by the entanglement entropy of the vacuum state
at late enough times.
These indicate that, at late enough times, almost all quantum degrees of freedom (entropy) are concentrated at the fixed point $x=X^1_f$:
At the fixed point $x=X^1_f$, a local excitation with as much information as
compatible with thermal entropy emerges
(Fig.\ \ref{evaporation_and_creation_of_BH_main}).
The high entropy state at the fixed point is ``holographic'' in the sense that 
the zero-dimensional fixed point carries the entire entropy 
of the system -- effectively, the system is reduced to a point
\cite{Susskind:1994vu,tHooft:1993dmi}.
This behavior is reminiscent of a black hole:
In quantum gravity theory, in the low-energy limit, almost all the degrees of
freedom are localized on the surface of a black hole \cite{articlePhysRevD., Hawking:1975vcx, Gibbons:1976ue}.
This similarity leads us to call the excitation concentrated at $x=X^1_f$ a
black hole-like excitation.
After this black hole-like excitation emerges at $x=X^1_f$,
the von Neumann entropy is well approximated by \eqref{frho}.
We argue that by the SSD quench we can simulate the formation process of black holes
in which the black hole-like excitation emerges.
As we will see below, the analogy between the high entropy state
at the fixed point and a black hole can be sharpened in holographic CFTs.
%

%


\section{The quasiparticle picture \label{Section:q-p}}

\begin{figure}
    \centering
    \includegraphics[width=0.35\textwidth]{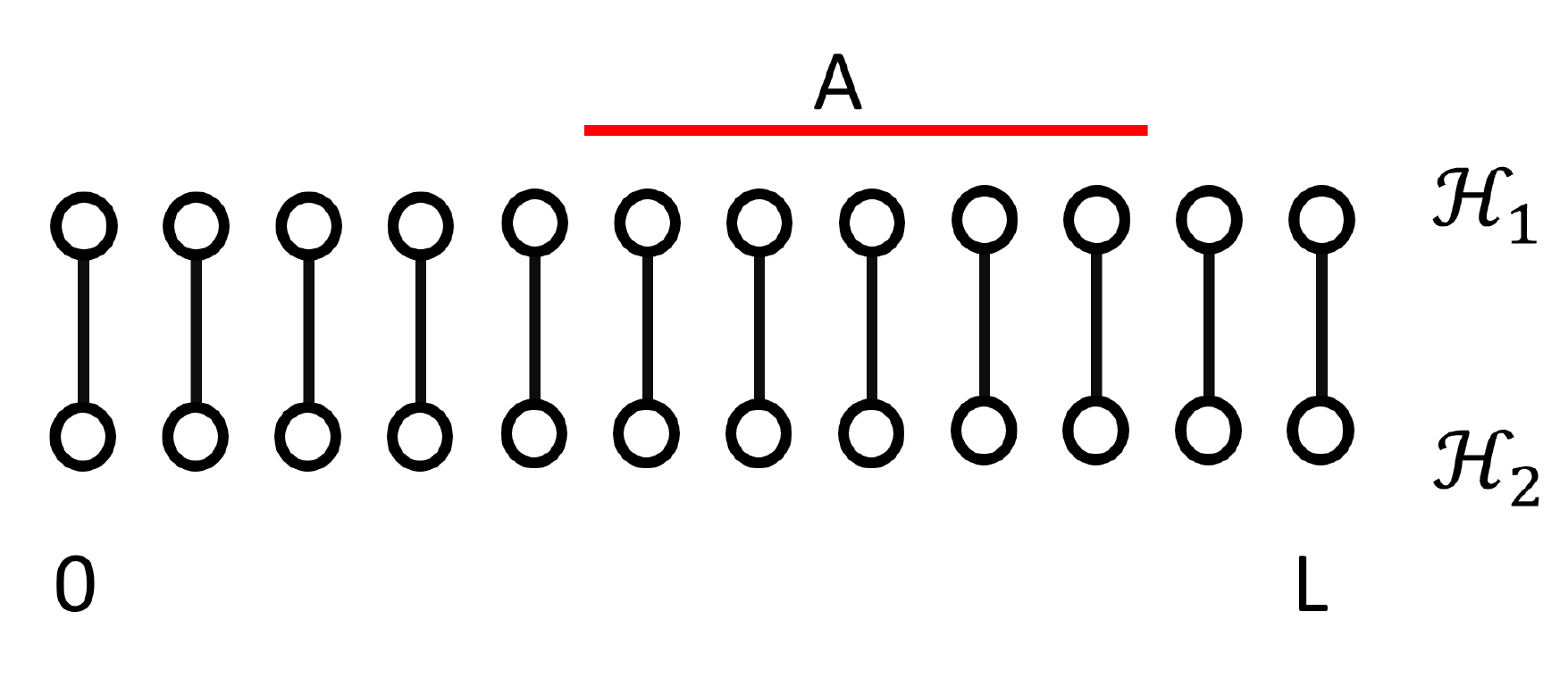}
    \includegraphics[width=0.35\textwidth]{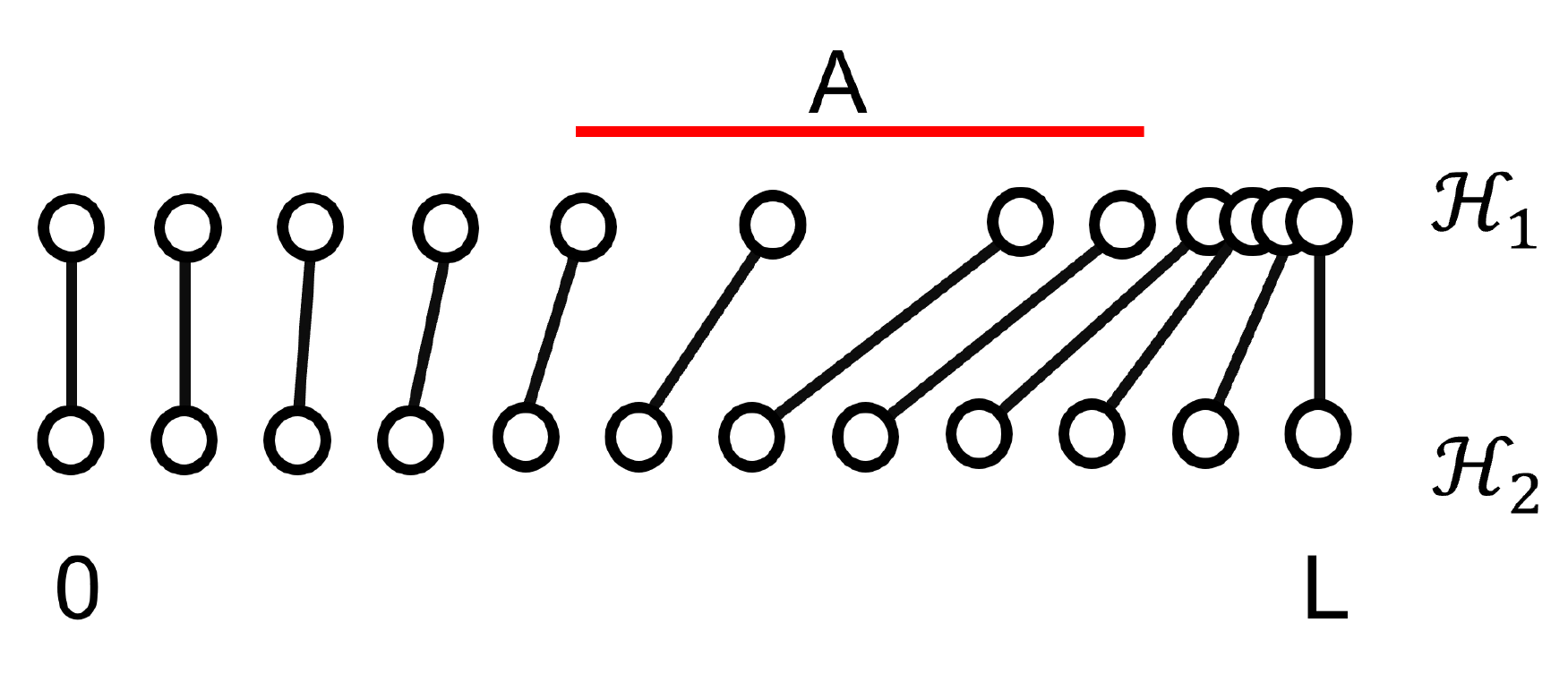}
    \caption{Plots of the right-moving bell pairs at the initial time (left) and at a later time (right)}
    \label{ThermalStateBellPairConfiguration}
\end{figure}

The R\'{e}nyi entanglement entropy associated with the reduced density matrix for a single interval turns out to be well-described by a quasiparticle picture, in line with expectations for an integrable theory. The origin of the quasiparticles comes from the purification of the density matrix via the introduction of a second Hilbert space
\begin{equation}
    e^{-i(H_\theta\otimes \mathbb{I})t}|\text{TFD}\rangle = \mathcal{N}e^{-i(H_\theta\otimes \mathbb{I})t} \sum_E e^{-\epsilon E} |E\rangle_{\mathcal{H}_1} |E\rangle^*_{\mathcal{H}_2}
\end{equation}
where $|E\rangle$ is an eigenstate of the uniform Hamiltonian $H_0$, 
$\mathcal{H}_1$ and $\mathcal{H}_2$ are the original and replicated Hilbert spaces respectively, and the normalization constant is related to the original partition function by $|\mathcal{N}|^2 = Z$. 
The second Hilbert space has also been CPT conjugated. In the limit where $\epsilon\rightarrow 0$, the state $|\text{TFD}\rangle$, which is also known as the thermofield double state, may be approximated by a product of Bell pair
\begin{equation}
    |\text{TFD}\rangle = \prod_x |\text{Bell}_x \rangle_L\otimes | \text{Bell}_x\rangle_R
\end{equation}
where $|\text{Bell}_x \rangle_i$ for $i=L,R$ is a Bell pair with one qubit in each Hilbert space located at spatial position $x$. Since the time evolution operator only acts on the first Hilbert space, the qubits in the first Hilbert space move with a velocity $v(x)=\pm f(x)$  while the qubits in the second Hilbert space remain stationary. The qubit belonging to the Bell pair $|\text{Bell}_x\rangle_R$ moves to the right while the qubit belonging to $|\text{Bell}_x\rangle_L$ moves to the left. Note that the left and right-moving modes are completely independent of one another. Plots of the right-moving Bell pairs before and after M\"{o}bius/SSD evolution are shown in Fig.\ \ref{ThermalStateBellPairConfiguration}. 
The configuration of the left-moving Bell pairs is given by a left-right reflection of the configuration of the right-moving Bell pairs.

Since the entanglement entropy of subsystem $A$ is proportional to the number of Bell pairs shared between $A$ and its complement, we only need to keep track of the qubits in the first Hilbert space to predict the entanglement entropy. These qubits are the quasiparticles that carry the information in the CFTs considered. It is worth pointing out that this picture holds in the limit where $\epsilon\rightarrow 0$ and only describes correlations between the two Hilbert spaces and not correlations within an individual Hilbert space. A detailed calculation of the number of quasiparticles contained in a subsystem $A=[X_2,X_1]$ can be found in appendix \ref{QuasiparticlePictureDetails}, and the resulting expression for the operator entanglement entropy is
\begin{equation}
    S_A(t) = \rho_0 
    \sum_{i=L,R} 
    \text{mod}
    \left[x_{0,i}(X_1,t)-x_{0,i}(X_2,t),L\right],
\end{equation}
where $x_{0,i}(X,t)$ is the initial position of a quasiparticle located at position $X$ at time $t$ and $i=R/L$ indicates the chirality of the quasiparticle. The physical interpretation of this formula is clear; the quasiparticles in the subsystem $[X_2,X_1]$ at time $t$ were initially in the interval $[x_{0,i}(X_2,t),x_{0,i}(X_1,t)]$.

The holographic entanglement entropy is plotted along with the quasiparticle prediction in Fig.\ \ref{time_ee_main} and is seen to be in excellent agreement. 
One small difference is that the holographic entanglement entropy of a subsystem centered about the midpoint $x=L/2$ decays to the ground state value of a 2d CFT under the SSD quench while the quasiparticle picture prediction decays towards zero since almost all the quasiparticles will be trapped near the fixed point $X^1_f$.

\section{Holographic picture}

\subsection{Deformed bulk horizon}

\begin{figure*}[tbp]
  \centering
  \includegraphics[keepaspectratio, scale=0.2]{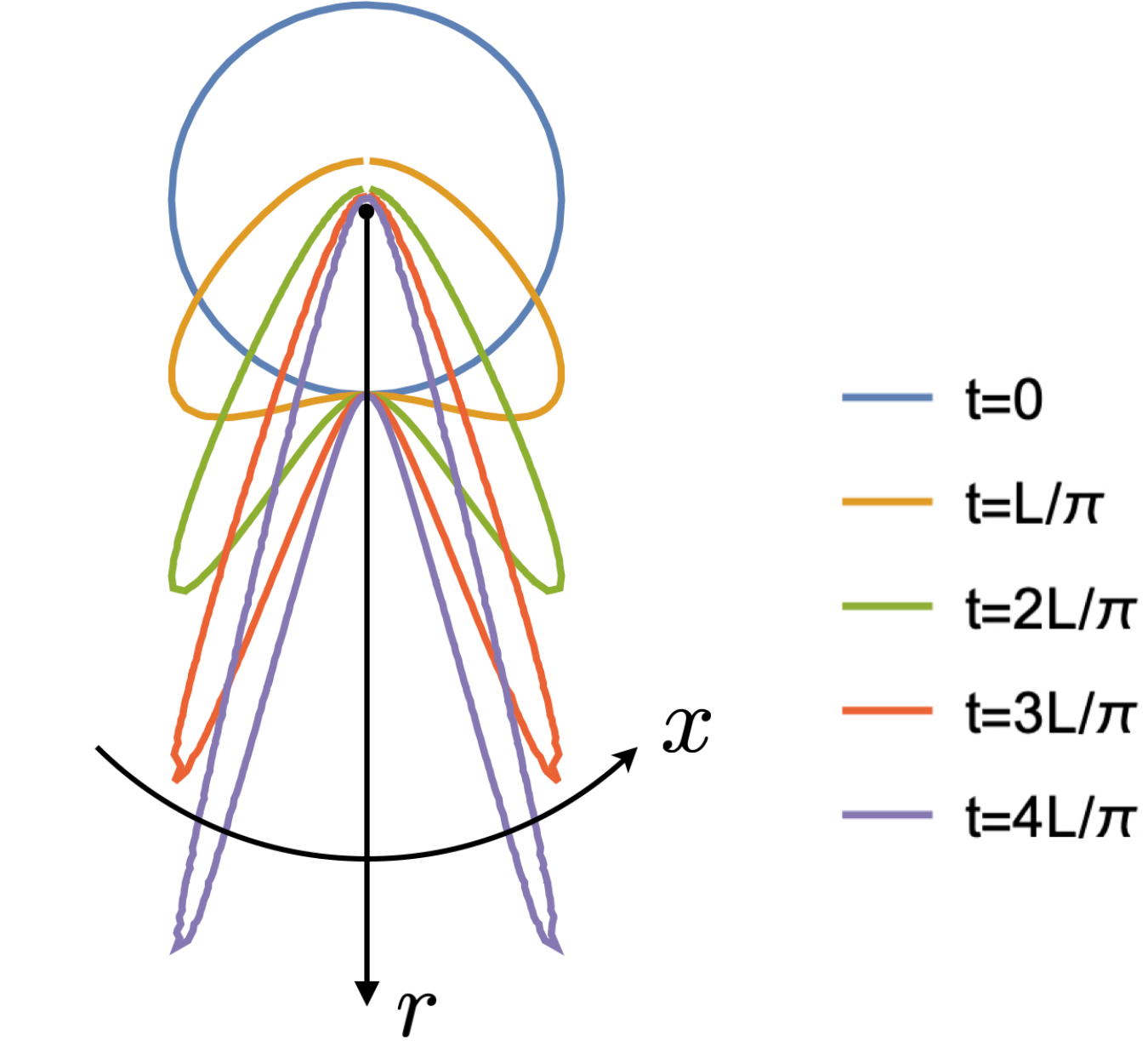}
  \hspace{1cm}
  \includegraphics[keepaspectratio, scale=0.3]{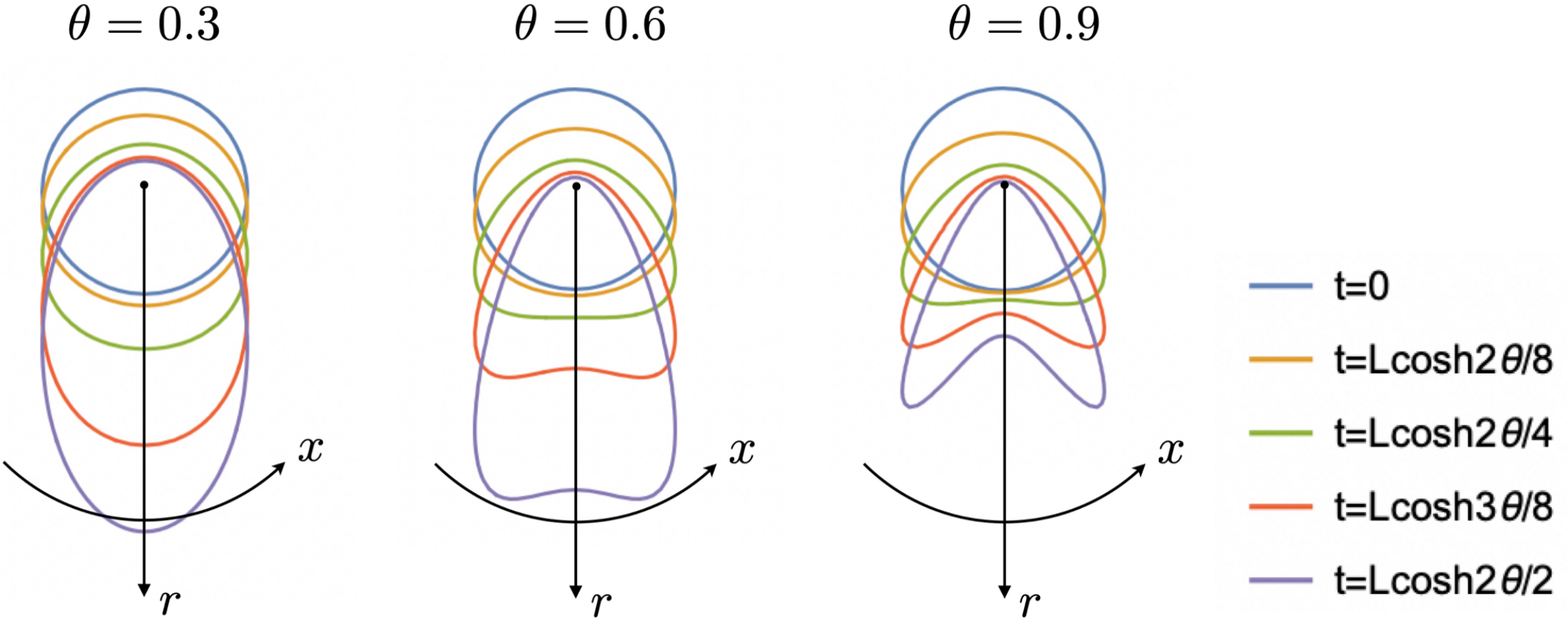}
  \caption{The time dependence of the
    bulk black hole horizon profile. Left: SSD Right: M\"{o}bius. We set $L=2\pi$.
    \label{horizonSSD2}}
\end{figure*}

In the Heisenberg picture,
the time evolution of the von Neumann entropy can be tracked by looking at
the time-dependence of geodesics in the static bulk geometry
(AdS$_3$ with the BTZ black hole).
On the other hand, we can adopt the Schr\"odinger picture
in which the bulk spacetime is time-dependent.
As discussed in \cite{Ban_ados_1999, Roberts:2012aq}, the gravitational dual can be constructed from the expectation value of the energy density after the quantum quench.
(This is equivalent to rewriting the geometry given by the static BTZ black hole in 
the $w^{\rm new}$ and $\bar{w}^{\rm new}$ 
coordinate system in terms of the original $X$ and time $t$ which parametrize the time evolution under the SSD Hamiltonian.
Here,
the $w^{\rm new}$ and $\bar{w}^{\rm new}$ coordinates
are defined/obtained from the (Heisenberg) time evolution of operators --  
see Appendix \ref{Time dependence of operators}
for more details.) 
The details of the calculations are presented in
Appendix \ref{shroginder_geometry}.

In Fig.\ \ref{horizonSSD2}
we plot the horizon for different values of $\theta$ in the global coordinate $(t,x,r)$
where $r$ represents the bulk radial coordinate
($r=\infty$ corresponds to the boundary).
At $t=0$, we have a circular horizon of the BTZ black hole. 
For $t>0$, the horizon is deformed by the quantum quench.
For the SSD quench $\theta \to \infty$, 
the center of the mass of the black hole moves towards the fixed point $X^1_f=0$ on the boundary.
Furthermore, the profile of the black hole horizon becomes highly non-circular.
It develops two peaks or ``spikes'' that stretch/are elongated towards the boundary.
These peaks on the boundary appear as the corresponding peaks in the energy-density profile.
In $t\to \infty$, the spikes merge and asymptotically touch the boundary --
a black hole-like excitation emerges at the boundary.
In this sense, the black hole-like excitation
can indeed be identified with the horizon.
For $0 < \theta < \infty$, 
the horizon is similarly deformed by the quantum quench. 
Compared to the SSD case, however,
the horizon exhibits an eternal oscillation.

\subsection{Time evolution of bulk excitations in the SSD/M\"obius quench \label{Sec:evolution-of-bulkexc}}

It is well known that the expectation value of the energy-momentum tensor for high-energy eigenstates can be well approximated by the expectation value for thermal states. 
(See Appendix \ref{sec:stress}.) 
%
Now we consider the time dependence of the bulk local excitation that corresponds to the energy eigenstates created by inserting spinless primary operators ${\cal O}_{h,\bar{h}}$
with conformal dimensions $h=\bar{h}$ on the vacuum state.
The details of analysis are described in Appendix \ref{App:evolution-of-bulkexc}.
It is known that a primary operator ${\cal O}_{h,\bar{h}}$
with large conformal dimension $h>\frac{c}{24}$ creates a black hole in AdS with temperature
$T=\frac{1}{2\pi}\sqrt{\frac{24h}{c}-1}$
while one with $c\gg h$ corresponds to a small bulk excitation on the pure AdS can be created by the bulk matter field dual to ${\cal O}_{h,\bar{h}}$.
In this section, we consider how this bulk excitation moves under the SSD and M\"{o}bius Hamiltonians. We consider a state
\begin{align}
    |\psi_{{\cal O}_{h,\bar{h}}}(t)\rangle=e^{-i H_{\text {SSD}} t}{\cal O}_{h,\bar{h}}(z=0,\bar{z}=0)|0\rangle\, ,\label{SSDquenched-main}
\end{align}
 where ${\cal O}_{h,\bar{h}}$is inserted at the center of the Euclidean plane, i.e., $\tau=-\infty$ in Euclidean time,
The strategy that we will use here is summarized in \cite{Goto:2017olq}, where they move the bulk excitation by the corresponding bulk $SL_2$ generators. Here we focus on SSD and describe the case of the M\"obius quench in Appendix \ref{App:evolution-of-bulkexc}.
As described in detail in (\ref{PoincaretoSSD}), one can show that the SSD Hamiltonian in the boundary global coordinate $(w,\overline{w})$ is equivalent to the uniform Hamiltonian in the Poincar\'e coordinate related to the global coordinate as $\left(iz_P,i\overline{z}_P\right)=\left(L\cot{\left(i \pi w/L\right)},L\cot{\left(i \pi \overline{w}/L\right)} \right)$, i.e.,
    $H_{P} =\frac{1}{2\pi}H_{\rm SSD}+\frac{c}{12L}$.
Therefore, the SSD Hamiltonian generates the time-flow in the Poincar\'e
coordinate. 
This indicates that a black hole (or the
bulk excitation) dual to the state (\ref{SSDquenched-main}) moves along the
Poincar\'e time direction.
Static objects in the bulk Poincar\'e coordinate are seen as ones falling to the
asymptotic boundary of the AdS over an infinitely long time from the perspective
of the boundary observer in the global coordinate. 
Thus the black
hole (or the bulk excitation) 
gets closer and closer to the AdS boundary during the time evolution by the SSD Hamiltonian. 
 
 To check this, let us consider the time evolution of  the profile of the bulk excitation corresponding to the SSD quenched state $|\psi_{{\cal O}_{h,\bar{h}}}(t)\rangle$ through the overlap with the state
$|\phi_{{\cal O}_{h,\bar{h}}}(\zeta, x_P)\rangle=\phi_{{\cal O}_{h,\bar{h}}}(\zeta, x_P)|0\rangle
$
whose excitation is localized at the bulk point $(\zeta, x_P)$ in the AdS, where $\zeta$ is the bulk direction in the Poincar\'e coordinate.

\begin{figure}[t!!]
\begin{center}
  \includegraphics[width=4cm]{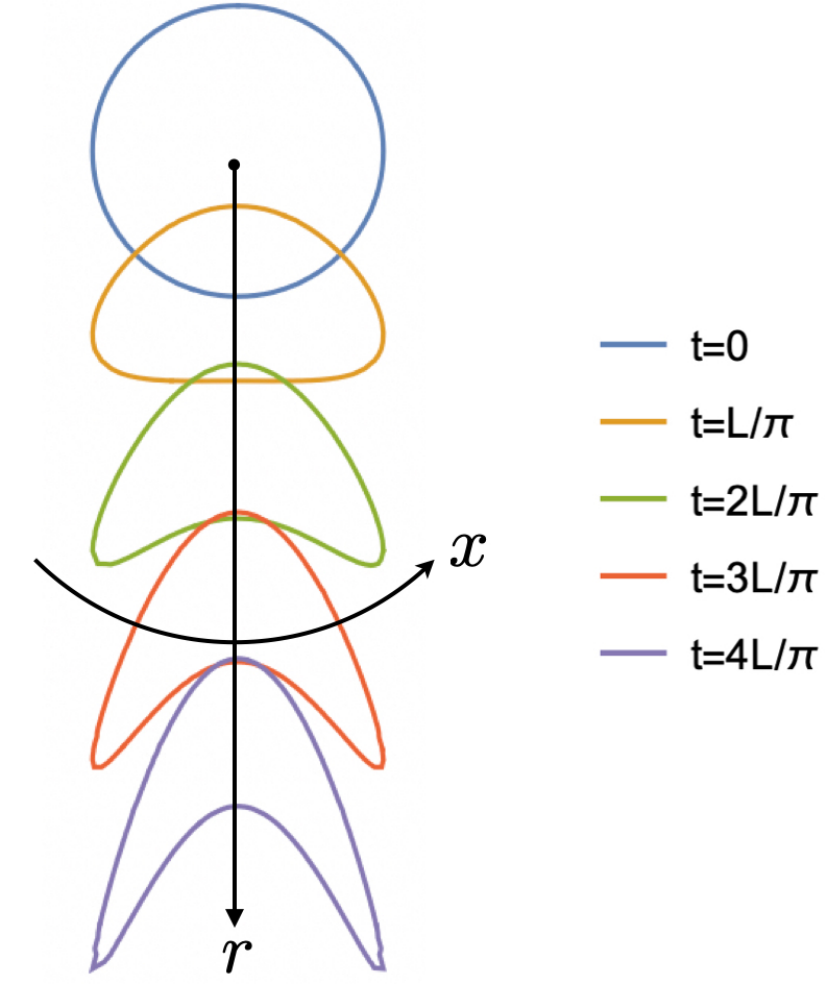}
  \caption{Contours corresponding to $|\langle\phi(r, x)|\psi_{{\cal O}_{h,\bar{h}}}^{\text {SSD}}(t)\rangle|=1$ for several values of $t$ depicted in the global coordinate. Here, we set $L=2\pi$ and $h=2$.
By properly shifting the center of each excitation using the AdS isometry, the contours nicely match those for the black hole horizon Fig.\ \ref{horizonSSD2} .}\label{BTZSSD}
\end{center}
\end{figure}

Notice that when ${\cal O}_{h,\bar{h}}$ is inserted at the origin $\tau_P=x_P=0$ in the boundary Poincar\'e coordinate, the overlap is just given by the usual bulk-to-boundary propagator
\begin{align}
    \left\langle\phi(\zeta, x_P)|{\cal O}_{h,\bar{h}}\left(\tau_{P}=0,x_P=0\right)\right\rangle=\frac{\zeta^{2 h}}{\left(\zeta^{2}+x_P^{2}\right)^{2 h}}\, .
\end{align}
Before the SSD quench, the primary operator ${\cal O}_{h,\bar{h}}$ sits at the origin of the Euclidean global coordinate $z=\bar{z}=0$, which corresponds to $z_P=\bar{z}_P=L$ in the Poincar\'e coordinate. Since the SSD Hamiltonian simply generates the Poincar\'e time flow,  
the SSD quenched state $|\psi_{{\cal O}_{h,\bar{h}}}^{\text {SSD }}(t)\rangle$ can be obtained by inserting the primary operator at $z_P=L-it,\bar{z}_P=L-it$. In the Lorentzian regime obtained by $\tau_P\rightarrow \tau_P-it_P$, the complex coordinate becomes $z_P=\tau_P-i(t_P-x_P), \bar{z}_P=\tau_P-i(t_P+x_P)$.
Therefore, we can regard the operator as being inserted at a complex time $t_p=t+iL$ in the Poincar\'e coordinate.
A simple modification to the bulk-to-boundary propagator yields
\begin{align}
   \langle\phi(\zeta, x_P)|\psi_{{\cal O}_{h,\bar{h}}}^{\text {SSD }}(t)\rangle=\frac{\zeta^{2 h}}{\left(\zeta^{2}+x_P^{2}-(t+iL)^2\right)^{2 h}}\, ,\label{overlap}
\end{align}
for the SSD quenched state $|\psi_{{\cal
    O}_{h,\bar{h}}}^{\text {SSD }}(t)\rangle$.  We plot the contours corresponding to $|\langle\phi(\zeta, x_P)|\psi_{{\cal O}_{h,\bar{h}}}^{\text {SSD }}(t)\rangle|=1$ for several values of $t$ in Fig.\ \ref{BTZSSD}.
As we expected, the bulk excitation approaches the fixed point as time evolves. Moreover, by properly shifting the center of the excitations using the AdS isometry, the contours nicely match those for the black hole horizon Fig.\ \ref{horizonSSD2}.

\section{Mutual information and finer structure of the late time density matrix} 
\label{Finer structure  of the late time density matrix} 

\begin{figure}[tbp]
  \centering
  \includegraphics[scale=0.3]{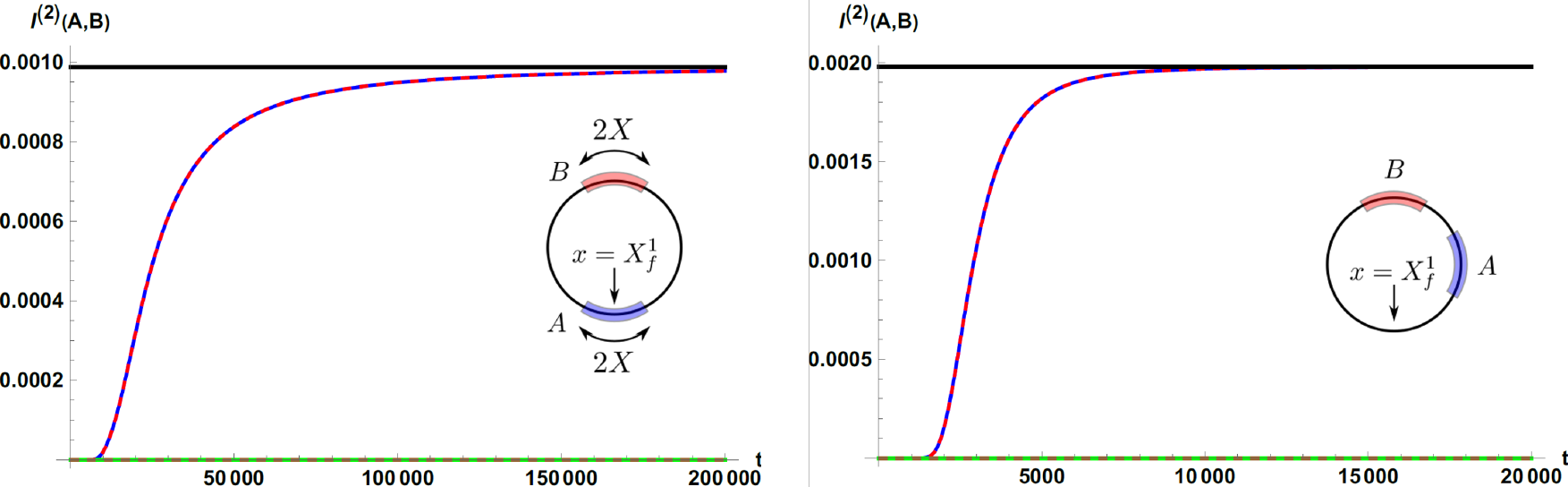}
  \caption{The time evolution
    of the (R\'enyi) mutual information
    for the free fermion CFT
    for two different configurations
    of the intervals $A$ and $B$.
    The black lines indicate
    the ground state value of the mutual information.
    \label{MI}
  }
\end{figure}

We now study the mutual information for two intervals ($A$ and $B$),
$I(A,B):= S_A + S_B - S_{A\cup B}$,
which can provide more information on the density matrix.
Unlike the von Neumann entropy for a single interval,
the mutual information depends on the details of the CFT
beyond the central charge
\cite{Furukawa_2009,Asplund:2015eha}.
We recall that in mutual information
leading order contributions
in $S_A$, $S_B$, $S_{A\cup B}$ 
cancel with each other.
Hence, the subleading (sub-extensive) terms in the von Neumann entropy contribute
to mutual information.
Here, we consider two kinds of CFTs, the free fermion CFT with $c=1$
and holographic CFT in the large $c$ limit. 
These represent two classes of dynamics --
the integrable dynamics that can be described by
the quasiparticle picture
\cite{2005JSMTE..04..010C},
and the quantum information scrambling dynamics
that can effectively be described by the
membrane picture
\cite{2017PhRvX...7c1016N,
2018PhRvX...8b1013V,
2018arXiv180300089J,
2018PhRvD..98j6025M,
  Nie:2018dfe, Kudler-Flam:2019wtv, Kudler-Flam:2020yml}.
Some calculation details can be found in Appendix \ref{Mutual information}.

Plotted in Fig.\ \ref{MI} is the time evolution
of the mutual information in the free fermion CFT
for two representative configurations of the intervals. 
For both configurations,
we found the mutual information between the two intervals
approaches to the ground state value at late times.
For the case when both intervals do not include
the fixed point $x=X^f_1$, this behavior confirms 
the late time approximation \eqref{frho}.
On the other hand,
even when 
one of the subsystems includes the fixed point $x=X^f_1$ while the other does not, 
the mutual information is still given by
the ground state value (Fig.\ \ref{MI}).
This behavior is not explained by
the leading order late time approximation \eqref{frho}.
Thus, beyond the leading order in $1/\epsilon$,
the late time density matrix 
deviates from \eqref{frho}.
Put differently, the above consideration shows that  
our state acquires (quantum) correlations
by the SSD evolution:
At high enough temperatures, the initial state $\rho(0)$ 
has very little quantum correlations, 
$\rho_{A\cup B} \approx \rho_A \otimes \rho_B$,
while at late enough times, 
the non-zero mutual information
suggests that 
$\rho_{A\cup B} \neq \rho_A \otimes \rho_B$,
i.e., 
a separable reduced density matrix can become entangled.

The mutual information can also be computed for
holographic CFT (using the Heisenberg picture mentioned above).
We confirmed that the late time mutual information is
given by the ground state value for the holographic case as well.  
Based on these results for the two types of CFTs, 
we conjecture that the same behavior can be found in any CFTs.

\section{Discussion and outlook}

We have studied inhomogeneous quantum quench using
the M\"obius and SSD Hamiltonians in (1+1)d CFT.
In the SSD quench, at late enough times,
a black hole-like excitation emerges at the fixed point 
with as much information as the total thermal entropy,
i.e., 
the density operator can be approximated by \eqref{frho}.
Before closing, we will further discuss our findings
in the M\"obius/SSD quench.

%

\paragraph{Simulation of formation and evaporation of a black hole}

Our setup can be readily realized in recent experimental platforms, such as IBM's online quantum computing platform known as IBM Q. A simple quench by the XX and XXZ spin chain Hamiltonians on this platform has been done in \cite{Smith2019}.
The flexibility of this system would allow us to create
inhomogeneous quantum many-body Hamiltonians with the M\"obius or SSD deformation.
Since we are interested in studying quantum dynamics in both integrable and quantum chaotic spin chains, a more appropriate spin chain to study would be the mixed-field Ising model
\begin{equation}
    H = -\sum_i J_i Z_i Z_{i+1}-\sum_i g_i X_i-\sum_i h_i Z_i
\end{equation}
where $X_i$ and $Z_i$ are Pauli matrices, $J_i$ is a nearest neighbour coupling, and $h_i$ and $g_i$ are magnetic fields. This model is integrable for certain choices of the parameters $J_i$, $h_i$, and $g_i$ and is chaotic for other choices of the parameters \cite{PhysRevLett.106.050405}. To make this an SSD Hamiltonian, simply set the values of the parameters according to \eqref{SSDDefinition}. Following \cite{Smith2019}, the unitary time evolution can be discretized as $e^{-iHt}=\left(e^{-iH\Delta t}\right)^M$ where $\Delta t = t/M$ and $M$ is the number of Trotter steps. Each discrete Trotter step can be approximated by a product of single-qubit and two-qubit gates
\begin{equation}\label{Trotterization}
    e^{-iH\Delta t}= \prod_i e^{i J_i Z_i Z_{i+1}\Delta t}\prod_i e^{i h_i Z_i\Delta t} \prod_i e^{i g_i X_i\Delta t}+\mathcal{O}(\Delta t^2)
\end{equation}
This can be implemented on IBM Q which is capable of implementing arbitrary single qubit gates as well as the CNOT gate. The two-qubit gates in \eqref{Trotterization} can be implemented with CNOT gates and single-qubit gates using the optimal decomposition in \cite{PhysRevA.69.032315}. Many of our findings can then be directly tested in experiments in principle.
In particular, the formation (and destruction) of a black hole-like excitation,
which has much resemblance with the formation and evaporation of a black hole,
can be tested in the lab. Consider, for example, the SSD quench process, which collects the degrees of freedom
to create a black hole-like excitation at the origin. This
can be thought of as a process of creating a black hole. 
In the holographic picture, near the origin,
the bulk horizon asymptotically approaches the boundary,
and hence the black hole ``expands''
from the point of view of a local subregion near the origin.
(Contrary, for subsystems not including the origin, the black hole ``shrinks'').
We can also consider the time-reversal of these processes,
i.e., $e^{+ i t H_{\mathrm{SSD}}}$ instead of $e^{- i t H_{\mathrm{SSD}}}$, 
where the black hole shrinks near the origin
and expands for regions away from the origin. 
Thus, our SSD quench can be used to test/simulate 
the formation and evaporation of a black hole
in the experimental systems mentioned above.
In this interpretation, the von Neumann entropy of
a subsystem including the fixed point $x=X^1_f$ 
is interpreted as the entanglement entropy
between late-time radiation and the black hole.
On the other hand,
for a subsystem not including the fixed point $x=X^1_f$ 
the von Neumann entropy
(at late times) is interpreted
as the entanglement entropy of early-time radiations
\footnote{
We note that the evolution by the SSD/Mobius Hamiltonian itself  
gives rise to Killing horizons:
Choosing these generators as the time-evolution operator corresponds
to choosing a particular Killing vector,
which may result in the presence of a Killing horizon
\cite{1999CQGra..16..363A}.
In the SSD limit, we have an extremal Killing horizon.
Killing horizons may also emerge effectively
by periodically driving the system using these generators,
i.e., at the level of Floquet Hamiltonians
\cite{Lapierre_2020_1,Lapierre_2020}.
Based on this observation, in 
\cite{Lapierre_2020_1,Lapierre_2020},
it was proposed that the optical lattice system can be used to simulate how excitations propagate in the space-time where a black hole exists.
In contrast, in this work, 
a non-equilibrium process in which a black hole itself emerges or evaporates can be simulated by exciting a thermal equilibrium state with an inhomogeneous quench.
In the holographic picture, we have a bulk black hole horizon
(in contrast to the Killing horizons)
since our initial state is a thermal state. 
%
}.

\paragraph{Measurement-induced transition}

We found the crossover from the volume- to area-law entanglement
for the subsystem not including the fixed point $x=X^1_f$.
This reminds us of the measurements-induced transition
in monitored quantum circuits 
\cite{Li_2018,Chan_2019,Skinner_2019,Li_2019,Chen_2020}.
Instead of introducing measurements,
in our setup, we control the amount of dissipation
by acting with the unitary $e^{ -i t H_{\mathrm{SSD}}}$.
In the holographic dual language, 
the unitary deforms the horizon of the BTZ black hole
and controls locally the distance between the horizon and
the boundary (the origin). 
We also note that in the volume-law phase of
the monitored quantum circuits
there is a sub-extensive term (logarithmic term)
in the entanglement entropy
that reflects non-trivial 
quantum error-correcting properties
\cite{Gullans:2019zdf,2021PhRvX..11a1030I,2021PhRvB.103j4306L,Yoshida:2021haf}. 
As discussed above,
we note that 
our late time state after the SSD
quench also exhibits a finer structure
-- in addition to the leading contribution
of the von Neumann entropy indicating
the formation of a black hole-like excitation,
there are subleading, logarithmic 
contributions
that contribute to the saturation value
of mutual information and
indicate the deviation from 
\eqref{frho}.
Investigating further the 
properties of the fine structure,
and its possible connection to quantum-error-correcting
properties, 
is an interesting future direction.

\paragraph{Creation of a low-entropy (low-temperature) state by local measurements}
From a slightly practical point of view,
our SSD quench protocol
can be used to heat/cool a particular local region of the system. 
Furthermore, 
for $t \gg t_{*,2}$, once a black hole-like excitation is created
at the origin,
it may be interesting to
``remove'' the black hole-like excitation
to cool the entire system.
This may be achieved by turning off the coupling between
the origin and the rest of the system. 
It is also interesting to 
perform a projective measurement at the origin:
If we perform the projective measurement \cite{Chitambar_2014} by the state of
$\ket{\Psi}_A$ in subsystem $A$, the state
transitions from
\eqref{frho}
to
$
  \rho \rightarrow \rho' =\ket{\Psi}_A \bra{\Psi}_A \otimes
  \mathrm{Tr}_A \ket{0}\bra{0}.
$
The entropy after this measurement is given by
\begin{align}
  S_{\text{thermal}}\approx S_A= \frac{c}{3}\log{\left[
  \frac{L}{\pi}
  \sin{\left(\frac{\pi l}{L}\right)}\right]},
\end{align}
which is at most a quantity of $\mathcal{O}(1)$
($l$ is the size of subsystem $A$). 
In this way, the SSD time evolution combined with
a local projection measurement induces a transition from a
high-entropy (high-temperature) to low-entropy (low-temperature) state. 



\begin{figure}[tbp]
 \begin{center}
  \includegraphics[width=80mm]{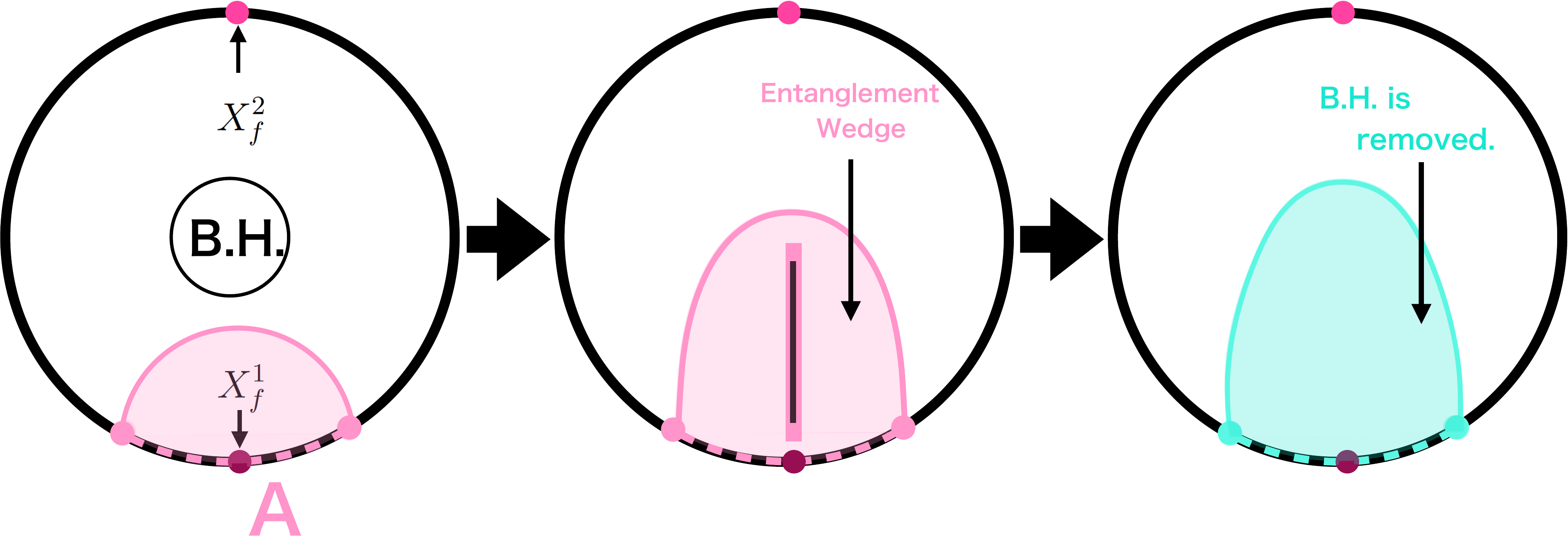}
 \end{center}
 \caption{
   The time evolution of a black hole after the SSD quench
   (left and middle)
   and subsequent projective measurement (right)
   in the Schr\"odinger picture.
   (Time flows from the left to right.)
   In the right image the black hole in the entanglement wedge is
   removed by the projective measurement.}
 \label{projection_measurement}
\end{figure}

This transition
can be also discussed in the gravity dual in the Schr\"odinger picture
(Fig.\ \ref{projection_measurement}).
At $t=0$, there is a spherically symmetric black
hole with its center at the origin of ${\it AdS}$ spacetime.
The SSD time evolution deforms its shape for $t>0$.
After enough time has passed, $t\gg t_{*,2}$, the black hole is deformed into
a black brane-like shape extending from the origin of
${\it AdS}$ spacetime to its boundary near $x=X^1_f$.
As a result, the black hole will be included in the bulk region that is 
dual to subsystem $A$ containing the fixed point $x=X^1_f$,
i.e., 
the entanglement wedge of $A$
\cite{Czech_2012, Wall_2014, Headrick:2014cta, Jafferis:2014lza}.
After this, the black hole can be removed by projective measurements in 
subsystem $A$, and the measurement-induced phase transition from the BTZ black
hole to the ``almost'' thermal ${\it AdS}_3$ can occur.


It may also be possible to transfer energy between the system
(the system quenched by the SSD evolution) and the observer
performing the projective measurement.
The energy for the total composite system (including both the CFT and the observer)
should be conserved immediately before and after the measurement.
The difference between the energy just before and after the measurement
$
\Delta E
\approx
\int_A dx \,\Tr \left(T_{00}(x) \rho_{\text{B.H.}}\right)
-
\int_A dx \bra{\Psi}_A T_{00}(x) \ket{\Psi}_A
$
is the energy ``extracted'' by the measurement.
The observer can thus obtain a large amount of energy of $\mathcal{O}\left({1}/{\epsilon}\right)$.


%

\paragraph{Outlook}

As we demonstrated, by controlling the inhomogeneity of the system during dynamics, 
interesting non-equilibrium many-body quantum states can be realized. 
Many of our findings can be directly tested by
recent experimental platforms for quantum simulators.
In particular, we proposed that the formation and evaporation processes of
a black hole can be simulated in the SSD quench.
The SSD quench can also be used as a method to create a low-temperature state.
Finally, we close by listing some future directions.

First of all, our study in this paper is limited to (1+1)d CFT.
It would be interesting to study inhomogeneous quenches
in a wider class of systems, i.e.,
those that are not described by (1+1)d CFT,
such as lattice spin systems away from critical points.
In particular, we studied holographic CFTs as one of our examples,
that exhibit strong quantum information scrambling
\cite{Liu:2013iza,Liu:2013qca,Asplund:2013zba,Caputa:2014vaa,Asplund:2014coa,Asplund:2015eha,Hosur:2015ylk,2017PhRvX...7c1016N,2018arXiv180300089J,Nie:2018dfe,Kudler-Flam:2019wtv,Kudler-Flam:2020yml,Kudler-Flam:2020xqu}.
It would be interesting to study other (non-CFT)
systems that also exhibit quantum information scrambling,
such as the chaotic quantum spin chain \cite{Ba_uls_2011},
and see the effects of inhomogeneity.
It would be also interesting to
study other types of dynamics, e.g., those that break ergodicity,
such as many-body localizing dynamics
\cite{Basko_2006, Serbyn_2013, Huse_2014, Nandkishore_2015},
and those that exhibit quantum many-body scars
\cite{2017Natur.551..579B,
2018PhRvB..98o5134T,
2018arXiv180609624M,
ChengLin2019,
Ho:2018rum,2018NatPh..14..745T,
2021arXiv210803460P,2021arXiv210900548M}.
Even within the context of (1+1)d CFT,
the effects of the inhomogeneity
on quantum information scrambling should be studied,
by looking at various ``indicators'',
such as the level statistics, spectral form factor, and out-of-time-order correlators.

Furthermore, integrable and chaotic dynamics are described by
different effective descriptions,
the quasiparticle and membrane (line-tension) pictures
\cite{2017PhRvX...7c1016N,
2018PhRvX...8b1013V,
2018arXiv180300089J,
2018PhRvD..98j6025M}.
It is interesting to study how one can use
these effective descriptions in the presence of inhomogeneity.
In the current work, we are able to describe
many (but not all) dynamical behaviors using the quasiparticle picture.
The time evolution of von Neumann entropy for a subsystem
not including a fixed point $x=X^{1}_f$ cannot be described
by the quasiparticle picture for late times.
It is thus interesting to construct an effective theory
that can describe this regime where the quasiparticle picture is invalid. 
It is also interesting to understand the mechanism by which the quasiparticle picture breaks down.


Second, it would be interesting to study a wider class of
inhomogeneous time-evolution operators.
For example, \cite{Fan_2021} studied the dynamics of inhomogeneous Hamiltonians
with an arbitrary smooth envelope function that can have more than one fixed point.
Adapted to our setup,  we expect that such dynamics can create
multi black hole-like excitations. 
This may allow us to construct the experimental systems that can simulate the process of
two (more than one) merging black holes, 
and one black hole splitting into several black holes.
Also, by engineering inhomogeneity,
we may be able to create
different kinds of non-equilibrium steady states,
for example, those that support a steady thermal gradient.
Another possible extension of the current work is to consider Floquet dynamics 
(and find its gravitational dual). For recent works involving these inhomogeneous quenches, see \cite{nozaki2023inhomogeneous, mao2024local}.

\section*{Acknowledgements}

We thank useful discussions with 
Jonah Kudler-Flam,
Shuta Nakajima,
Tokiro Numasawa,
and
Tadashi Takayanagi.
K.G.~is supported by JSPS KAKENHI Grant-in-Aid for Early-Career Scientists (21K13930) and Research Fellowships of Japan Society for the Promotion of Science for Young Scientists (22J00663).
M.N.~is supported by JSPS Grant-in-Aid for Early-Career Scientists 19K14724. 
K.T.~is supported by JSPS Grant-in-Aid for Early-Career Scientists 21K13920. This material is based upon work supported by the National Science Foundation under Grant No. NSF-DMR 2018358 and by an
appointment to the YST Program at the APCTP through the Science and Technology Promotion Fund and Lottery Fund of the Korean Government, as well as the Korean Local Governments -
Gyeongsangbuk-do Province and Pohang City (MT).
S.R.~is supported by the National Science Foundation under 
Award No.\ DMR-2001181, and by a Simons Investigator Grant from
the Simons Foundation (Award No.~566116).
This work is supported by
the Gordon and Betty Moore Foundation through Grant
GBMF8685 toward the Princeton theory program.

\begin{widetext}

\appendix

\section{The time evolution of the density matrix}
\label{Time-evolution by Mobius and SSD quench in 2d CFT}

Let us first calculate
$
\rho(t)=
e^{-i t H_{\theta} }
\rho(0)
e^{+i t H_{\theta} }
$
directly.
To this end,
we recall that the M\"obius Hamiltonian
is written in terms of the Virasoro generators
$\{L_{0,\pm 1},\bar{L}_{0,\pm 1}\}$
as
\begin{align}
  H_0
  &=\oint\frac{dw}{2i\pi}T(w)
  +\oint \frac{d\bar{w}}{2i\pi}\bar{T}(\bar{w})
    =\frac{2\pi}{L}
    \left(
    L_0 + \bar{L}_0
    \right)
    -\frac{\pi c}{6 L},
  \nonumber \\
  H_{\pm}
  &=\oint\frac{dw}{2i\pi}e^{\pm\frac{2\pi w}{L}}T(w)
    +\oint\frac{d\bar{w}}{2i\pi}e^{\mp\frac{2\pi \bar{w}}{L}}\bar{T}(\bar{w})
  =
    \frac{2\pi}{L} (L_{\pm 1}+\bar{L}_{\pm 1}),
\end{align}
where we introduce the complex coordinates, 
$w=\tau+ix$ and $\bar{w}=\tau-i x$,
with $\tau$ and $x$ coordinatizing
the (Euclidian) temporal and spatial directions,
respectively,
and
$T(w)$ and $\bar{T}(\bar{w})$ are the holomorphic and anti-holomorphic
parts of the energy-momentum tensor.
These generators form
the ${\it sl}(2,\mathbb{R})$ algebra,
\begin{align}
  &
  X = L_0,
    \quad
    Y= \frac{1}{2}(L_{-1}-L_{+1}),
  \quad
    Z= \frac{1}{2}(L_{-1}+L_{+1}),
    \nonumber \\
  &
  [X,Y]=Z,
  \quad
  [X,Z]=Y,
  \quad 
  [Z,Y]=X.
\end{align}
This algebraic structure allows us to compute
$
\rho(t)=
e^{-i t H_{\theta} }
\rho(0)
e^{+i t H_{\theta} }
$
explicitly.
For the presentational simplicity,
let us focus on the holomorphic sector only. 
Then, for the M\"{o}bius Hamiltonian
$H_{\theta} = 
(2\pi/L) (X-c/24) 
-
\tanh(2\theta)
(2\pi/L) Z
$, 
it is straight forward to show
%
(see \cite{2020JHEP...04..094C} for a similar calculation)
\begin{align}
\label{transformed X}
  \tilde{X}
  &=
e^{-it H_{\theta} }
X
e^{+it H_{\theta} }
    \nonumber \\
&=
\left[ 
\cosh^2(2\theta)
-
\sinh^2(2\theta)
\cos (\Omega t) 
\right]X
\nonumber \\
&\quad
-
\cosh (2\theta)
\sinh (2\theta)
[1- \cos(\Omega t)] Z
- \sinh(2\theta) \sin(\Omega t) i Y.
\end{align}

Thus, the state oscillates with
the frequency $\Omega$ defined in \eqref{frequency}.
The oscillatory behavior after the M\"obius quench
can be understood from 
the discrete energy spectrum of the M\"obius Hamiltonian
\cite{Okunishi_2016, 2016PhRvB..93w5119W}.
The M\"{o}bius Hamiltonian,
in a proper coordinate system
$z,\bar{z}$,
can be written down by the Virasoro generator as 
$
H_{\theta}
 =
 \Omega
 \big(L_0^{\tilde{z}}+\bar{L}_0^{\overline{\tilde{z}}}
 \big)
-\frac{c\pi}{6L}.
$
\footnote{
Here, we use
the energy-momentum tensor defined in the coordinate system
$\tilde{z}$ and $\bar{\tilde{z}}$,
defined by
\begin{align}
  e^{\frac{2\pi w}{L}}
    =z
    =\frac{\cosh{\theta}\tilde{z}+\sinh{\theta}}{\sinh{\theta}\tilde{z}+\cosh{\theta}},
    \quad
    e^{\frac{2\pi \bar{w}}{L}}
    =\bar{z}
    =\frac{\cosh{\theta}\bar{\tilde{z}}+\sinh{\theta}}{\sinh{\theta}\bar{\tilde{z}}+\cosh{\theta}}.
\end{align}
}
The ``regularity'' or ``integrability''
of the energy spectrum within each tower of states
is responsible for the oscillation:
The matrix elements of the density matrix
in terms of the eigenstates $\ket{n}_\theta$
of the M\"{o}bius Hamiltonian,
$\rho_{mn}(t)
=
\bra{m}_{\theta} \rho(t) \ket{n}_{\theta}
=
e^{it (E_m-E_n)}
\bra{m}_{\theta} \rho(0) \ket{n}_{\theta}
$,
are periodic in time
within each tower of states,
since
the energy difference
$E_m-E_n$
is
an integer multiple of 
$
\Omega
$.
The periodicity of the oscillation
is set by $2\pi/\Omega$
\cite{PhysRevLett.44.1323, PhysRevA.42.6308, Haque_2014}.
In the later sections, we will investigate this oscillation more closely. 
\footnote{
Furthermore,
since the gap in the spectrum is smaller
for larger $\theta$,
$\bra{n}_{\theta}e^{-2\epsilon H_0}\ket{{n\neq m}}_{\theta}$
is larger.
As a result, the peak values of the von Neumann entropy and the two-point
correlation functions are larger
because of the larger contribution from the coherent term.
}

On the other hand,
since the system size is effectively infinite in the SSD limit, 
the periodic behavior does not occur.
Taking the SSD
limit $\theta \to +\infty$
in \eqref{transformed X},
\begin{align}
&
\cosh^2(2\theta)
-
\sinh^2(2\theta)
\cos (\Omega t) 
\to 
1+ 2\pi^2 t^2/L^2,
\nonumber \\
&
-
\cosh (2\theta)
\sinh (2\theta)
[1- \cos(\Omega t)] 
\to 
-2\pi^2 t^2/L^2,
\nonumber \\
&
- \sinh(2\theta) \sin(\Omega t) 
\to
-i 2\pi t/L.
\end{align}
Further taking the limit $t\to \infty$,
%
$
  e^{ -i t H_{\mathrm{SSD}}} H_0
  e^{+ it H_{ \mathrm{SSD}}}
  \sim  
  \frac{2\pi^2 t^2}{L^2} H_{\mathrm{SSD}}+
{\it const}.
$
We then conclude at late times,
$\rho(t)$ would be given by
$
  \rho(t)
  \sim
  e^{ - (\epsilon 2\pi^2 t^2/L^2 ) H_{\mathrm{SSD}}}.
$
When the ground state of $H_{\mathrm{SSD}}$
is 
the same as the ground state of $H_0$, which is guaranteed for CFTs with an additional Kac-Moody symmetry \cite{2012JPhA...45k5003K},
at late enough times, 
we expect $\rho(t)$ would be approximated 
by the ground state of $H_0$.
In the next section, we will confirm
this expectation by studying the von Neumann  
entropy defined for single intervals.
When the intervals do not include $x=0$, 
we will see that the von Neumann entropy
at late enough times is given by 
entanglement entropy of the ground state.

On the other hand, when 
the interval includes $x=0$, 
we will see that 
the von Neumann entropy is not given 
by the ground state value, but by the 
total thermal entropy (once again at 
late enough times);
the above expectation 
$\rho(t)\sim 
  e^{ - (\epsilon 2\pi^2 t^2/L^2 ) H_{\mathrm{SSD}}}
$
breaks down around the origin $x=0$.
We defer the detailed discussion 
for later sections.
However, we note that
taking the $t\to \infty$ limit is
somewhat subtle around the origin $x=0$.
We go back to \eqref{transformed X},
and 
look at the transformed $X$
more closely.
Recalling $(2\pi/L)L_{\pm}= \int dx e^{i 2\pi x/L}h(x)$, 
where $h(x)$ is the Hamiltonian density,
$\tilde{X}$ can be written as
$
  \tilde{X}
  =
  \int dx\,
  \tilde{f}(x) h(x),
$
with the envelope function given by 
\begin{align}
\tilde{f}(x)
&=
\frac{\pi}{L}
     \Big[
     1+\cosh (4 \theta )
     - \sinh (4
   \theta ) \cos \left(\frac{2 \pi  x}{L}\right)
   [1 - \cos (\Omega t)]
   \nonumber \\
   &
   \quad 
   \quad
   -2 \sinh^2 (2 \theta ) 
   \cos (\Omega t)
   -2 \sinh (2 \theta ) 
   \sin
   \left(\frac{2 \pi  x}{L}\right) 
   \sin (\Omega t)
   \Big].
\end{align}
For a given $x$, 
the envelop function in the SSD limit
is given by
\begin{align}
\tilde{f}(x) \to
\frac{2\pi}{L}
   \left\{
   1
   -\frac{2 \pi   t}{L} \sin \left(\frac{2 \pi x}{L}\right) 
   + \frac{2 \pi ^2 t^2}{L^2} 
   \left[1-\cos \left(\frac{2 \pi  x}{L}\right)
   \right]
   \right\}.
\end{align}
For generic $x\neq 0$, the envelop function is
quadratic in $t^2$, in agreement with the discussion above. 
On the other hand, for $x=0$,
$f(x=0) \to \frac{2\pi}{L}$
and hence we do not have $t^2$ dependence at late times. 
This indicates that 
the density operator $\rho(t)$ near the origin
should not be approximated as 
$e^{ - (\epsilon 2\pi^2 t^2/L^2 ) H_{\mathrm{SSD}}}$.

%
%

\section{Observables in the Heisenberg picture}
\label{Time dependence of operators}

\begin{figure*}[bp]
  \begin{tabular}{cc}
    \begin{minipage}[t]{0.5\hsize}
      \centering
      \includegraphics[keepaspectratio, scale=0.1]{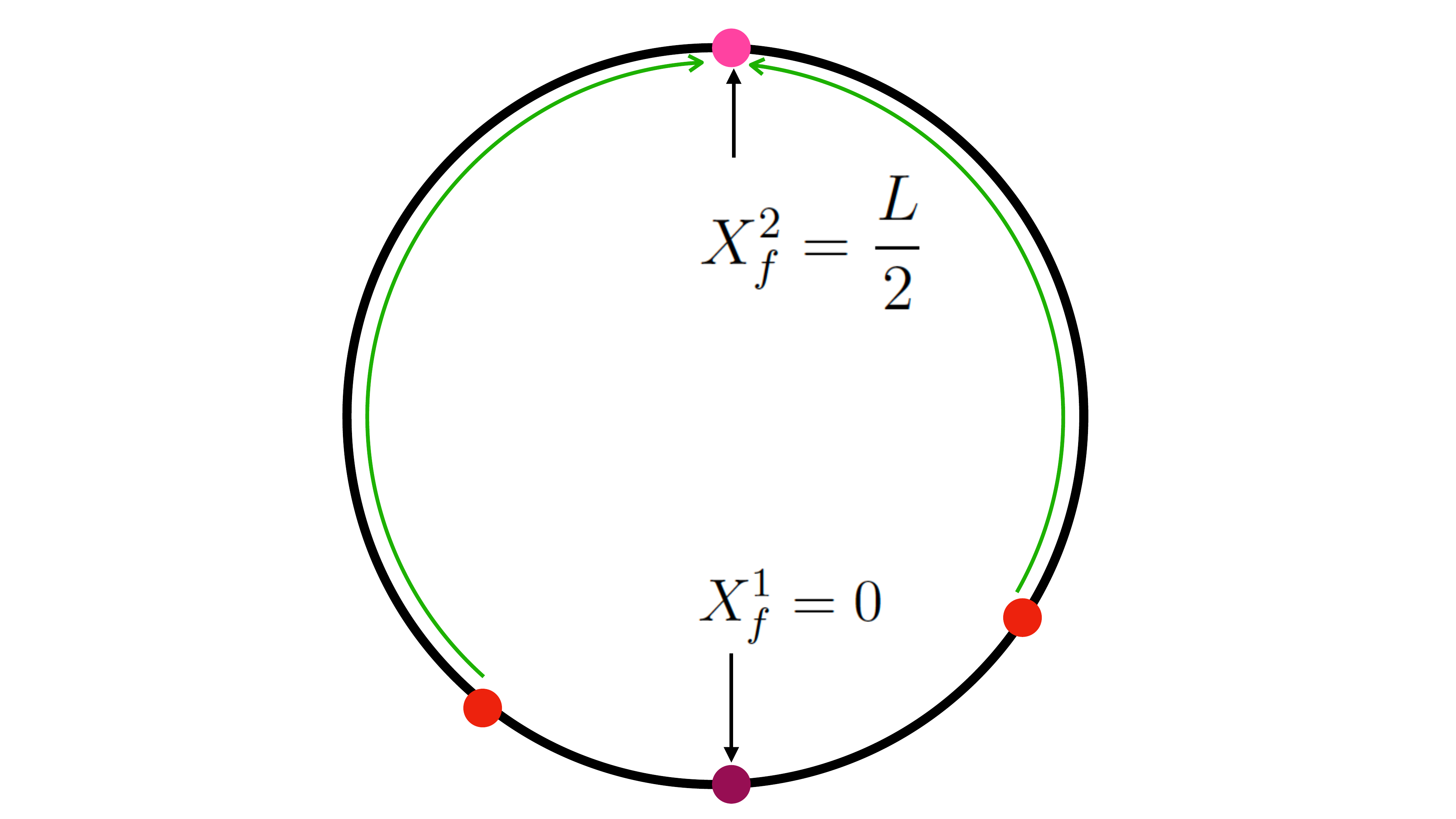}
      \\
      (a) The SSD time evolution 
    \end{minipage}   
    \begin{minipage}[t]{0.5\hsize}
      \centering
      \includegraphics[keepaspectratio, scale=0.1]{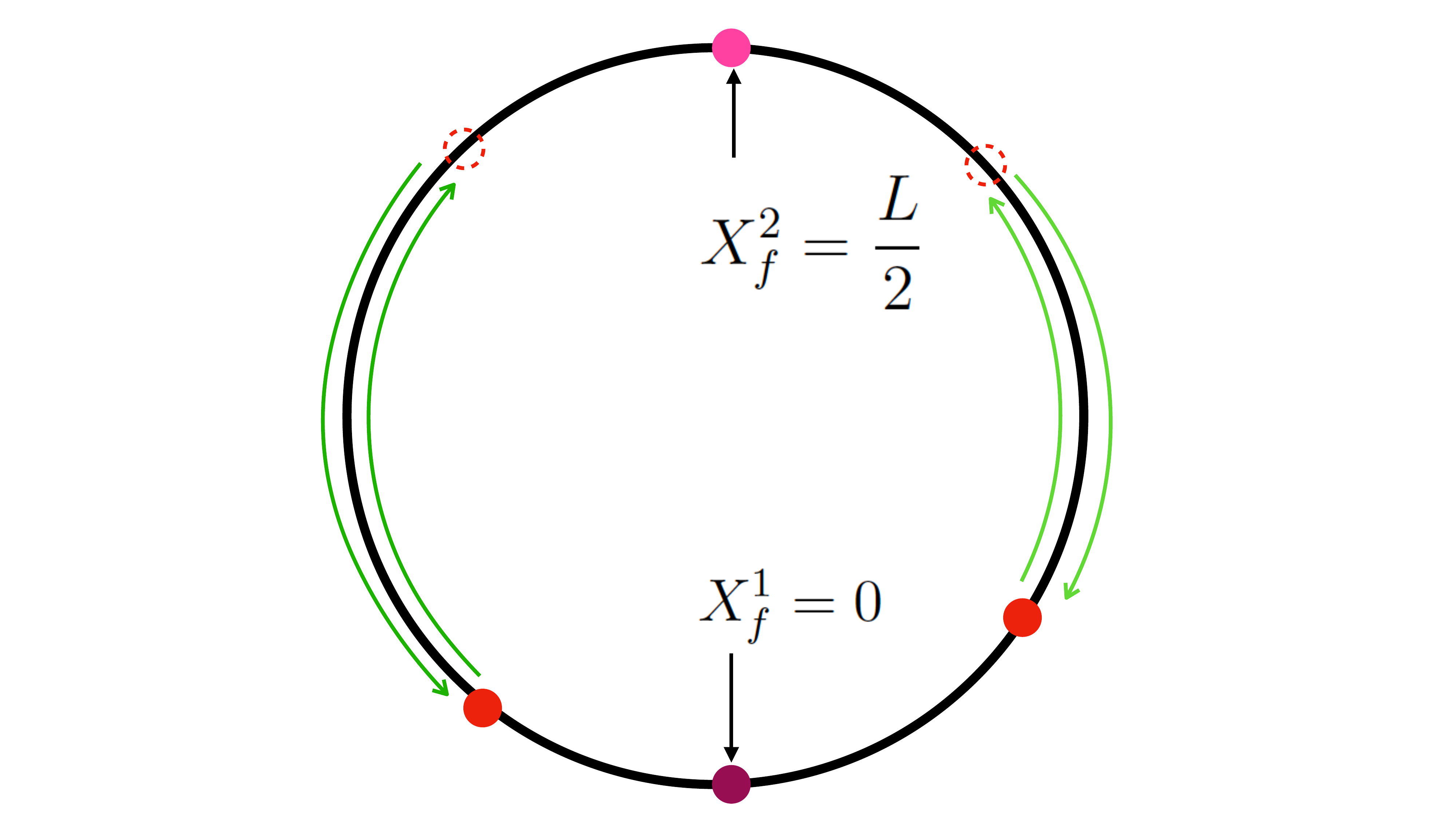}
      \\
      (b) The M\"{o}bius time evolution 
    \end{minipage} 
  \end{tabular}
  \caption{A sketch of
    how the spatial locations of an operator evolve
    in the Heisenberg picture
    for (a) the SSD and (b) M\"obius time evolutions. 
    The initial insertion points of 
    the operator are marked by red.
    The two fixed points $X^{1}_f=0$ and $X^2_f=L/2$
    are marked by purple. For the M\"obius case, the dashed red circles illustrate the turning point of the evolved operators.}
  \label{location_of_Operator}
\end{figure*}

Instead of the following
the time-dependence of the density matrix
$\rho(t)$,
the time-dependence of correlation functions
$
\mathrm{Tr}\,
\left[
\mathcal{O}_1(X_1) \mathcal{O}_2(X_2) \cdots 
\rho(t)
\right]
$
can be followed by using the Heisenberg picture,
\begin{align}
  \label{Heisenberg op}
\mathcal{O}_i(X,t)
=
e^{ + \epsilon H_0}
e^{-i t H_{\theta} }
\mathcal{O}_i(X)
e^{+i t H_{\theta} }
e^{ - \epsilon H_0}.
\end{align}
Here, 
$\mathcal{O}_i(X)$
is a (primary) operator
located at $X$ on the circle.
For a primary operator $\mathcal{O}$ at $X$
with conformal dimension $(h,\bar{h})$,
its Heisenberg evolution can be computed
explicitly as
\begin{align}
\label{transformation_by_SSD}
  e^{\epsilon H_0}e^{it H_{\theta}}
  \mathcal{O}(w_{X},\bar{w}_{X})
  e^{-it H_{\theta}}
  e^{-\epsilon H_0}=
  \left(\frac{d w_{X}^{\text{new}}}{dw_{X}}\right)^h
  \left(\frac{d \bar{w}_{X}^{\text{new}}}{d\bar{w}_{X}}\right)^{\bar{h}}
  \mathcal{O}(w^{\text{new}}_{X},\bar{w}_{X}^{\text{new}}),
\end{align}
where
$w_X=i X, \bar{w}_X=-i X$,
and
$w^{\text{new}}_{X}$ and $\bar{w}^{\text{new}}_{X}$ are
given by
\begin{align}
 \label{def_of_w_new}
  &w^{\text{new}}_{X} = \frac{L}{2\pi}
  \log{\left(\lambda_0
      \left[ \frac{[(1-\lambda)\cosh{(2\theta)-(\lambda+1)}]z
          +(\lambda-1)\sinh{(2\theta)}}{(1-\lambda)\sinh{(2\theta)}z+[(\lambda-1)\cosh{(2\theta)}-(\lambda+1)]}\right]
    \right)},
  \\
  &\bar{w}^{\text{new}}_{X} =  \frac{L}{2\pi} \log{\left(\lambda_0
      \left[ \frac{[(1-\lambda)\cosh{(2\theta)-(\lambda+1)}]\bar{z}
          +(\lambda-1)\sinh{(2\theta)}}{(1-\lambda)\sinh{(2\theta)}
          \bar{z}+[(\lambda-1)\cosh{(2\theta)}-(\lambda+1)]}\right] \right)},
\end{align}
where
$z=e^{\frac{2\pi w}{L}}$,
$\bar{z}=e^{\frac{2\pi \bar{w}}{L}}$,
$\lambda_0=e^{\frac{2\pi \epsilon}{L}}$,
and
$\lambda = e^{i\Omega t}$.
In the SSD limit, $w^{\text{new}}_{X}$ and $\bar{w}^{\text{new}}_{X}$ reduce to
\begin{align}
\label{w_new_and_bw_new}
\begin{split}
  &w^{\text{new}}_{X}
  \underset{\theta \rightarrow \infty}{\approx}
  \frac{L}{2\pi}
  \log{\left(
      \lambda_0\left[\frac{i \pi (1-z) t -L z}{i \pi (1-z)
          t-L}\right]\right)},
  \\
  &\bar{w}^{\text{new}}_{X}
  \underset{\theta \rightarrow \infty}{\approx}
  \frac{L}{2\pi}
  \log{\left(\lambda_0\left[\frac{i \pi (1-\bar{z})
          t -L \bar{z}}{i \pi (1-\bar{z}) t-L}\right]\right)}.
\end{split}
  \end{align}
Denoting
the real and imaginary parts of 
$w^{\text{new}}_{X}, \bar{w}^{\text{new}}_{X}$
as
$w^{\text{new}}_{X}
=\epsilon + i \varphi L/\pi$,
$\bar{w}^{\text{new}}_{X}
=\epsilon + i \bar{\varphi} L/\pi
$
($-\pi \le -\varphi, \bar{\varphi}\le 0$), 
the spatial and temporal locations,
$X^{\text{new}}$, $\tau^{\text{new}}$,
of
the transformed operator can be identified
as
\begin{align}
 \label{def_of_X_tau}
X^{\text{new}}
&=\frac{w_{X}^{\text{new}}-\bar{w}_{X}^{\text{new}}}{2i}=
\frac{L(\varphi-\bar{\varphi})}{2\pi},
\\
\tau^{\text{new}}
&=\frac{w_{X}^{\text{new}}+\bar{w}_{X}^{\text{new}}}{2}
 =\epsilon +i\frac{L\left(\varphi+\bar{\varphi}\right)}{2\pi}.
\end{align}
In this coordinate system, $\tau^{\text{new}}$ is a complex function of $X_i$ and $t$, while $X^{\text{new}}$ is a real function of $X_i$ and $t$.
The time evolution operator moves the operator along the spatial
and imaginary time directions
$X^{\text{new}}$ and $\tau^{\text{new}}$.

There are two fixed points that are left 
invariant under the M\"obius and SSD evolutions:
\begin{align}
X^1_{f}=0, \qquad X^2_{f}=\frac{L}{2}.
\end{align}
For the M\"obius evolution,
if an operator is inserted at a point other than the fixed points,
$X^{\text{new}}$ and $\tau^{\text{new}}$ undergo
periodic motion with period
$2\pi/\Omega$.
(Fig.\ \ref{location_of_Operator}(b)).
For the SSD evolution, 
and
operators inserted at other points than $X^1_f$
flow to $X^2_{f}$
(Fig.\ \ref{location_of_Operator}(a)).

\section{Von Neumann entropy for single intervals}
\label{Entanglement entropy for single intervals}

The von Neumann entropy
for a given subsystem $A$,
$
S_A=
\lim_{n \rightarrow 1} \frac{1}{1-n}\log{\left[
    \Tr_A\left( \rho_A\right)^n\right]}
$,
can be calculated
by using the twist operator formalism
\cite{Calabrese_2004,Calabrese_2009}.
For 
a single interval $[X_1, X_2]$, 
\begin{align}
\begin{split}
  S_A=\lim_{n\rightarrow 1} \frac{1}{1-n}
  \log{\left \langle
      \mathcal{T}_n(w_{X_1},\bar{w}_{X_1})
      \bar{\mathcal{T}}_n(w_{X_2},\bar{w}_{X_2}) \right \rangle},
\end{split}
\end{align}
where
$\mathcal{T}_n$ and $\bar{\mathcal{T}}_n$
are the twist and anti-twist operators in
the Heisenberg picture, \eqref{Heisenberg op}.
In terms of the original twist and anti-twist operators 
$S_A$
is given by
\cite{Wen_2018,wen2018floquet,Han_2020,Fan_2021},
\begin{align}
  \label{EE_single}
\begin{split}
  S_{A}&= -\frac{c}{12}\log{\left[\Pi_{i=1,2}
      \left(\frac{dw_{X_i}^{\text{new}}}{dw_{X_i}}
        \frac{d\bar{w}_{X_i}^{\text{new}}}{dw_{X_i}}\right)
    \right]}
\\
&\quad 
+ \lim_{n\rightarrow 1}\frac{1}{1-n}\log{\left[\left \langle \mathcal{T}_n(w^{\text{new}}_{X_1},\bar{w}^{\text{new}}_{X_1})\bar{\mathcal{T}}_n(w^{\text{new}}_{X_2},\bar{w}^{\text{new}}_{X_2}) \right\rangle_{2\epsilon}\right]},
\end{split}
     \end{align}
where the last term of (\ref{EE_single}) is given as the von Neumann entropy of
a thermal state at inverse temperature $2\epsilon$
on a compact spacetime.
We note that,
since
$w_{X_i}^{\text{new}}$ and $\bar{w}_{X_i}^{\text{new}}$
vary in time
in the Heisenberg picture,
the subsystem size varies in
the M\"obius/SSD time evolution.

Since there is no translation symmetry in
our inhomogeneous quenches,
the von Neumann entropy $S_A$ depends not only on the
size of subsystem $A$ but also on the location of $A$.
In the following,
we will work with the following three choices
of subsystem $A$:
\begin{align}
\begin{split}
A =\begin{cases}
\left\{x\big{|}0 \le x \le X, L-X \le x \le L \right\} & \text{Case 1} \\
\left\{x\big{|}\f{L}{4}-X \le x \le \f{L}{4}+X\right\} & \text{Case 2} \\
\left\{x\big{|}\f{L}{2}-X \le x \le \f{L}{2}+X\right\} & \text{Case 3} \\
\end{cases}.
\end{split}
\end{align}
In Case 1, the center of subsystem $A$ is $X^1_f$,
one of the fixed points,
and in Case 3 the center is the other fixed point $X^2_f$.
In Case 2, the center of subsystem $A$ is the midpoint
between $X^1_f$ and $X^2_f$.

We will study
both a CFT with a gravity dual (holographic CFT)
and a free fermion CFT.
However, 
for the von Neumann entropy for a single interval,
there is essentially no difference between
these two cases.
We therefore focus on the holographic CFT here.
On the other hand, mutual information defined for
two (disjoint) intervals probes the details of CFTs,
as we will see in  Sec.\ \ref{Mutual information}.
In holographic CFTs,
in the coarse-grained limit,
i.e., the limit where all parameters are sufficiently larger than $\epsilon$,
the final term in \eqref{EE_single}
can be computed from the gravity dual which is
the BTZ black hole \cite{Hawking:1983vo}.
As in \cite{Ryu_2006,Headrick_2014},
it is given by
\begin{align}
\label{NU_hol}
\begin{split}
  &\lim_{n\rightarrow 1}\frac{1}{1-n}
  \log{
  \left
        \langle \mathcal{T}_n(w^{\text{new}}_{X_1},\bar{w}^{\text{new}}_{X_1})
        \bar{\mathcal{T}}_n(w^{\text{new}}_{X_2},\bar{w}^{\text{new}}_{X_2})
      \right\rangle_{2\epsilon}
      }\\
  &
  \approx \frac{c}{3}
  \log{\left(\frac{2 \epsilon}{\pi }\right)}
  \\
  &
  \quad
  +
  \left\{
    \begin{array}{l}
      \text{Min}\bigg{[}
      \frac{c\pi L}{6\epsilon}
      +\frac{c}{6}\log{
      \left|
      \sin{
      \left[\frac{\pi}{2\epsilon}(w^{\text{new}}_{X_1}-w^{\text{new}}_{X_2})\right]}\right|^2},
      \frac{c}{6}\log{
      \left|
      \sin{\left[\frac{\pi}{2\epsilon}(w^{\text{new}}_{X_1}-w^{\text{new}}_{X_2}
      \pm iL)\right]}\right|^2} \bigg{]}
      \\
      \qquad \qquad \qquad \text{for Case 1}
      \\
      \text{Min}\bigg{[} \frac{c}{6}\log{
      \left|
      \sin{\left[\frac{\pi}{2\epsilon}(w^{\text{new}}_{X_1}-w^{\text{new}}_{X_2})\right]}\right|^2},
      \frac{c\pi L}{6\epsilon}
      +\frac{c}{6}\log{\left|
      \sin{\left[\frac{\pi}{2\epsilon}(w^{\text{new}}_{X_1}-w^{\text{new}}_{X_2} \pm
      iL)\right]}\right|^2} \bigg{]}
      \\
      \qquad \qquad \qquad \text{for Case 2 and 3}
    \end{array}
  \right.
\end{split}
\end{align}
Here, $\frac{c\pi L}{6\epsilon}$ is the entropy of the black hole, i.e., the
thermal entropy, and
all lengths are measured in the unit of
some UV cutoff (lattice spacing).


\section{Entanglement dynamics from the quasiparticle picture}\label{QuasiparticlePictureDetails}
Let $\rho^{(n)}_R(x,t)$ and $\rho^{(n)}_L(x,t)$ denote the right-moving and left-moving quasiparticle densities at position $x$ and time $t$ respectively, where $n$ is the R\'{e}nyi index which will determine the quasiparticle density. 
If the quasiparticles are conserved, then their densities must satisfy the continuity equation
\begin{equation}\label{ThermalStateContinuityEquation}
    \frac{\partial \rho^{(n)}_i(x,t)}{\partial t} = \mp \frac{\partial}{\partial x} \rho^{(n)}_i(x,t) v(x)
\end{equation}
where the minus sign is for $i=R$ and the plus sign is for $i=L$. Since the thermofield double state in the high temperature limit is approximately a product of uniformly distributed Bell pairs, the initial density is
\begin{equation}
    \rho^{(n)}_i(x,0) = \rho^{(n)}_0 
\end{equation}
for $i=R,L$ where $\rho_0$ is the initial quasiparticle density in subsystem A. A solution to \eqref{ThermalStateContinuityEquation} depends on the trajectories of the quasiparticles which are determined by the inhomogeneous velocities due to the inhomogeneous Hamiltonian. If a quasiparticle is initially located at $x_0$ at time $t_0$, its position $x(t)$ at a later time $t$ is 
\begin{equation}
    dt =\pm \frac{dx}{v(x)} \Rightarrow t-t_0 = \pm \int_{x_0}^{x(t)} \frac{dx'}{1-\tanh{2\theta}\cos{\frac{2\pi x'}{L}}}
\end{equation}
where the plus sign refers to right-moving quasiparticles while the minus sign refers to left-moving quasiparticles. Performing the integral gives an implicit relation between the quasiparticles initial position $x_0$ and its position $x(t)$ after a time $t-t_0$ has elapsed:
\begin{align}\label{QuasiparticleTrajectory}
\frac{\pi (t-t_0)}{L\cosh{2\theta}}  
=\pm\left[\tan^{-1}\left(e^{2\theta}\tan\frac{\pi x(t)}{L}\right) - \tan^{-1}\left(e^{2\theta}\tan\frac{\pi x_0}{L}\right)\right]
\end{align}

Note that if $x(t) = x_0+k L$ where $k\in \mathbb{Z}$, then $t-t_0 = m L \cosh{2\theta}$ with $m\in \mathbb{Z}$ is a solution as well so the quasiparticle trajectories are periodic with period $L\cosh{2\theta}$ which is consistent with the other physical quantities considered in this paper.

A general solution for \eqref{ThermalStateContinuityEquation} can be written in terms of these trajectories \cite{doi:10.1098/rspa.1978.0190}. The quasiparticles that are at position $x$ at time $t$ were initially located at $x_{i,0}(x,t)$ where the initial position can be written as a function of the current position and time via \eqref{QuasiparticleTrajectory} and the subscript $i=L/R$ refers to the chirality of the quasiparticles. Assuming that the number of quasiparticles is conserved, the number of quasiparticles at $x$ at time $t$, $\rho^{(n)}(x,t)dx$ must be equal to the number of quasiparticles initially located at $x_{i,0}(x,t)$, $\rho^{(n)}(x_{i,0}(x,t),0)dx_{i,0}$. Therefore, the solution to the continuity equation \eqref{ThermalStateContinuityEquation} is\begin{equation}
    \rho^{(n)}_i(x,t) = \rho^{(n)}_i(x_{i,0}(x,t),0) \frac{\partial x_{i,0} (x,t)}{\partial x}
\end{equation}
for $i = L,R$. The R\'{e}nyi entropy is given by the number of quasiparticles contained in the subsystem
\begin{equation}
    S^{(n)}_A(t) = \int_{x\in A}dx \rho^{(n)}_L(x,t)+ \int_{x\in A}dx\rho^{(n)}_R(x,t).
\end{equation} 
The initial quasiparticle density $\rho^{(n)}_0$ can be fixed by equating the initial R\'{e}nyi entropy to that of the thermal entropy for subsystem $A$ since we are performing a quench from the uniform thermal state. When the subsystem is taken to be much larger than the regulator $\epsilon$, the initial quasiparticle density is found to be
\begin{equation}
    \rho^{(n)}_0 = \frac{n+1}{24n} \frac{\pi}{\epsilon}
\end{equation}
which depends on the replica index $n$. Since $\rho_i(x,0)=\rho_0$ is uniform, the integral for the entanglement entropy of a single interval $[X_2,X_1]$ is easily carried out and when $x_{0,i}(X_1,t)>x_{0,i}(X_2,t)$, the integral is simply given by $x_{0,i}(X_1,t)-x_{0,i}(X_2,t)$. This simply states that the interval $[x_{0,i}(X_2,t),x_{0,i}(X_1,t)]$ flows to $[X_2,X_1]$ at time $t$. Since $X_1 > X_2$ and quasiparticles cannot overtake one another, $x_{0,i}(X_1,t)$ is "to the right" of $x_{0,i}(X_2,t)$ for both chiralities. Thus, when $x_{0,i}(X_1,t)<x_{0,i}(X_2,t)$, the correct value due to the spatial periodicity of the system is given by $L-(x_{0,i}(X_2,t)-x_{0,i}(X_1,t))$. Therefore, the entanglement entropy for a single interval is 
\begin{equation}
    S_A(t) = \rho_0 
    \sum_{i=L,R} 
    \text{mod}
    \left[x_{0,i}(X_1,t)-x_{0,i}(X_2,t),L\right] 
\end{equation}
where the modulo operation takes the periodicity of the system into account.

\begin{figure}
    \centering
    \includegraphics[width=0.32\textwidth]{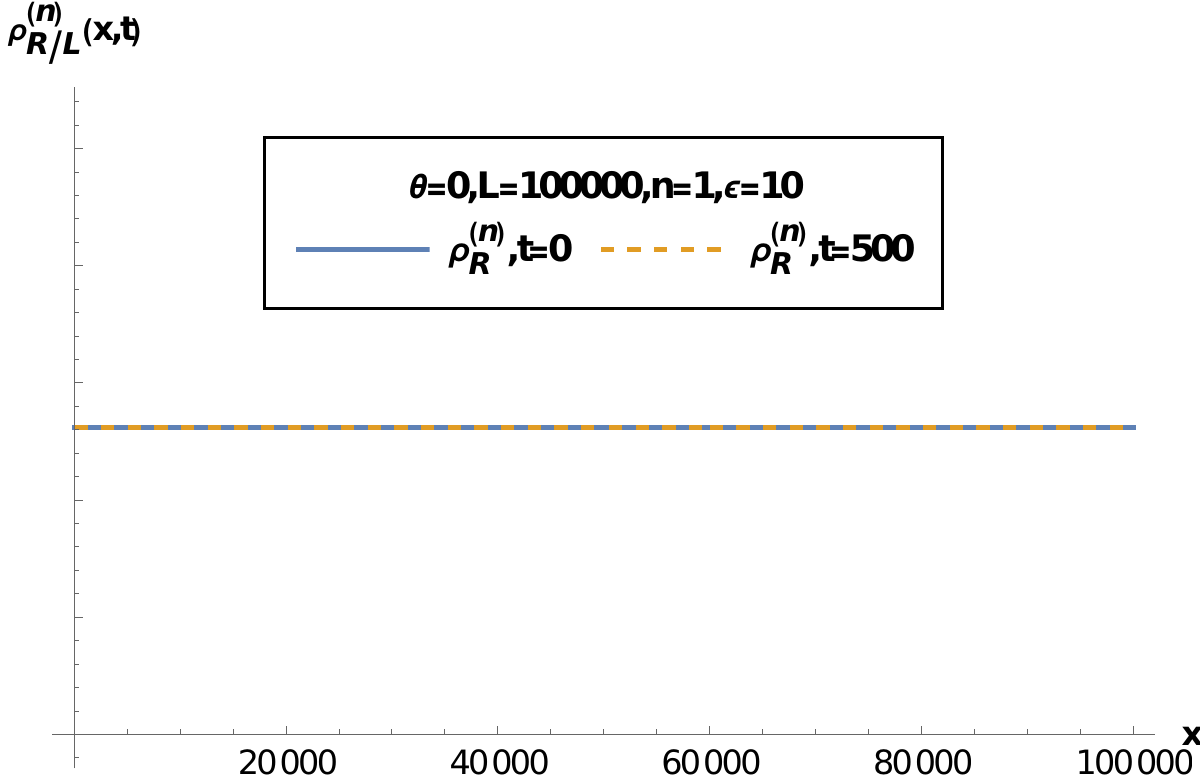}
    \hspace{1cm}
    \includegraphics[width=0.32\textwidth]{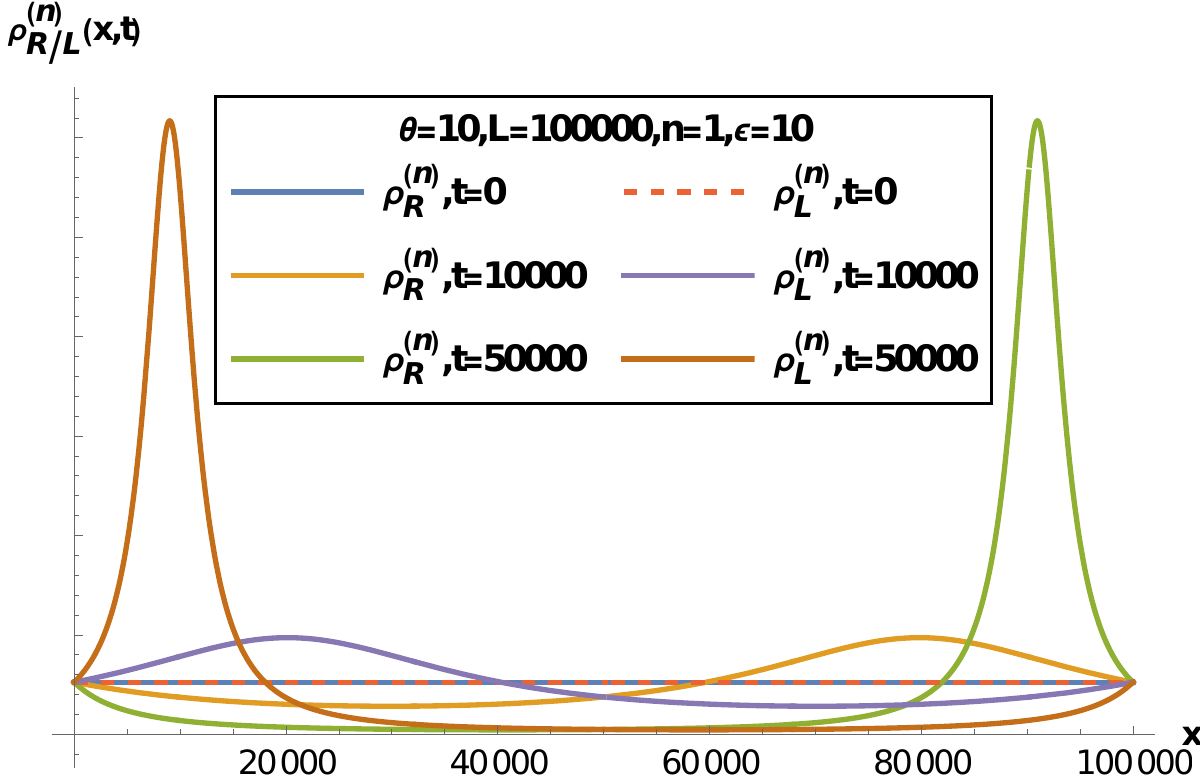}
    \caption{Plots of the quasiparticle densities $\rho^{(n)}_{R}(x,t)$ and $\rho^{(n)}_{L}(x,t)$ for $L=1000$ and $\epsilon=n=1$ for $\theta=0,10$. The plots for $t=L\cosh{2\theta}$ are identical to the ones for $t=0$ in agreement with the periodicity of the density matrix. For the uniform $\theta=0$ case, only the right-moving quasiparticle density is shown since the plots for the left-moving quasiparticle density are identical.}
    \label{ThermalStateQuenchQuasiparticleDensity}
\end{figure}
Plots of the quasiparticle densities are shown in figure \ref{ThermalStateQuenchQuasiparticleDensity}. Initially, the quasiparticles are uniformly distributed throughout the system. When the time-evolution Hamiltonian is the uniform one, the quasiparticle density remains uniform and constant. This does not mean that the quasiparticles are stationary. Instead, the quasiparticles are moving with uniform velocity throughout the entire system so the number of quasiparticles that are moving towards any given point equals the number of quasiparticles that are moving away from that point and the system remains in a steady thermal state. When the time-evolution Hamiltonian is inhomogeneous, the velocity profile of the quasiparticles is inhomogeneous and the quasiparticle density is no longer constant. Instead, the quasiparticles begin to accumulate near the SSD fixed points. This gives rise to the appearance of the black hole like excitation in the free fermion CFT just like how the spikes in the black hole's event horizon account for the appearance of the black-hole like excitation in the holographic theories. For finite values of $\theta$, the quasiparticles densities return to the initial uniform distribution after a period of $L\cosh{2\theta}$.

\section{Energy-momentum tensor and energy current}\label{sec:stress}

The time-evolution 
of the von Neumann entropy studied in 
the main body suggests that  
under the SSD quench
a black hole-like excitation propagates 
and localizes at the origin at late times.
In this appendix,
we examine 
the expectation value of 
the energy-momentum tensor and 
local energy current to probe 
the dynamics of the black hole-like excitation.
%
The 
energy-momentum tensor profile is
also useful for studying the holographic dual of the SSD/M\"obius quench
(Sec.\ \ref{shroginder_geometry}).

The transformation law of
the energy-momentum tensor 
under SSD/M\"{o}bius deformation is the same as that of local operators discussed in 
\eqref{transformation_by_SSD}, except for the contribution from the Schwarzian derivative.
For the holomorphic part, it is given by
\begin{align}
T^{\theta}(w)&\equiv e^{iH_{\theta} t}
T(w)e^{-iH_{\theta} t}
\nn\\
&=\left(\dfrac{dw^{\textrm{new}}_X(w)}{dw}\right)^2T(w^{\textrm{new}}_{X})+\dfrac{c}{12}\mathrm{Sch}(w^{\textrm{new}}_{X}(w),w).
\end{align}
where $w=iX$ (see around Eq.\ \eqref{transformation_by_SSD}) and we define the Schwarzian derivative as
\be
\mathrm{Sch}(f(w),w)=\dfrac{f^{\prime\prime\prime}(w)}{f^\prime(w)}-\dfrac{3}{2}\left(\dfrac{f^{\prime\prime}(w)}{f^\prime(w)}\right)^2.
\ee
The Schwarzian term is a consequence of the Weyl anomaly and explains the contribution from the Casimir energy.
\subsection{Pure state approximation}

It is well known that the expectation value of the energy-momentum tensor for high-energy eigenstates can be well approximated by the expectation value for thermal states. Based on this fact, instead of the thermal state itself, we first estimate the expectation value of the energy-momentum tensor for a high-energy eigenstate without angular momentum.
We will later confirm 
that this approximation precisely reproduces 
the time-dependent part of 
the energy-momentum tensor of the free fermion CFT,
which will be derived without approximation. 

\paragraph{The SSD quench}
In the SSD quench, we have 
\begin{align}
T^{\textrm{SSD}}(X,t)&\equiv\langle \psi_{{\cal O}_h}|T^{
\theta\to \infty} (w)|\psi_{{\cal O}_h}\rangle\nonumber\\
&=\left(\dfrac{2\pi}{L}\right)^2\left[\dfrac{4h}{\left(\left(\frac{2\pi t}{L}\right)^2\left(1- \cos\left(\frac{2\pi}{L}X\right)\right)-2 \left(\frac{2\pi t}{L}\right) \sin\left(\frac{2\pi}{L}X\right)+2\right)^2}-\dfrac{c}{24}\right].
\label{TSSD}
\end{align}
The second term comes from the Casimir energy. Here we introduced $|\psi_{{\cal O}_h}\rangle$ as a spinless primary state with the conformal dimension $h_L=h_R=h$. Namely, the total energy at $t=0$ is given by $\frac{4\pi h}{L}$ up to the Casimir energy\footnote{Since we are interested in a large system with a finite energy density, i.e. the thermodynamic (large-$L$) limit with fixed $h/L^2$, the Casimir energy can be negligible.}.  
We can also obtain $\langle \psi_{{\cal O}_h}|\bar{T}^{\theta \to \infty}(\bar{w})|\psi_{{\cal O}_h}\rangle$ by exchanging $X\rightarrow-X$.
Note that one can relate the energy with (inverse) temperature $\beta$ by using the relation (see \cite{Datta:2019jeo}, for example),
\be
\dfrac{h}{L^2}=\dfrac{c}{24\beta^2}.
\ee
We plot
the energy-momentum tensor profile Eq.\ \eqref{TSSD}
in Fig.\ \ref{fig:SSDT_3dplot}. 
\begin{figure}[tbp]
        \centering
      \includegraphics[keepaspectratio, scale=0.6]{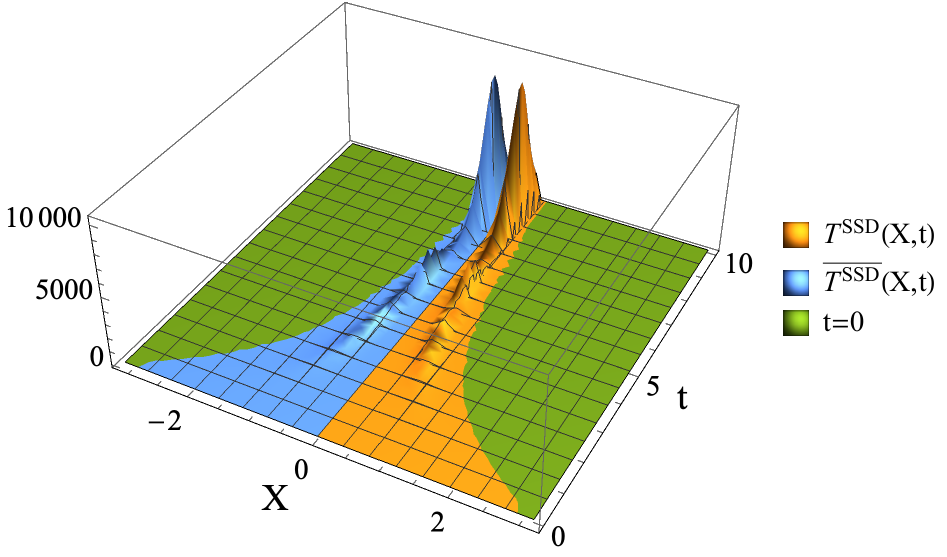} \caption{
      The holomorphic (orange) and anti-holomorphic (blue)
      parts of the energy-momentum tensor. The sum of these two parts gives the local energy density, while the difference gives a local energy current. 
      In this figure, we set $L=2\pi$, and $c=\beta=1$. 
      Also plotted are their initial values (green). 
      These two peaks are both converging to a fixed point of the SSD quench, $X=X^1_f=0$.}
    \label{fig:SSDT_3dplot}
  \end{figure}
We can see that both holomorphic and anti-holomorphic energy-momentum tensors are gathered towards the fixed point of the SSD transformation. Similarly, we obtain a local energy current as 
\begin{align}
J^{\textrm{SSD}}(X,t)&\equiv\langle \psi_{{\cal O}_h}|
T^{\theta\to \infty}(w)-\bar{T}^{\theta\to \infty}(\bar{w}) 
|\psi_{{\cal O}_h}\rangle 
\nonumber \\
&=\left(\dfrac{2\pi}{L}\right)^2\left[\dfrac{4h}{\left(\left(\frac{2\pi t}{L}\right)^2\left(1- \cos\left(\frac{2\pi}{L}X\right)\right)-2 \left(\frac{2\pi t}{L}\right) \sin\left(\frac{2\pi}{L}X\right)+2\right)^2}-(X\rightarrow-X)\right].
\end{align}
If we set $X=X^1_f=0$, both holomorphic and anti-holomorphic energy-momentum tensor take a constant value, hence there are no local energy flow, i.e.  $J^{\textrm{SSD}}(0,t)=0$.

The location where the energy-momentum tensor takes its maximum value at each fixed time is given by
\begin{align}
X^{(T)}_{\textrm{max}}(t)=\left(\dfrac{L}{\pi}\right)\tan^{-1}\left(\dfrac{L}{2\pi t}\right)\underset{t\rightarrow\infty}{\longrightarrow}0, \label{eq:XTmax}
\end{align}
which is obtained from $\partial_t T^{\textrm{SSD}}(X,t)=0$. We will compare this with the peak of the black hole horizon in the next section. 

These results, along with ones for the von Neumann entropy, provide further evidence that there are local black hole-like excitations that propagate towards the fixed point of the SSD Hamiltonian and account for the thermal entropy.

\paragraph{The M\"{o}bius quench} 
A similar analysis can be done for the M\"{o}bius  
quench with generic $\theta$.
The energy-momentum tensor and the local energy current will be denoted by $T^M$ and $J^M$ respectively. 
As in the von Neumann entropy, the M\"{o}bius deformed energy-momentum tensor has the $2\pi/\Omega= L\cosh2\theta$ periodicity. For this reason, the M\"{o}bius deformed local energy current also acquires the $\pi/\Omega$ anti-periodicity. We defer plots for $T^M$ and $J^M$ to the next section as these are identical to ones for the free fermion. 


\subsection{The free fermion CFT}
The one-point correlation function of the energy-momentum tensor on a torus is given by a derivative of the logarithm of the partition function with respect to the modular parameter \cite{francesco1997conformal,PhysRevD.102.026023}. 
For the $c=1$ free Dirac fermion theory, the partition function for spin-structure $\nu$ is 
$
    Z_\nu = \frac{1}{2}\left|\frac{\theta_\nu(\tau)}{\eta(\tau)}\right|^2
$\cite{Herzog2013}.
Therefore, the expectation value of the energy-momentum tensor for spin-structure $\nu$ is 
\begin{equation}\label{EnergyMomentumTensorTorusOnePoint}
    \langle T(w_X^\text{new})\rangle_\nu =2i\pi \partial_\tau \ln \left|\frac{\theta_\nu(\tau)}{\eta(\tau)}\right|^2
\end{equation}
Using the product representation of the elliptic theta function and the Dedekind eta functions, the expectation value is found to be
\begin{equation}
    \langle T(w_X^\text{new})\rangle_\nu=\frac{\pi^2}{3}-16\pi^2 (-1)^{\nu+1}\sum_{m=1}^\infty \frac{(m-\frac{1}{2})e^{2\pi i(m-\frac{1}{2})\tau}}{1+(-1)^{\nu+1}e^{2\pi i(m-\frac{1}{2})\tau}}
\end{equation}
Since the imaginary part of the modular parameter $\text{Im}\tau = \frac{L}{2\epsilon}\gg 1$, 
the sum is negligible,
so the expectation value of the energy-momentum tensor under a M\"{o}bius evolution is
\begin{equation}\label{MobiusEnergyMomentumTensor}
    \langle T^{\theta}(w)\rangle = \left(\frac{dw_X^\text{new}}{dw}\right)^2 \frac{\pi^2}{3} + \frac{c}{12} \text{Sch}(w_X^\text{new},w).
\end{equation}
The one-point function for the energy-momentum tensor on the torus depends only on the modular parameter so \eqref{EnergyMomentumTensorTorusOnePoint} is the same for the anti-holomorphic part. In fact, the anti-holomorphic coordinates have the exact same expression as the holomorphic coordinates with the replacement $w\rightarrow \bar{w}$. Therefore, $\bar{T}^{\theta}(\bar{w})$ can be obtained from $T^{\theta }(w)$ by making the replacement $X\rightarrow -X$.

Plots of the spatial profile of the expectation value of the energy-momentum tensor \eqref{MobiusEnergyMomentumTensor} as well as the energy current at various instances in time for the SSD and M\"{o}bius quenches are shown in Fig. \ref{SSD_EnergyMomentum} and \ref{Mobius_EnergyMomentum}, respectively.

\begin{figure}
    \centering
    \includegraphics[width=0.35\textwidth]{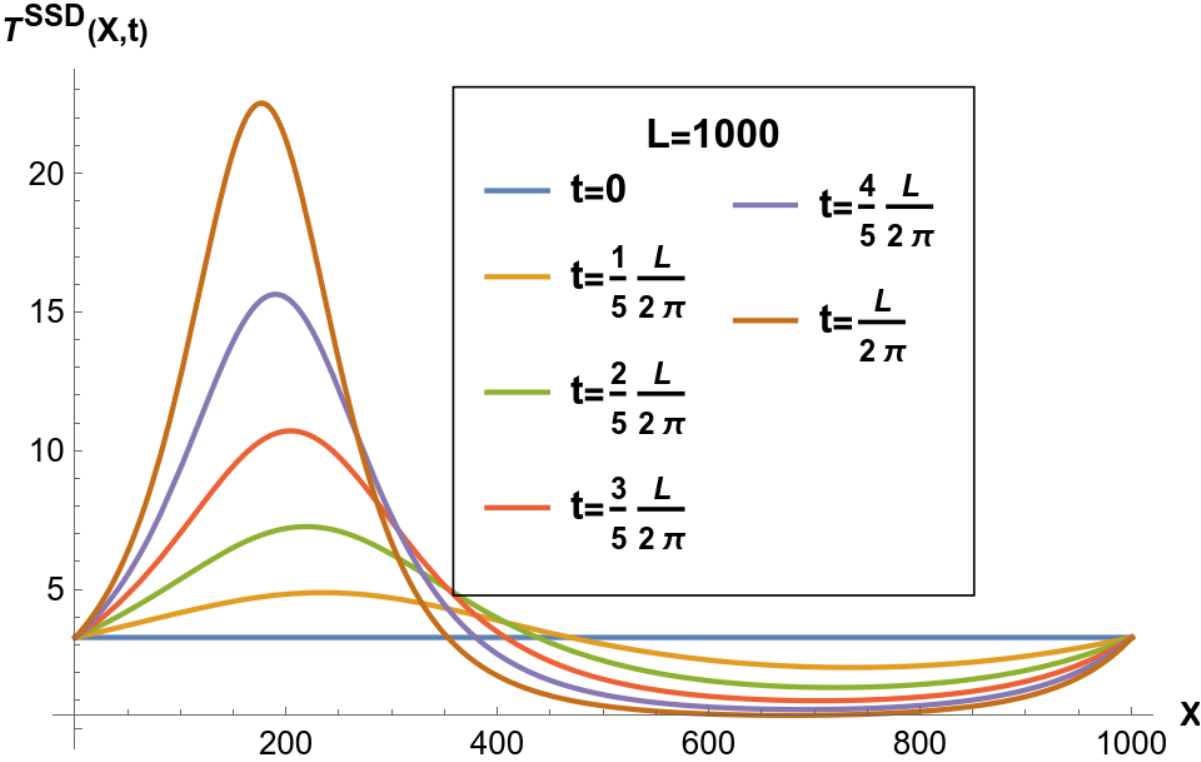}
    \hspace{0.7cm}
    \includegraphics[width=0.35\textwidth]{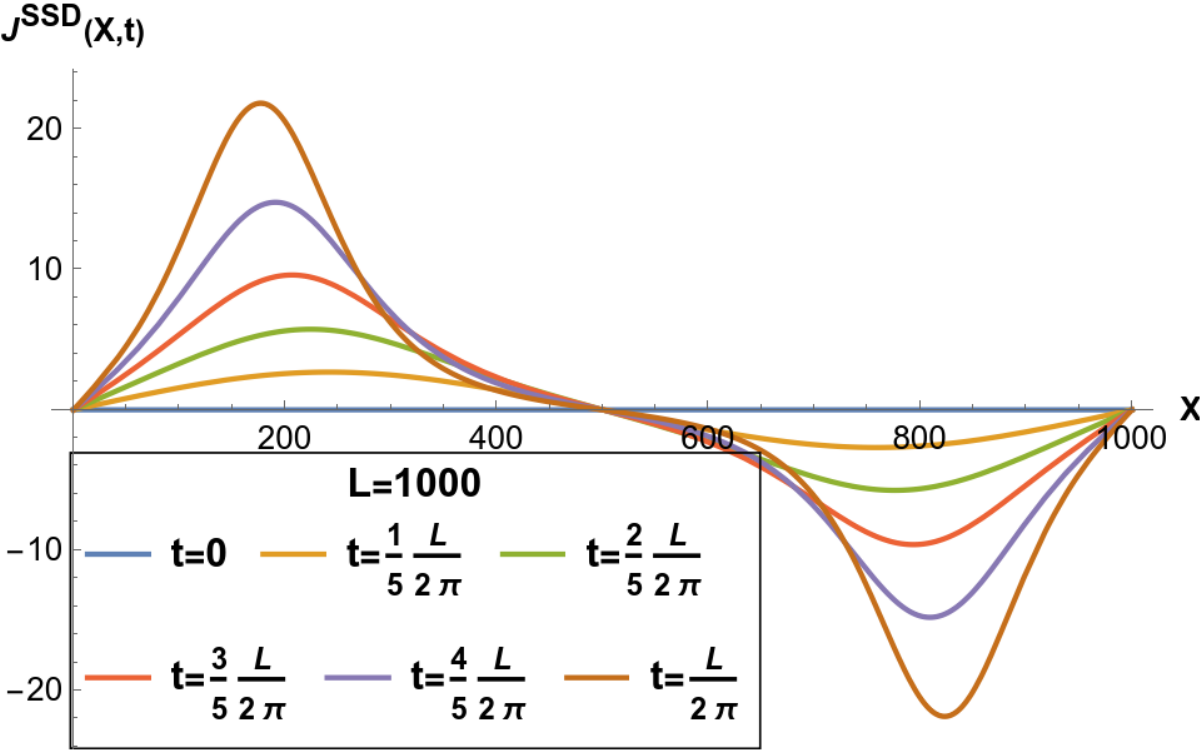}
    \caption{
    The spatial profile of the expectation value of the energy-momentum tensor \eqref{MobiusEnergyMomentumTensor} as well as the energy current 
    at different times
    in the SSD limit for a total system size of $L=1000$.}
    \label{SSD_EnergyMomentum}
\end{figure}

\begin{figure}
    \centering
    \includegraphics[width=0.3\textwidth]{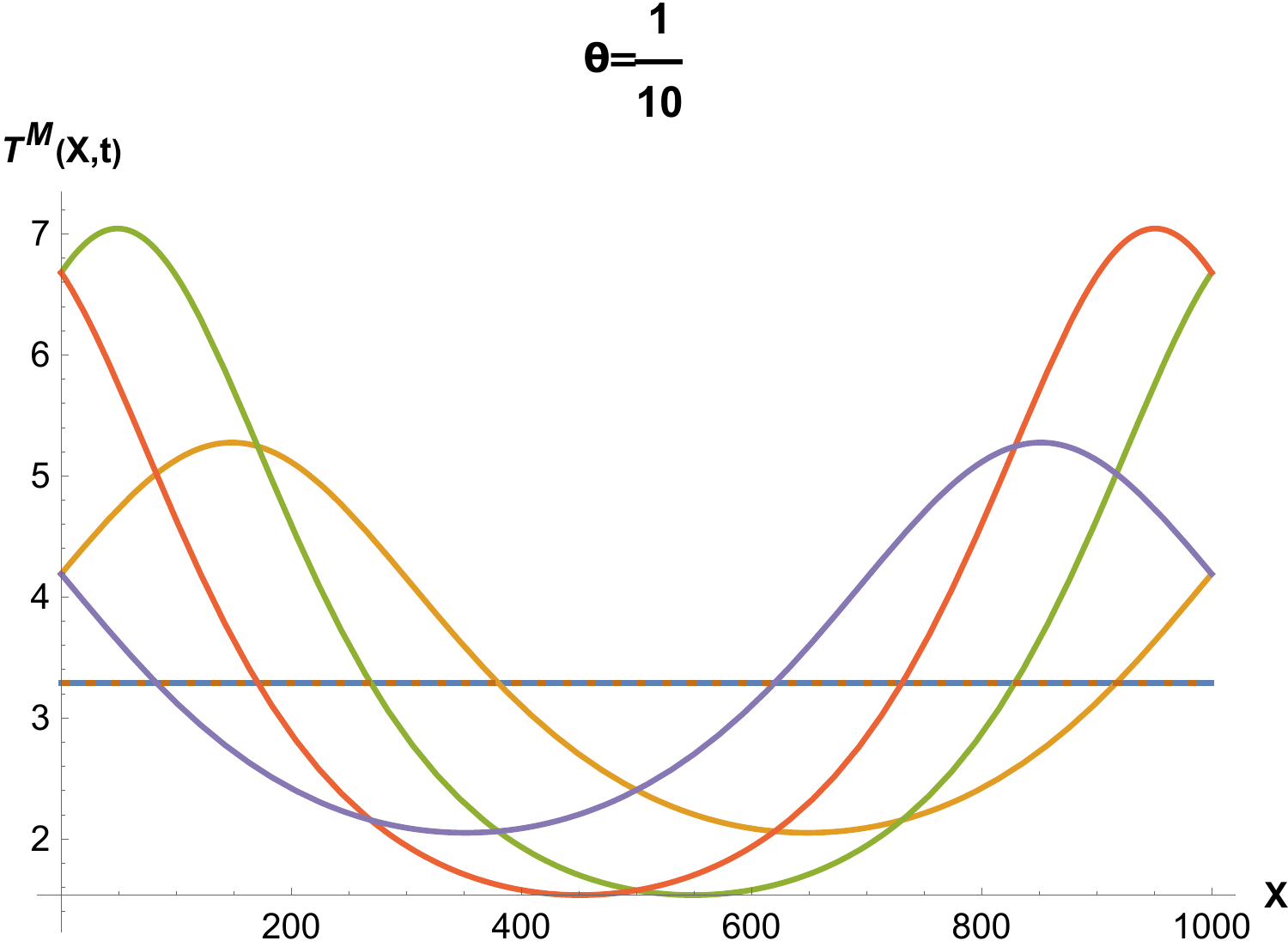}
    \hspace{0.5cm}
    \includegraphics[width=0.3\textwidth]{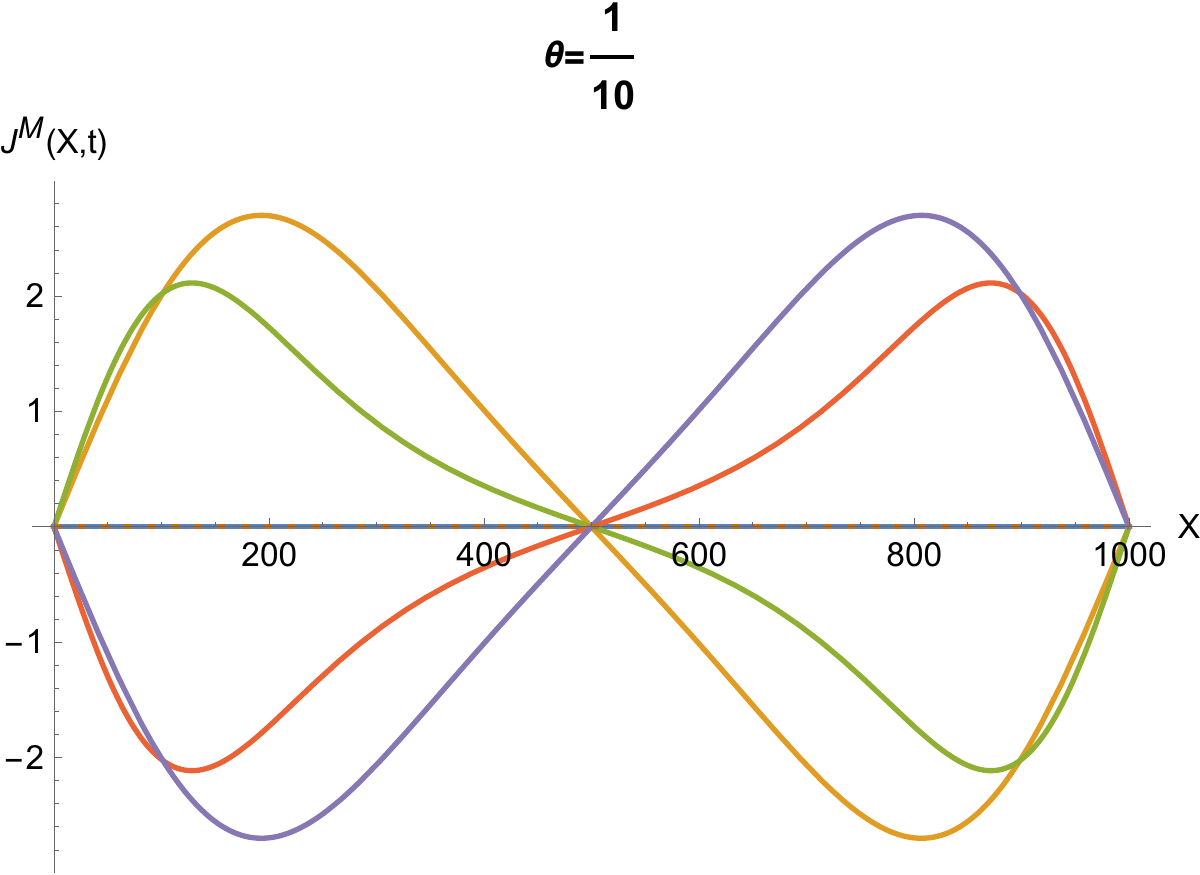}
    \hspace{0.5cm}
    \includegraphics[width=0.3\textwidth]{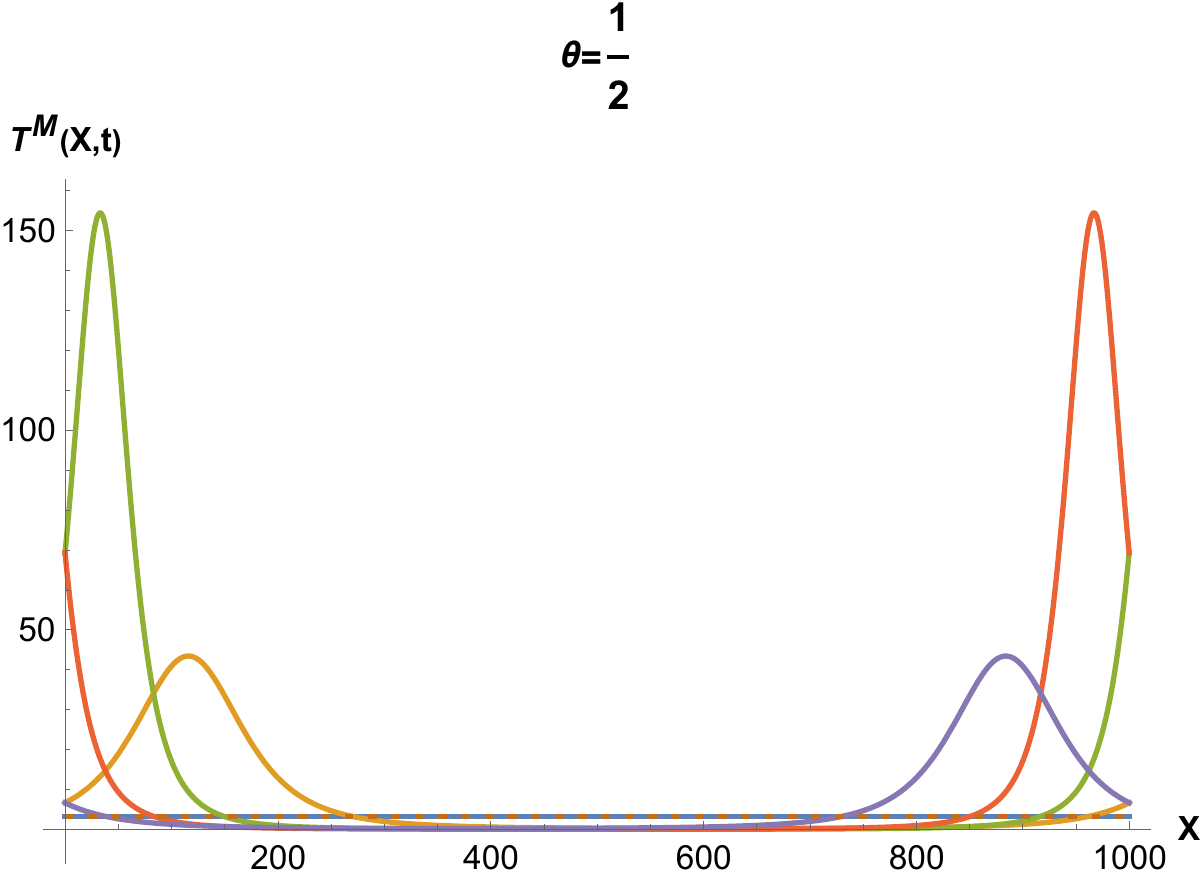}
    \hspace{0.5cm}
    \includegraphics[width=0.3\textwidth]{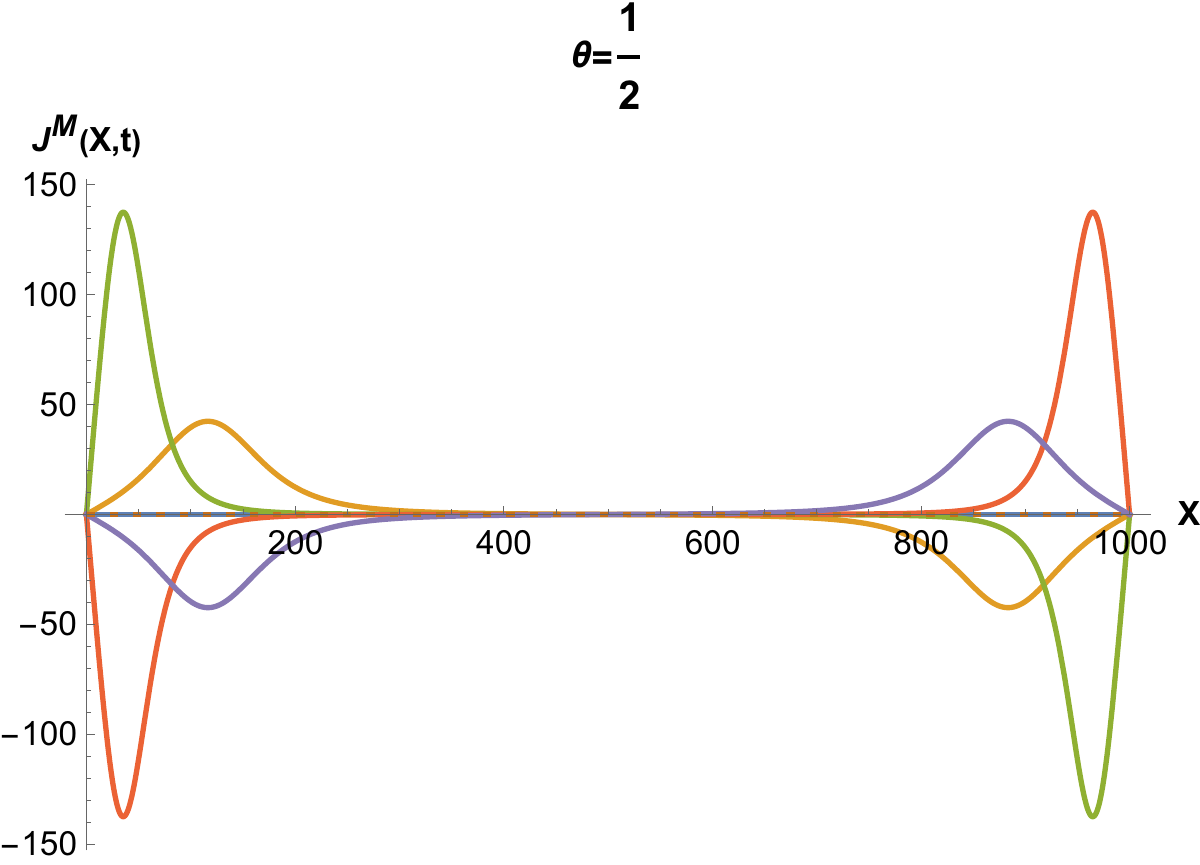}
    \hspace{0.5cm}
    \includegraphics[width=0.3\textwidth]{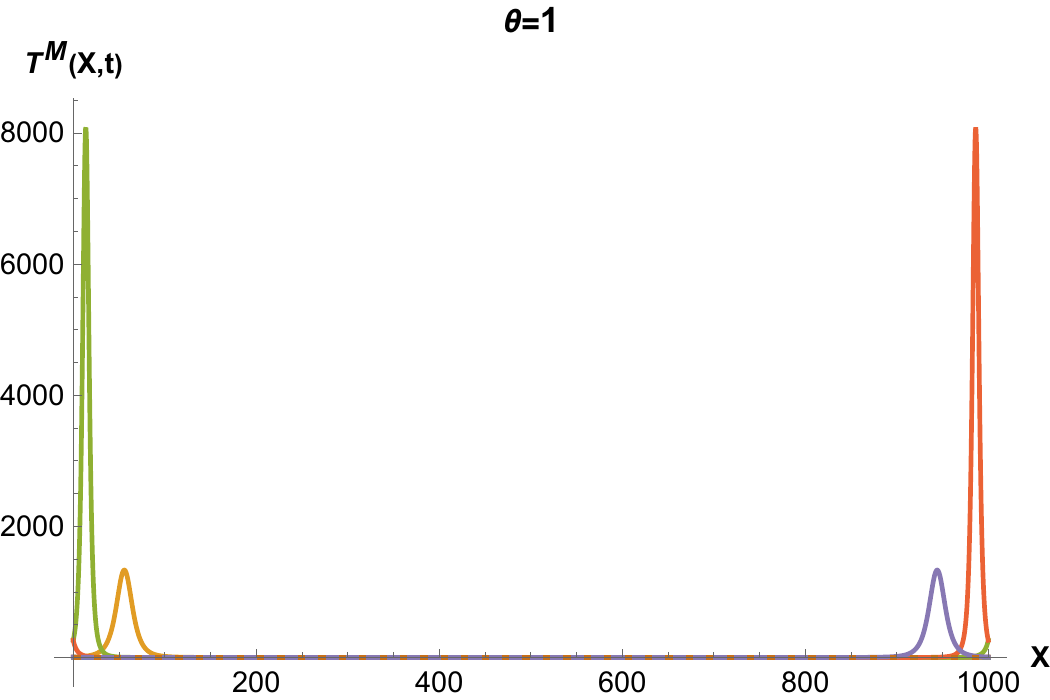}
    \hspace{0.5cm}
    \includegraphics[width=0.3\textwidth]{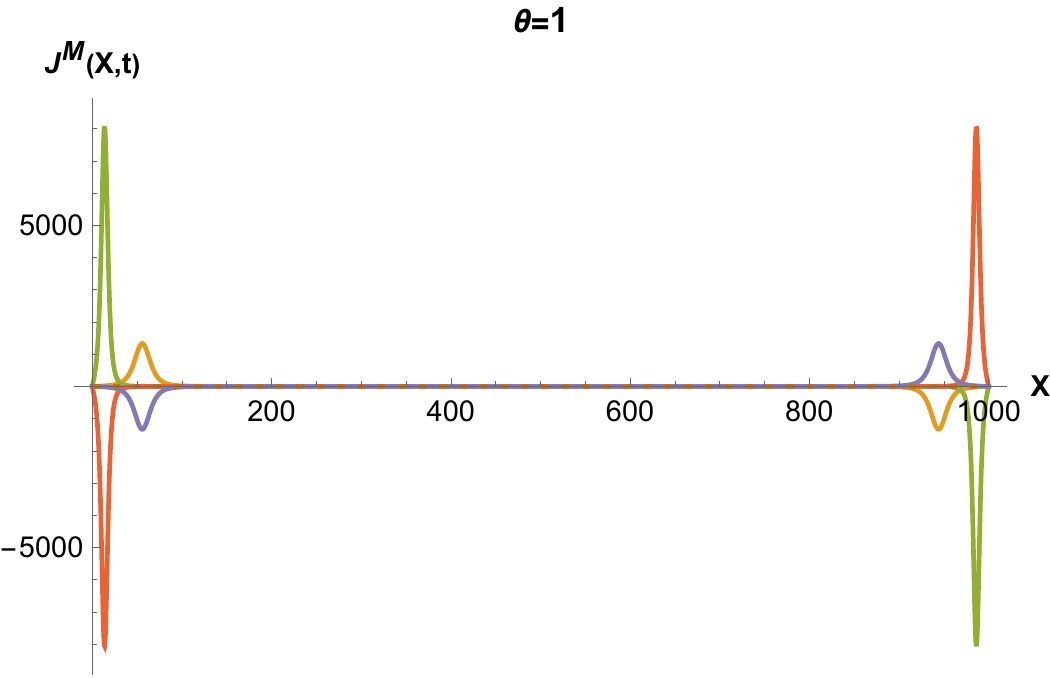}
    \hspace{0.5cm}
    \includegraphics[scale=0.3]{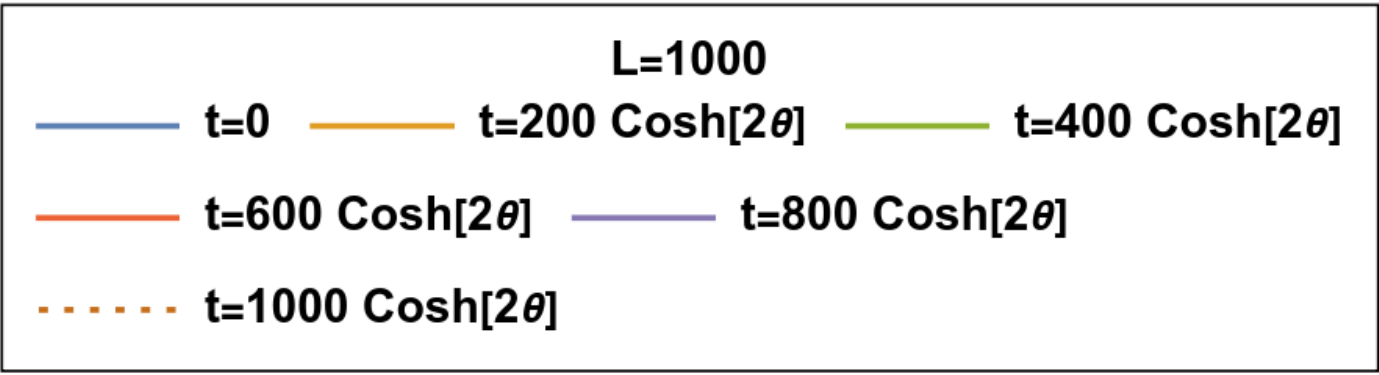}
\caption{Plots of the spatial profile of the energy-momentum tensor and heat current after a M\"{o}bius quench for $\theta=\frac{1}{10},\frac{1}{2},1$ at $t = 0,\frac{L}{5}\cosh{2\theta},\frac{2L}{5}\cosh{2\theta},\frac{3L}{5}\cosh{2\theta},\frac{4L}{5}\cosh{2\theta},L\cosh{2\theta}$ with the total system size fixed at $L=1000$.}
    \label{Mobius_EnergyMomentum}
\end{figure}

The spatial profiles for the energy-momentum tensor and the energy current for the M\"{o}bius and SSD quenches in the free fermion CFT are very similar to the pure state case. That is because the Schwarzian derivative in \eqref{MobiusEnergyMomentumTensor} turns out to be negligible compared to the conformal factor that comes from the Heisenberg evolution. The energy-momentum tensor is thus approximately determined by the first term in \eqref{MobiusEnergyMomentumTensor} where the conformal factor is theory-independent. The theory dependence only comes in through the one-point function of the energy-momentum tensor which is a time-independent quantity that only depends on the modular parameter of the torus. Therefore, the energy-momentum tensor for different CFTs differs only by a proportionality constant.

\section{Bulk geometry in the Schr\"{o}dinger picture}
\label{shroginder_geometry}

We have used the Heisenberg picture where the operators
transformed under the M\"obius/SSD quench
while the state remained unchanged from the original thermal state.
In this section, we discuss the Schr\"{o}dinger picture
where the state transforms under the inhomogeneous quench,
and study the gravitational dual of the SSD-quenched state.

\subsection{The SSD quench \label{geometry_SSD}}

As discussed in \cite{Ban_ados_1999, Roberts:2012aq}, the gravitational dual can be constructed from the expectation value of the energy density 
after the quantum quench.
This is equivalent to rewriting the geometry given by the static BTZ black hole in 
the $w^{\rm new}$ and $\bar{w}^{\rm new}$ coordinate system in terms of the original $X$ and time $t$ which parametrize the time evolution under 
the SSD Hamiltonian. The metric is given by
\be \label{metric_SSD}
\begin{split}
ds^2&= L^2\left[\f{dr^2}{r^2-r_0^2}-\left(f_{tt,1} r^2-f_{tt,2} \f{r_0^2}{4}\right)dt^2+\left(-f_{xx,1} r^2+f_{xx,2}\f{r_0^2}{4}\right)dx^2+r^2_0f_{tx}dtdx\right],\\
\end{split}
\ee
where the details of functions, $f_{xx,i=1,2},f_{tt,i=1,2}$ and $f_{tx}$, are reported in \cite{Goto:toappear}. 
We introduce a new radial coordinate $r'=r\sqrt{-f_{xx,1}}$. The geometry asymptotically approaches 
\be
ds^2\approx L^2\left[\f{dr'^2}{r'^2}+r'^2\left[-4\sin^4{\left(\f{\pi x}{L}\right)}dt^2+dx^2\right]\right]\, ,
\ee
as $r'\rightarrow \infty$, where the dual CFT lives. Notice that  this boundary metric is sine-square 
deformed from the usual flat metric \cite{MacCormack_2019,wen2018floquet}.
Since the horizon sits at $r=r_0$,
we identify
the location of the horizon by
\be \label{time_dep_horizon}
r'_{\text{horizon}}=r_0\sqrt{-f_{xx,1}}
\ee
in the $r'$ coordinate.
The position of the horizon depends on the spatial coordinate $X$, and has a peak at
\be\label{KGSSD}
t=\frac{L \sqrt{1-4\sin^2 \left(\frac{\pi  X}{L}\right)}}{2 \pi  \sin \left(\frac{\pi  X}{L}\right)}\, ,
\ee
where $0<X<L/6, 5L/6<X<L$. This is obtained by solving the equation
$\partial_Xr'_{\rm horizon}=0$ with respect to $t$.
See 
Fig.\ \ref{horizonSSD2} for 
the plot of the profile of the horizon.

By plugging this into (\ref{time_dep_horizon}), we obtain the $X$ dependence of the horizon at time (\ref{KGSSD})
\be
r'_{\text{horizon}}=\f{r_0}{2\left|\sin{\left(\f{\pi X}{L}\right)}\right|}\, .
\ee
The position of the peak at time $t$ is given by 
\be\label{Xhorizon}
\begin{split}
    X=\begin{cases}
        \f{L}{\pi} \tan^{-1}{\left(\frac{L}{\sqrt{3L^2+4\pi^2 t^2}}\right)} & \f{L}{4}\le X\le0\\
        - \f{L}{\pi} \tan^{-1}{\left(\frac{L}{\sqrt{3L^2+4\pi^2 t^2}}\right)} & L\le X\le\f{3L}{4}\\
    \end{cases}\, .
\end{split}
\ee
This is obtained by solving  the equation $\partial_Xr'_{\rm horizon}=0$ with
respect to $X$. 
Notice that when $L\ll t$, the peak is located at $X=0$. The time-dependence of the horizon at $x=X$ is obtained by plugging  this into (\ref{time_dep_horizon})
\be\label{KGhorSSD}
r'_{\text{horizon}}=\frac{r_0\sqrt{L^2+\pi ^2 t^2}}{L}\, .
\ee
The value of this peak can be approximated by $r'_{\text{horizon}}\approx \f{r_0 t}{L}$ at late times $L\ll t$, and it grows linearly with time.
In the spatial region $\f{L}{6}<X<\f{5L}{6}$, the size of the horizon decreases monotonically. 

The peak of the horizon in \eqref{Xhorizon} can be compared
with the peak of the boundary energy-momentum tensor in \eqref{eq:XTmax}
(Fig.\ \ref{fig:peakcomp}). Clearly, the positions of these peaks coincide at late times. 
\begin{figure}[tbp]
        \centering
      \includegraphics[keepaspectratio, scale=0.5]{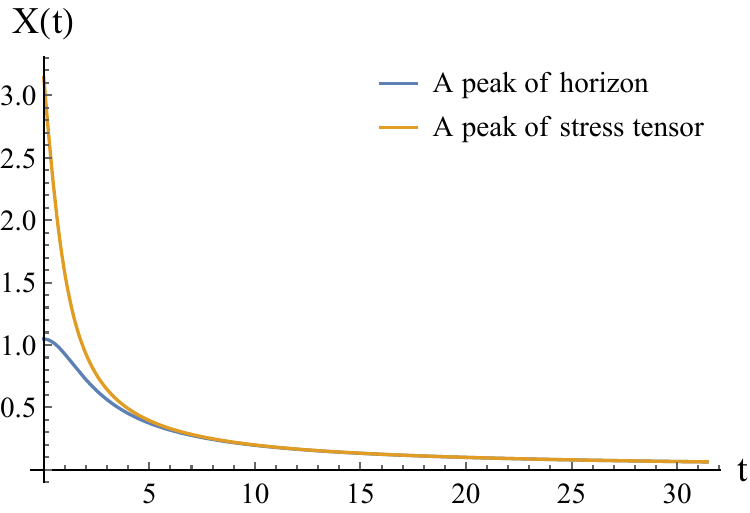} \caption{
      The peaks of the horizon and the energy-momentum tensor (we set $L=2\pi$). At late times, the location of one of the peak of 
      the horizon coincides with the propagation of the holomorphic energy-momentum tensor. Another peak coincides with anti-holomorphic one. The peak of energy-momentum tensor at $t=0$ coincides with $X=L/2$.}
    \label{fig:peakcomp}
  \end{figure}

\subsection{The M\"obius quench \label{geometry_mobius}}
The metric for the state quenched by the M\"obius Hamiltonian can be discussed similarly.
Introduce a new radial
coordinate $r'=r\sqrt{-f_{xx,1}}$ similarly to the SSD case. The position of the
horizon in $r'$ coordinate is shown in Fig.\ \ref{horizonSSD2}. The time
dependence of the position of this horizon has the periodicity
$2\pi/\Omega = L \cosh{2\theta}$.

The time when the horizon saddle appears at position $X$ is given by
\be\label{KGMobius}
t=\frac{L\cosh{2\theta}}{\pi} \tan^{-1}\frac{\sqrt{2 \tanh 2\theta \cos
    \left(\frac{2 \pi  X}{L}\right)-1}}{ \sqrt{\cosh 4\theta-\sinh 4\theta \cos
    \left(\frac{2 \pi  X}{L}\right)}},\,
\quad
t= m L \cosh{2\theta}
\ee
where $m$ are integers. This is given by solving  the equation $\partial_Xr'_{\rm horizon}=0$ with respect to $t$. This approaches (\ref{KGSSD}) in the SSD limit $\theta\rightarrow \infty$. By plugging this into $r'_{\rm horizon}=r_0\sqrt{-f_{xx,1}}$, we obtain the $X$ dependence of the horizon at time (\ref{KGMobius}),
\begin{align}
r'_{\text{horizon}}=\f{2r_0\sinh 2\theta}{\sqrt{\cosh4\theta-\cos\frac{2\pi X}{L}\sinh 4\theta}}.
\end{align}

The position of the peak at time $t$ is given by
\begin{align}
  X=\f{L}{\pi} \tan^{-1}e^{-2 \theta }
  \sqrt{\frac{(\sinh 4\theta-2)+(\cosh 4 \theta-1)
  \cos \left(
  \Omega t
  \right)}{(\sinh 4\theta+2)-(\cosh 4 \theta-1)
  \cos \left(
  \Omega t
  \right)}}\, .
\end{align}
This is given by solving  the equation $\partial_tr'_{\rm horizon}=0$ with respect to $X$.
 The time dependence of  the radius of the horizon at $x=X$ is given by plugging  this into $r'_{\rm horizon}=r_0\sqrt{-f_{xx,1}}$ as
\be
r'_{\text{horizon}}=r_0 
\sqrt{\cosh ^2 2\theta \sec ^2\left(
t \Omega/2
\right)-\sinh ^22\theta}\, .
\ee

\subsection{Time evolution of bulk excitations in the SSD/M\"obius quench  \label{App:evolution-of-bulkexc}}

We studied the time-evolution of the black hole horizon
in the SSD and M\"{o}bius quenches.
It is known that thermal states are well
approximated by the high-energy eigenstates. In this section, we consider the states created by inserting spinless primary operators ${\cal O}_h$
with conformal dimensions $h_L=h_R=h$ on the vacuum state.
A primary operator ${\cal O}_h$
with large dimension $h>\frac{c}{24}$ creates a black hole in AdS with temperature $
 T=\frac{1}{2\pi}\sqrt{\frac{24h}{c}-1}$ while one with small dimension $c\gg h$ creates a small bulk excitation on the pure AdS created by the matter fields dual to ${\cal O}_{h}$.
In this section, we consider how this small bulk excitation time evolves under the SSD and M\"{o}bius quenches.

 Let us insert a primary operator at the center of the Euclidean plane $z=\bar{z}=0$, i.e., $\tau=-\infty$ in Euclidean time,
\begin{align}
    |\psi_{{\cal O}_h}\rangle={\cal O}_{h,\bar{h}}(z=0,\bar{z}=0)|0\rangle\, .
\end{align}
This is dual to a bulk excitation centered at the origin of the AdS spacetime.
We are interested in how this excitation moves as it is time-evolved by the SSD
or  M\"obius Hamiltonian. The dual CFT state we will consider is given by
\begin{align}
    |\psi_{{\cal O}_{h,\bar{h}}}(t)\rangle=e^{-i H_{\text {SSD }} t}{\cal O}_{h,\bar{h}}(z=0,\bar{z}=0)|0\rangle\, .\label{SSDquenched}
\end{align}
The strategy that we will use here is summarized in \cite{Goto:2017olq}, where they move the bulk excitation by acting with the corresponding bulk $SL_2$ generators. We will explain the details below.
Let us assume the primary operator is decomposed into products of chiral and anti-chiral parts $
    {\cal O}_{h,\bar{h}}(z,\bar{z})=\sum_{i}{\cal O}_{h}^i(z){\cal O}_{\bar{h}}^i(\bar{z})$
and consider the chiral state created only by ${\cal O}_{h}^i(z)$ for simplicity $
|\psi_{{\cal O}_L}\lb={\cal O}_{h}(z=0)|0\lb$. It corresponds to a spinning BTZ black hole (or a small spinning excitation in the pure AdS) with mass $m =\s{h(h-2)}$ and spin $s=h$.

\paragraph{ The SSD Hamiltonian as the time-translation in the Poincar\'e coordinate} 
Before considering the evolution of the black hole or the bulk excitation
generated by the SSD/M\"{o}bius Hamiltonian,
we introduce new coordinates in which actions of these Hamiltonians
become simple.

The evolution  under the SSD Hamiltonian is simplified by introducing the boundary Poincar\'e coordinate $(z_P,\bar{z}_P)$. The boundary global coordinate $(w,\bar{w})$ and the boundary Poincar\'e coordinate $(z_P,\bar{z}_P)$ are related as
\begin{align}
    iz_P=L\cot\left(\frac{i\pi w}{L}\right)\, ,\quad  i\bar{z}_P=L\cot\left(\frac{i\pi \bar{w}}{L}\right)\, ,\label{poincare}
\end{align}
where $z_P=\tau_P+ix_P$ and $\bar{z}_P=\tau_P-ix_P$ are the complex coordinates in the Poincar\'e coordinate. The symbol $\tau_P$ is the Euclidean time coordinate, and $x_P$ is the spatial  coordinate ($-\infty<x_P<\infty$) in the plane where the Poincar\'e coordinate is defined. The two fixed points of the SSD Hamiltonian is located at the origin and the spatial infinity. 

Now let us see how the Poincar\'e coordinate simplifies the translation under the SSD Hamiltonian.
The flow of the Poincar\'e time is generated by the following Hamiltonian
\begin{align}
    H_P=\int^\infty_{-\infty}dx_P T_{\tau_P\tau_P}(x_P)=-i\int dz_P T(z_P)-i\int d\bar{z}_P \overline{T}(\bar{z}_P)\, .
\end{align}
We use the usual transformation rule for the energy-momentum tensor
$\left(\frac{d z_{P}}{d w}\right)^{2} T\left(z_{P}\right)
  =
    T(w)-\frac{c}{24 \pi}
    \mathrm{Sch}\left(z_{P}, w \right)$
with $\frac{d z_{P}}{d w}=\frac{\pi}{L\sin^2\left(\frac{i\pi w}{L}\right) }$ and move to the original global coordinate $(w,\bar{w})$ as
\begin{align}
H_{P} &=\oint \frac{d w}{iL}\left(\frac{d w}{d z_{P}}\right)\left(T(w)+\frac{\pi c}{12L^2}\right)+\oint \frac{d \bar{w} }{iL}\left(\frac{d \bar{w}}{d \bar{z}_{P}}\right)\left(\overline{T}(\bar{w})+\frac{\pi c}{12L^2}\right) =\int \frac{d x}{2\pi }2\sin^2\left(\frac{\pi x}{L}\right)T_{\tau\tau}(x)+\frac{c}{12L}
\nonumber\\
&=\frac{1}{2\pi}H_{\rm SSD}+\frac{c}{12L}\, .\label{PoincaretoSSD}
\end{align}
Therefore, the SSD Hamiltonian generates the time-flow in the Poincar\'e
coordinate defined as (\ref{poincare}). This indicates that a black hole (or the
bulk excitation) dual to the state (\ref{SSDquenched}) moves along the
Poincar\'e time direction.
Static objects in the bulk Poincar\'e coordinate are seen as ones falling to the
asymptotic boundary of the AdS over an infinitely long time from the perspective
of the boundary observer in the global coordinate. Thus, we expect that the black
hole (or the bulk excitation) after the SSD quench
gets closer and closer to the AdS boundary as time evolves. We will justify this expectation by explicit computations in the following. 

\paragraph{General M\"{o}bius case}
The general M\"{o}bius Hamitonians can also be identified to the generators of time directions in new coordinate systems $(z_\theta,\bar{z}_\theta)$. The relation to the original global coordinate is given by 
\begin{align}
    \tan \frac{iz_\theta}{2L\cosh 2 \theta}=e^{-2 \theta} \cot\left(\frac{i\pi w}{L}\right)\, ,\quad \tan \frac{i\bar{z}_\theta}{2L \cosh 2 \theta}=e^{-2 \theta} \cot\left(\frac{i\pi \bar{w}}{L}\right)\, . \label{Mobiusmap}
\end{align}
$(z_\theta,\bar{z}_\theta)$ approaches the Poincar\'e coordinate $(z_P,\bar{z}_P)$ (\ref{poincare}) as we send $\theta\rightarrow \infty$.
Let us check the Hamiltonian $H_{\theta}$ associated to this new coordinate $(z_\theta,\bar{z}_\theta)$ indeed gives the M\"{o}bius Hamitonian.
The Hamiltonian $H_{\theta}$ is given by 
\begin{align}
H_{\theta}=-i\int dz_\theta T(z_\theta)-i\int dz_\theta \overline{T}(\bar{z}_\theta)\, .
\end{align}
The transformation
$\left(\frac{d z_{\theta}}{d w}\right)^{2}T(z_{\theta})
  =T(w)+\frac{\pi c}{12L^2}-\frac{\pi c}{12L^2\cosh^2\theta}\frac{1}{(1-\tanh2\theta\cos\left(\frac{2i\pi w}{L}\right))^2
}$
with $
    \frac{d z_{\theta}}{d w}=\frac{2\pi}{1-\tanh2\theta\cos\left(\frac{2\pi x}{L}\right)}$ yields
\begin{align}
H_{\theta}&=\int \frac{dx}{2\pi} \left(1-\tanh2\theta\cos\left(\frac{2i\pi w}{L}\right)\right) T_{\tau\tau}(x)+\frac{c}{12L}\, \nonumber\\
&=\frac{1}{2\pi}H_{
\text{M\"obius}}+\frac{c}{12L}\, .
\end{align}
Therefore, the M\"obius Hamiltonian generates the time-flow in the coordinate defined as (\ref{Mobiusmap}).

\paragraph{Map between the boundary and the bulk}
We now return to the time-evolution of the CFT state dual to the black hole or the bulk excitation. First, we simply consider how the chiral part is time-evolved by the SSD Hamiltonian
\begin{align}
    \left|\psi_{{\cal O}_h}^{\text {SSD }}(t)\right\rangle=e^{-i H_{\text{SSD}} t}{\cal O}_h(z=0)|0\lb\, ,\label{SSDquenched2}
\end{align}
and see how the corresponding bulk excitation moves as time evolves.
Let us insert a primary operator at the infinite past in the Euclidean global
coordinate $w=-\infty$, i.e.,
$z=0$ in the plane coordinate given by the exponential map
$z=e^{\frac{2\pi w}{L}}$ from $w$.
This corresponds to inserting the operator at $z_P=L$ ($\tau_P=L$)
in the Poincar\'e coordinate.
This insertion of the primary operator creates a bulk excitation (or a black
hole) centered at the origin of the AdS at $t=0$.
This corresponds to $(\zeta,x_P)=(1,0)$ at the same time slice in the bulk Poincar\'e coordinate, where $\zeta$ is the coordinate corresponding to the bulk direction.

The action of $e^{-i H_{\text {SSD }} t}$ moves the operator to a new point $z_{\rm new}$ on the Euclidean boundary.
This creates the bulk excitation centered at the corresponding bulk point,
which is away from the origin as depicted in Fig.\ \ref{fig:geodesics-main}.
The bulk point is determined by the intersection between $t=0$ slice of the AdS and the geodesic in the Euclidean AdS starting from $z_{\rm new}$ as pointed out in \cite{Goto:2017olq}. Let us remind ourselves that the evolution generated by the SSD Hamiltonian
gives the time evolution in the boundary Poincar\'e coordinate.
Therefore, the chiral primary operator ${\cal O}_h$ inserted at the origin of the
$z$ coordinate (equivalently at $z_P=L$ in the Poincar\'e coordinate)
moves as
\begin{align}
    e^{-i H_{\mathrm{SSD}} t} {\cal O}_h(z_P=L) e^{i H_{\mathrm{SSD}} t}={\cal O}_h(z_P=L-it).
\end{align}
Since $z_P=\tau+ix_P$, we can interpret it that in
the Euclidean regime, the insertion point moves from $x_P=0$ to $x_P=-t$ on the
$\tau_P=L$ slice. This is schematically drawn as Fig.\ \ref{fig:geodesics-main}.
Correspondingly, the bulk excitation, which is originally centered at the origin
of the AdS: $(\zeta,x_P)=(1,0)$,
moves to $(\zeta,x_P)=(1,-t)$ by action of the SSD Hamiltonian as depicted in
Fig.\ \ref{Fig:excitations-main}.
Thus, the bulk excitation corresponding to the chiral part of the original
primary operator approaches to the fixed point of the SSD quench,
i.e., $x_P=\infty$ in the Poincar\'e coordinate,
$x=0$ in the global coordinate while rotating in the negative direction of $x$ (and $x_P$). Similarly the anti-chiral part moves as
\begin{align}
    e^{-i H_{\mathrm{SSD}} t} {\cal O}_{\bar{h}}(\bar{z}_P=L) e^{i H_{\mathrm{SSD}} t}={\cal O}_{\bar{h}}(\bar{z}_P=L-it)\, ,
\end{align}
thus the corresponding bulk excitation moves from $(\zeta,x_P)=(1,0)$ to $(\zeta,x_P)=(1,t)$. That is, it approaches the fixed point of the SSD quench while rotating in the direction of positive $x$ (and $x_P$). Since the original scalar primary operator ${\cal O}_{h,\bar{h}}$ is created by the products of the chiral and anti-chiral parts, the bulk excitation corresponding to ${\cal O}_{h,\bar{h}}$ just approaches the fixed point along $x_P=0$ without rotation.

To see the profile of the time-evolved bulk excitation more explicitly, let us compute the overlap between the state 
$|\phi_{{\cal O}_{h,\bar{h}}}(\zeta, x_P)\rangle=\phi_{{\cal O}_{h,\bar{h}}}(\zeta, x_P)|0\rangle
$
excited by the bulk local operator $\phi_{{\cal O}_{h,\bar{h}}}$
dual to ${\cal O}_{h,\bar{h}}$ and the excited states
evolved under the SSD Hamiltonian $|\psi_{{\cal O}_{h,\bar{h}}}^{\text {SSD
  }}(t)\rangle$ (\ref{SSDquenched}).
The excitation of $|\phi_{{\cal O}_{h,\bar{h}}}(\zeta, x_P)\rangle
$ is localized at the bulk point $(\zeta, x_P)$.

When the primary operator is inserted at the origin $(\tau_P=0,x_P=0)$ in the boundary Poincar\'e coordinate, the overlap is just given by the usual bulk-to-boundary propagator
\begin{align}
    \left\langle\phi(\zeta, x_P)|{\cal O}_{h,\bar{h}}\left(\tau_{P}=0,x_P=0\right)\right\rangle=\frac{\zeta^{2 h}}{\left(\zeta^{2}+x_P^{2}\right)^{2 h}}\, .
\end{align}
The SSD quenched state $|\psi_{{\cal O}_{h,\bar{h}}}^{\text {SSD }}(t)\rangle$ can be obtained by inserting the primary operator at $z_P=L-it,\bar{z}_P=L-it$. In the Lorentzian regime obtained by $\tau_P\rightarrow \tau_P-it_P$, the complex coordinate becomes $z_P=\tau_P-i(t_P-x_P), \bar{z}_P=\tau_P-i(t_P+x_P)$.
Therefore, we can regard the operator as being inserted  at a complex time $t_p=t+iL$ in the Poincar\'e coordinate.
Thus, simple modifications to the bulk-to-boundary propagator above lead to
\begin{align}
   \langle\phi(\zeta, x_P)|\psi_{{\cal O}_{h,\bar{h}}}^{\text {SSD }}(t)\rangle=\frac{\zeta^{2 h}}{\left(\zeta^{2}+x_P^{2}-(t+iL)^2\right)^{2 h}}\, ,\label{overlap}
\end{align}
for the overlap between the bulk locally excited state $|\phi_{{\cal
    O}_{h,\bar{h}}}(\zeta, x_P)\rangle$ and the SSD quenched state $|\psi_{{\cal
    O}_{h,\bar{h}}}^{\text {SSD }}(t)\rangle$.  We plot the contours corresponding to $|\langle\phi(\zeta, x_P)|\psi_{{\cal O}_{h,\bar{h}}}^{\text {SSD }}(t)\rangle|=1$ for several values of $t$ in Fig.\ \ref{BTZSSD}.
As we expected, the bulk excitation approaches the fixed point without rotation as time evolves. Moreover, by properly shifting the center of each excitation using the AdS isometry, the contours nicely match those for the black hole horizon Fig.\ \ref{horizonSSD2}.

\begin{figure}[t!!]
\begin{center}
  \includegraphics[width=10cm]{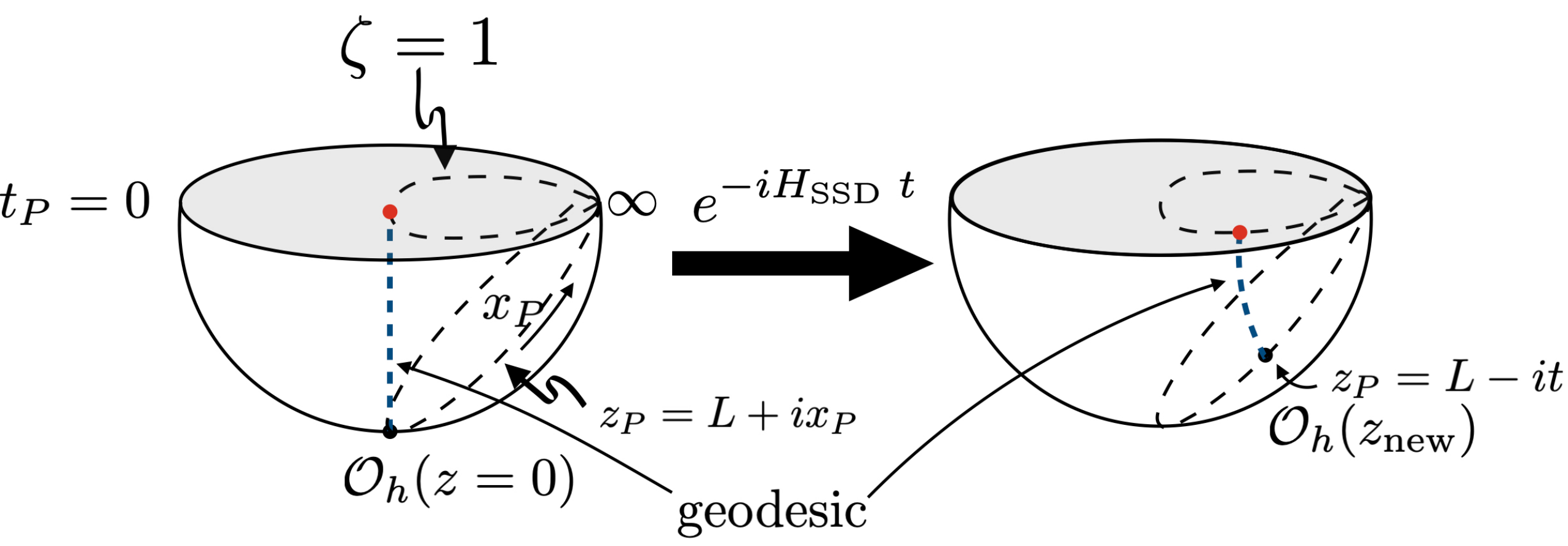}
  \caption{The gray disks correspond to the $t_P=0$ in the Poincar\'e coordinate (i.e.,  $t=0$ in the global coordinate) slices of the bulk AdS spacetime. The hemisphere attached to each disk represents the Hyperbolic disk corresponding to $-\infty<\tau<0$ ($|z|<1$) in the Euclidean AdS boundary. Left: Inserting a chiral primary operator ${\cal O}_h$ at the origin of the hyperbolic disk $z=0$ (i.e., $\tau=-\infty$) creates the bulk excitation around the center of the AdS, i.e., $\zeta=1, x_P=0$ on the $t_P=0$ slice in the bulk Poincar\'e coordinate. Right: The evolution under the SSD Hamiltonian $e^{-iH_{\rm SSD}t}$ moves the position of the primary operator along $z_P=L-it$ in the Poincar\'e coordinate. Correspondingly, the center of the bulk excitation also moves to $\zeta=1, x_P=-t$ on the $t_P=0$ slice. The map between the boundary and the bulk is simply represented by the geodesic connecting them.  }\label{fig:geodesics-main}
\end{center}
\end{figure}
\begin{figure}[t!!]
\begin{center}
  \includegraphics[width=5.5cm]{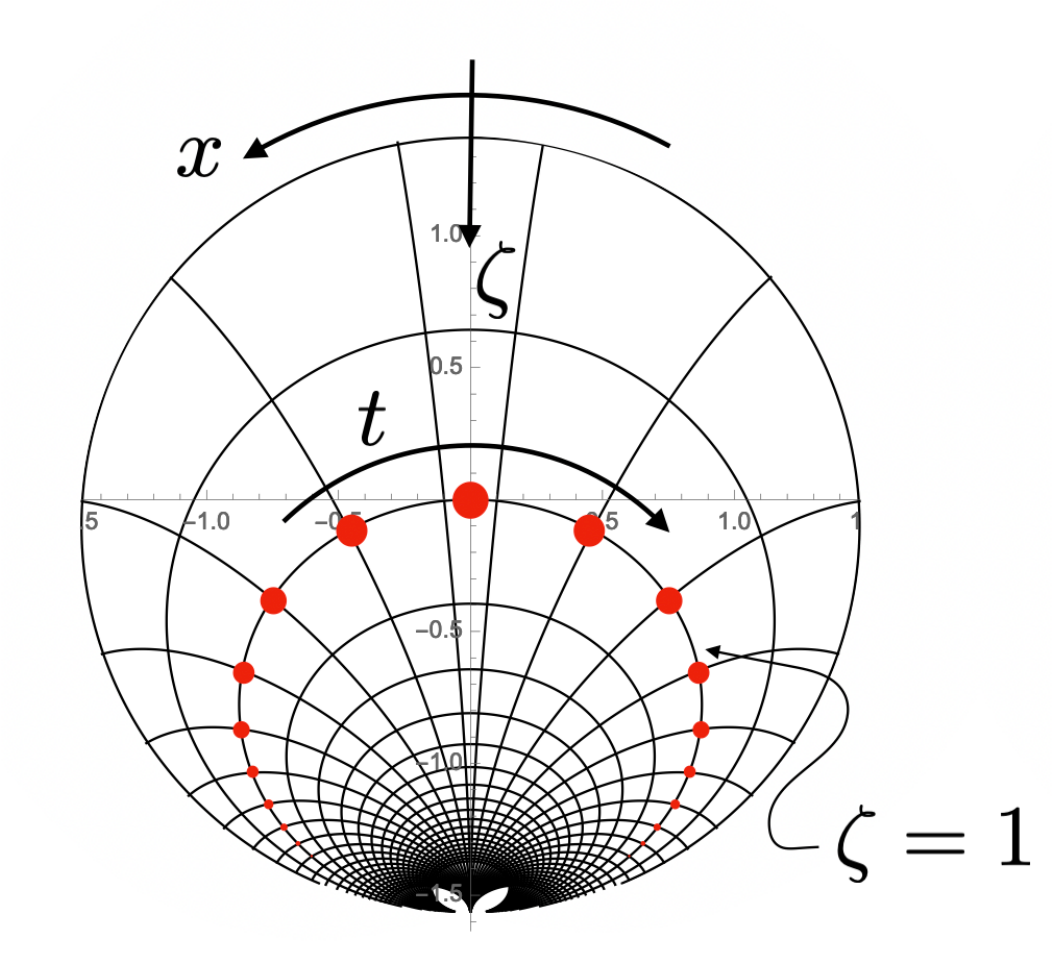}
  \caption{The bulk excitation corresponding to the time-evolved CFT chiral primary operator $e^{-i H_{\text{SSD}} t}{\cal O}_{h}(z=0)e^{i H_{\text{SSD}} t}$ drawn in the global AdS. The outermost circle corresponds to the boundary of the AdS, where $\zeta =0$ in the Poincar\'e coordinate. The point at the bottom of the circle corresponds to infinity in the Poicar\'e coordinate: $\zeta =\infty$. The distorted circles with $\zeta =\infty$ as a fixed point are the constant $\zeta$ slices. The curves orthogonal to them corresponds to the constant $x_P$ slices. The center of the bulk excitation at each time $t$ is depicted as a red point. It moves along the $\zeta =1$ slice as time evolves. }\label{Fig:excitations-main}
\end{center}
\end{figure}

\section{Mutual information}
\label{Mutual information}

The von Neumann entropy for single intervals
(Sec.\ \ref{Entanglement entropy for single intervals})
and 
the energy density/current
(Sec.\ \ref{sec:stress})
are found to be insensitive to 
the details of CFTs.
However, the theory dependence should show up
in more complex probes 
such as
mutual information defined for two intervals
and higher-point correlation functions.
In this section, we consider mutual information
for the free fermion CFT and holographic CFT.

We first recall that for two subsystems $A$ and $B$, the mutual information $I_{A,B}$ is defined by a
linear combination of entanglement entropy (von Neumann entropy):
\be
I_{A,B}=S_A+S_B-S_{A\cup B}.
\ee
We note that the mutual information
is free from the UV divergence when $A \cap B=0$,
i.e., $I_{A,B}$ is finite even if the lattice spacing is $0$,
while keeping $\epsilon$ (inverse temperature) finite.
Our choices of the subsystems (subintervals) will be given below.

For both free fermion and holographic CFTs,
we will find that
the mutual information after the SSD quench,
at late enough times,
is essentially given by
the mutual information of the uniform ground state
(except in some special cases
where one of the endpoints of the subsystems is on the fixed point): 
The difference between $I_{A,B}$ for
our state
$\rho(t)= e^{-i t H_{\mathrm{SSD}}}\rho(0) e^{i t H_{\mathrm{SSD}}}$
and the state in \eqref{frho}
is at most $\mathcal{O}(\epsilon)$ in the coarse-grain limit, $\epsilon \ll 1$. We will thus confirm 
that the density operator at late times
can be approximated by \eqref{frho}.
Namely,
``reverse'' thermalization occurs by the SSD quench 
where
the correlations between subsystems of the initial thermal state
undergo
a crossover to those of the ground state without any non-unitary operations. 
We will also discuss a finer structure of the late time state beyond the approximation \eqref{frho}. 

\subsection{Holographic CFT}

\begin{figure}[bp]
    \begin{tabular}{cc}
      \begin{minipage}[t]{0.5\hsize}
        \centering
        \includegraphics[keepaspectratio, scale=0.12]{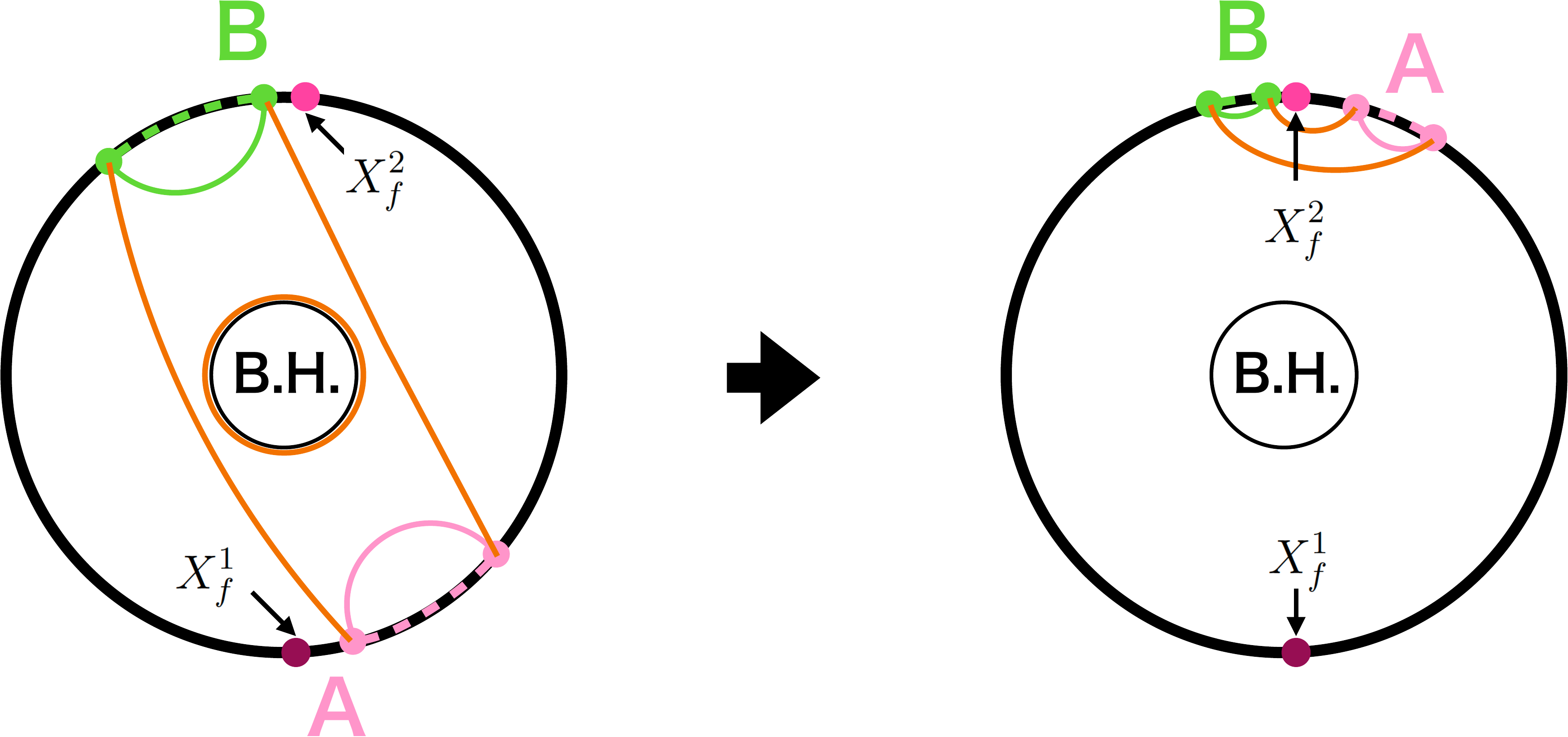}
        \\
        Case (a)
      \end{minipage} &
      \begin{minipage}[t]{0.5\hsize}
        \centering
        \includegraphics[keepaspectratio, scale=0.12]{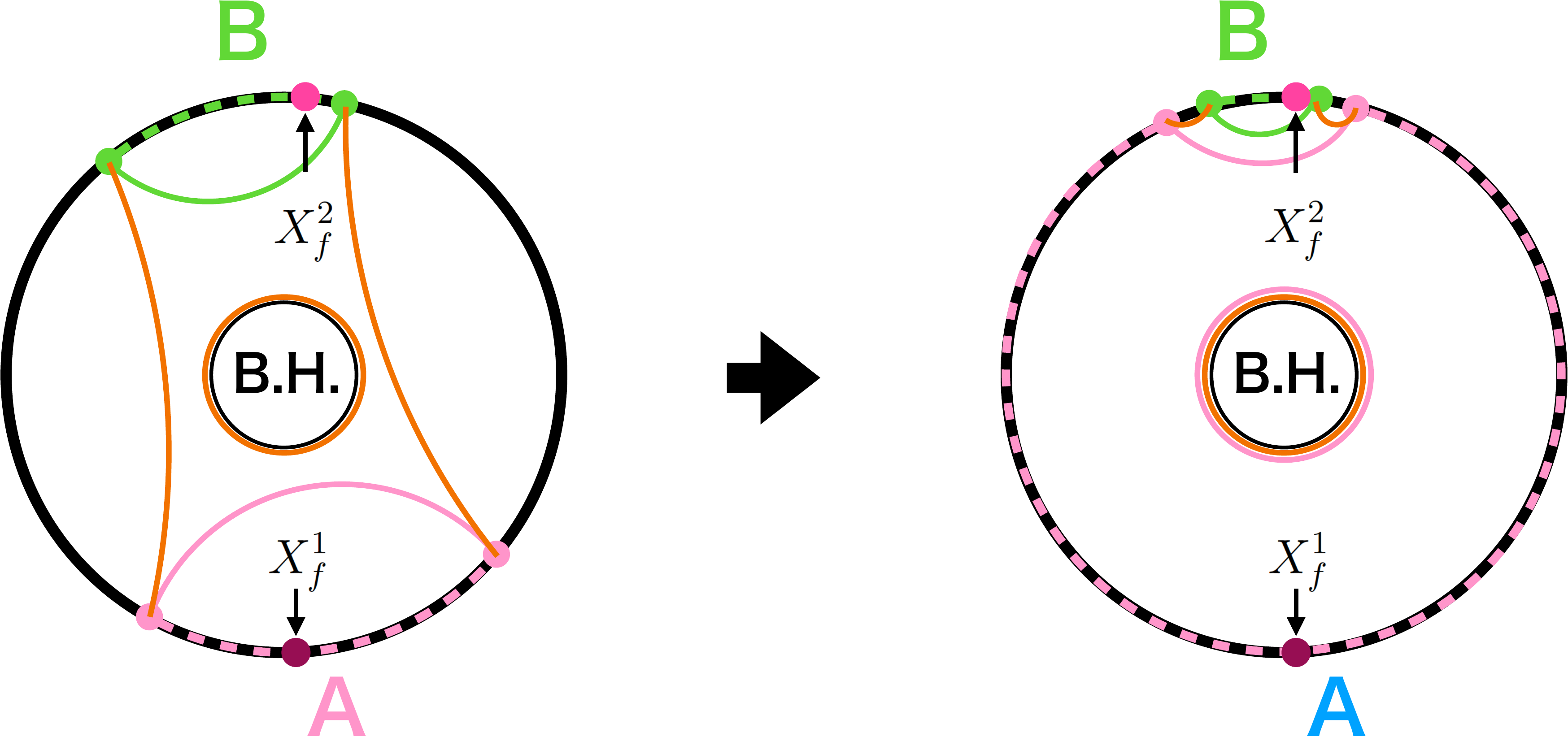}
        \\
        Case (b)
      \end{minipage} \\
  
    \end{tabular}
    \caption{
      A sketch of the time evolution of the minimal surfaces
      for $S_A$ (pink solid), $S_B$ (green solid)
      and $S_{A\cup B}$ (orange solid)
      in the Heisenberg picture.
      The subsystems $A$ and $B$ are represented by
      pink dotted and green dotted lines, respectively.
      \label{time_ev_MI}}
  \end{figure}

Let us first discuss
the time-evolution of the mutual information in holographic CFT
after the SSD quench.
We consider two cases and take subsystems $A$ and $B$ as follows.
In case (a), the fixed point $x=X^1_f$ is included
neither in subsystem $A$ nor in $B$ (Fig.\ \ref{time_ev_MI}(a)).
On the other hand, in case (b), the fixed point $x=X^1_f$ 
is included in subsystem $A$ but not in $B$ (Fig.\ \ref{time_ev_MI}(b)).
In the Heisenberg picture,
the twist and anti-twist operators defining the subsystems
flow and,
after enough time has passed,
meet  
at the other fixed point $x=X^2_f$
(both in case (a) and (b) -- see Fig.\ \ref{time_ev_MI}).
As a result, the minimal surface for $S_B$ leaves the black hole,
so $S_B$ becomes independent of temperature
and can be approximated by the entanglement entropy of the ground state
$S^{\text{vac}}_B$.
In case (a), after enough time has passed,
the minimal surfaces for $S_A$ and $S_{A\cup B}$ are
far enough away
from the black hole so that $S_A$ and $S_{A\cup B}$ can be approximated
by the entanglement entropy of the ground state,
$S^{\text{vac}}_A$ and $S^{\text{vac}}_{A\cup B}$, respectively.
On the other hand,
in case (b),
at late times,
the minimal surface for $S_A$
wraps around the black hole and
the minimal surfaces for $S_{A\cup B}$
are located near the boundary of the AdS.
Consequently,
$S_A$ and $S_{A\cup B}$ are given
by the thermal entropy $S_{\text{thermal}}$
and the entanglement entropy of the ground state,
$S^{\text{vac}}_A$ and $S^{\text{vac}}_{A\cup B}$, respectively.
To summarize,
in both cases (1) and (2), after enough time has passed,
$I_{A,B}$ is given by the mutual information $I^{\text{vac}}_{A,B}$
of the ground state, i.e.,
\be
I_{A,B}\approx I^{\text{vac}}_{A,B}= \text{Max}\left[0, S^{\text{vac}}_A+S^{\text{vac}}_B-S^{\text{vac}}_{\text{Con.}}\right],
\ee
where $S^{\text{vac}}_{\text{Con.}}$ is the area of the minimal surface connecting the subsystems in the vacuum state.

\subsection{The free fermion CFT}

The mutual information for the case of free fermion CFT, 
for both SSD and M\"obius quenches, 
can be computed explicitly (using the bosonization approach,
that is also used in the computation of the von Neumann entropy).
It can be expressed 
as a sum of the spin-structure 
independent and dependent terms as
\begin{equation}\label{FreeFermionTotalMI}
I_{A\cup B,\nu\\}^{(N)} = I_{A\cup B,\text{univ.}}^{(N)}+ 	I_{A\cup B,\nu,\text{non-univ.}}^{(N)}
\end{equation}
where 
\begin{align}\label{FreeFermionMIUnivAndNonUniv} I_{A\cup B,\text{univ.}}^{(n)}=&
	\frac{n+1}{12n}
	\log\left|\frac{\vartheta_1\left(\frac{w_{X_2}^{\text{new}}-w_{X_4}^{\text{new}}}{2\epsilon}|\tau\right)
		\vartheta_1\left(\frac{w_{X_1}^{\text{new}}-w_{X_3}^{\text{new}}}{2\epsilon}|\tau\right)
	}{\vartheta_1\left(\frac{w_{X_2}^{\text{new}}-w_{X_3}^{\text{new}}}{2\epsilon}|\tau\right)
		\vartheta_1\left(\frac{w_{X_4}^{\text{new}}-w_{X_1}^{\text{new}}}{2\epsilon}|\tau\right)
	}\right|^2,
	\\ \nonumber
	I_{A\cup B,\nu,\text{non-univ.}}^{(n)}=&
	\frac{1}{1-n}\sum_{k=-\frac{n-1}{2}}^{\frac{n-1}{2}} \log\left|
	\frac{
		\vartheta_\nu\left(\frac{k}{n}\frac{w_{X_1}^{\text{new}}-w_{X_2}^{\text{new}}}{2\epsilon}\bigg|\tau\right)
		\vartheta_\nu\left(\frac{k}{n}\frac{w_{X_3}^{\text{new}}-w_{X_4}^{\text{new}}}{2\epsilon}\bigg|\tau\right)
	}{\vartheta_\nu(0|\tau)\vartheta_\nu\left(\frac{k}{n}\frac{w_{X_1}^{\text{new}}-w_{X_2}^{\text{new}}+w_{X_3}^{\text{new}}-w_{X_4}^{\text{new}}}{2\epsilon}\bigg|\tau\right)}\right|^2
\end{align}

\begin{figure}
	\centering
	\includegraphics[scale=0.32]{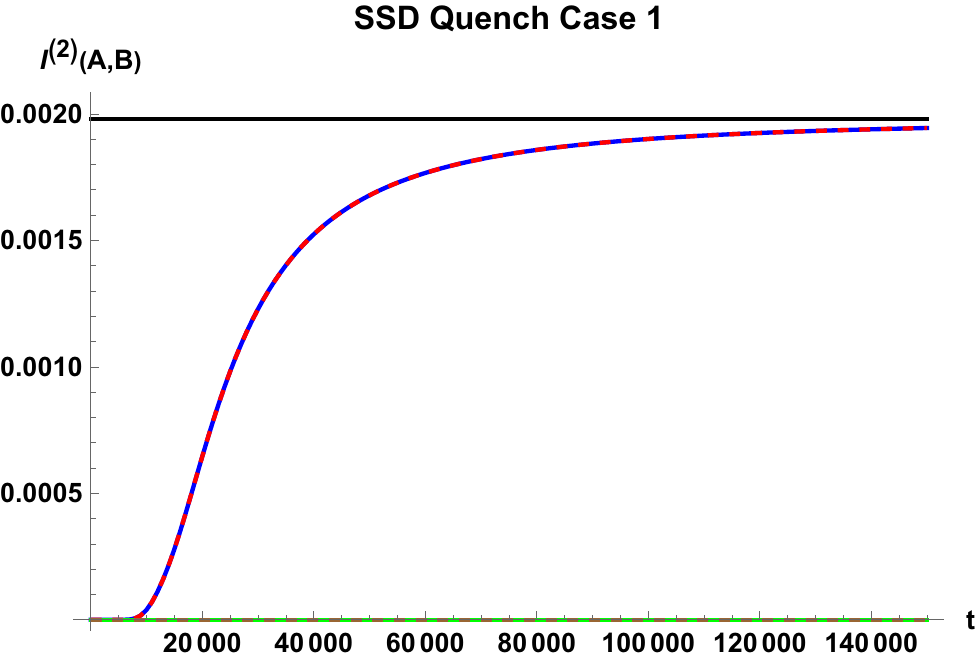}
	\includegraphics[scale=0.32]{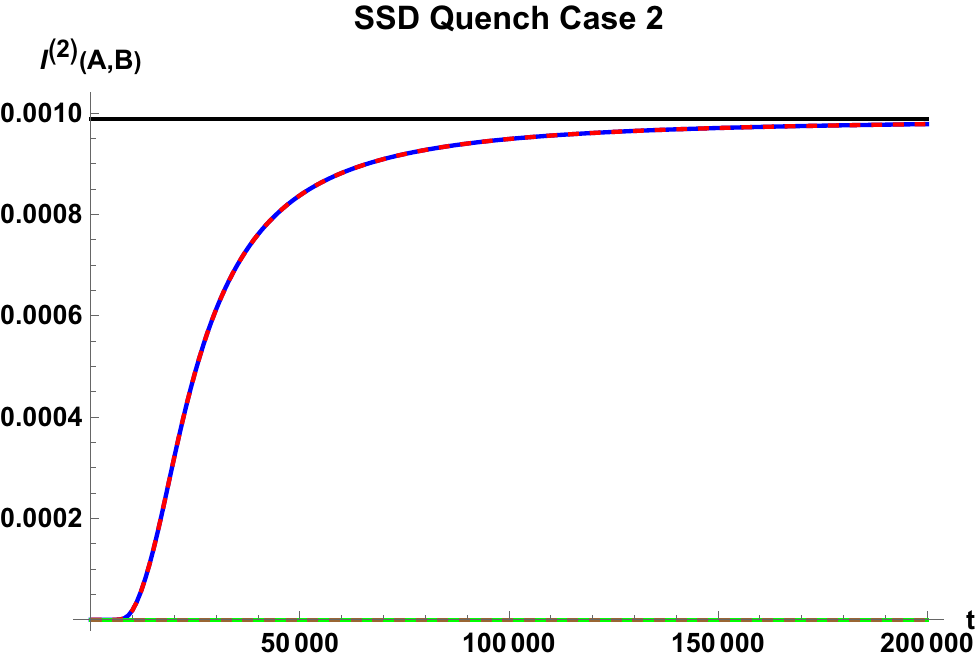}
	\includegraphics[scale=0.32]{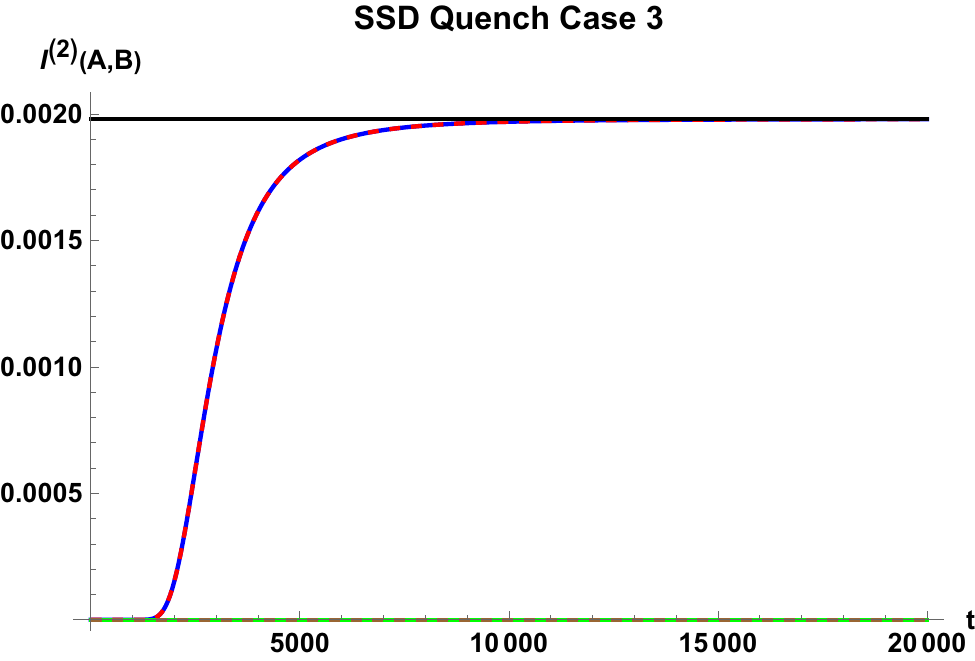}
	\includegraphics[scale=0.3]{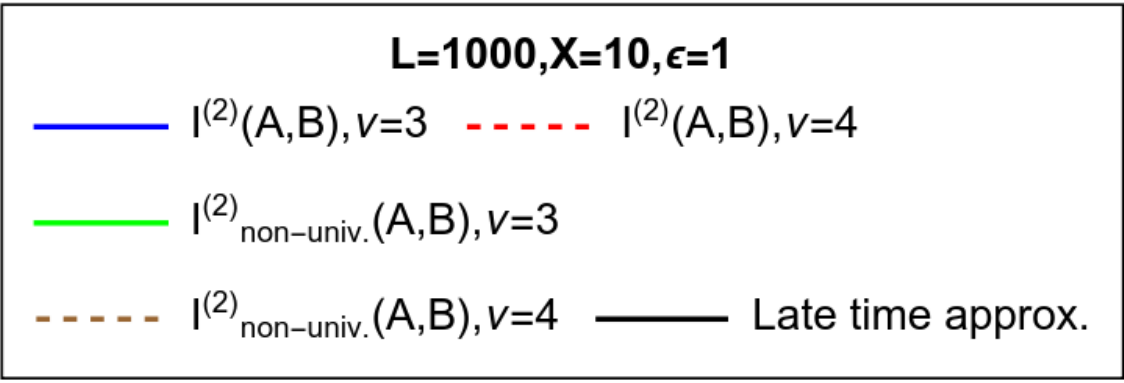}
	\caption{
	The second R\'{e}nyi mutual information of the thermal state after the SSD quench for the three cases listed in \eqref{setup:MI} with $L=1000$, $X=10$ and $\epsilon=1$ for the two physical spin structures $\nu=3,4$. 
	The blue and red curves correspond to the total mutual information \eqref{FreeFermionTotalMI} while the green and brown curves correspond to the non-universal spin-structure term in  \eqref{FreeFermionMIUnivAndNonUniv}.
		The black line is the late time approximation to the second R\'{e}nyi mutual information \eqref{SSDQuenchLateTimeValue}.}
	\label{SSD_ThermalQuench_BOMI}
\end{figure}
In Fig.\ \ref{SSD_ThermalQuench_BOMI},
the mutual information after the SSD quench is plotted 
for the following three choices of 
the subsystems:
\be \label{setup:MI}
\begin{split}
&A= \left\{x\big{|}0\le x \le X, L-X \le x \le L\right\}, ~~B= \begin{cases}
\left\{x\big{|}\f{L}{4}- X \le x \le \f{L}{4}+X \right\} &~ \text{Case~1} \\
\left\{x\big{|}\f{L}{2}- X \le x \le \f{L}{2}+X \right\} &~ \text{Case~2} \\
\end{cases},\\
&A= \left\{x\bigg{|}\f{L}{4}- X \le x \le \f{L}{4}+X, \right\}, ~~B=  \left\{x\bigg{|}\f{L}{2}- X \le x \le \f{L}{2}+X\right\} ~ \text{Case~3},
\end{split}
\ee
for the physical spin structures $\nu=3,4$ which are identical. 
In all these cases with 
$\epsilon$ much smaller than 
the other length scales, 
the non-universal piece is essentially zero.
In the following,
we will focus on the universal spin structure independent term in \eqref{FreeFermionMIUnivAndNonUniv}.
In all three cases, the mutual information simply grows and saturates at a late time value that we will discuss momentarily.
The most salient features of these plots are the saturation values of the mutual information as well as the time it takes 
for the saturation to occur.
The saturation values for cases 1 and 3 are identical and greater than the saturation value in case 2.
As we will see 
momentarily,
this is because the separation between the pair of intervals is the same for cases 1 and 3 which is smaller than the separation between the pair of intervals in case 2. 
The mutual information saturates much faster in case 3 than in cases 1 and 2, and marginally faster in case 2 than in case 1. This is likely due to the fact that the pair of intervals in case 3 is situated away from the SSD fixed point $x=X^1_f$ while one of the intervals in cases 1 and 2 contains this fixed point where the envelope function of the SSD Hamiltonian vanishes. 
Saturation of the mutual information is achieved in case 2 slightly earlier than in case 1 likely due to the smaller separation between the two intervals.

The late time saturation value can be 
studied analytically.
At late time $t\gg L \gg \epsilon$, the universal part of the mutual information \eqref{FreeFermionMIUnivAndNonUniv} is approximately given by
\begin{align}\label{SSDQuenchLateTimeValue}
	\lim_{t\rightarrow\infty}I_{A\cup B,\text{univ.}}^{(n)} 
	&= \frac{n+1}{6n} \log 
	\left|\frac{\left(\tan\frac{\pi X_1}{L}-\tan\frac{\pi X_3}{L} \right)\left(\tan\frac{\pi X_2}{L}-\tan\frac{\pi X_4}{L} \right)}{\left(\tan\frac{\pi X_1}{L}-\tan\frac{\pi X_4}{L} \right)\left(\tan\frac{\pi X_2}{L}-\tan\frac{\pi X_3}{L} \right)}\right| 
	\nonumber\\
		&=  \frac{n+1}{6n} \log 
		\left|
		\frac{\sin{\frac{\pi(d+L_1)}{L}}\sin{\frac{\pi(d+L_2)}{L}}}{\sin{\frac{\pi d}{L}}\sin{\frac{\pi(d+L_1+L_2)}{L}}} 
		\right|
\end{align}
for all three cases 1, 2 and 3.\footnote{To derive this we note, at late times $t \gg L$,
$r_i \approx \bar{r}_i \approx 2\pi t \sin{\frac{\pi X_i}{L}}$,
and hence
\begin{equation}
	\varphi_i-\varphi_j \approx -(\overline{\varphi}_i-\overline{\varphi}_j) \approx \frac{L^2}{4\pi^2 t^2}\left(\frac{1}{\tan\frac{\pi X_j}{L}}-\frac{1}{\tan\frac{\pi X_i}{L}}\right) \ll 1
\end{equation}
The universal part of the free fermion mutual information \eqref{FreeFermionMIUnivAndNonUniv} can be further simplified applying the following approximation
$\vartheta_1(z|\tau) \approx 2\pi z e^{-\frac{\pi |\tau|}{4}} $
for $z\to 0$ and $\tau\rightarrow i\infty$
(the two limits commute).
}
In the second line, 
we introduce the lengths of the two subsystems,
$L_1=X_1 - X_2$ and $L_2 = X_3-X_4$,
and the separation  $d = X_2 - X_3$. 
This late time saturation value of the mutual information \eqref{SSDQuenchLateTimeValue} is 
precisely the mutual information of the vacuum state \cite{Calabrese_2004,Ryu_2006,2006JHEP...08..045R,Furukawa_2009}. 
Furthermore, 
in the limit of well-separated small intervals, i.e., 
$L_1,L_2 \ll d$ where the separation $d$ is on the same order as the total system size $L$,
\begin{align}
	\frac{\sin{\frac{\pi(d+L_1)}{L}}\sin{\frac{\pi(d+L_2)}{L}}}{\sin{\frac{\pi d}{L}}\sin{\frac{\pi(d+L_1+L_2)}{L}}} 
	\approx 
	1 + \frac{\pi^2 L_1L_2}{L^2 \sin^2 \frac{\pi d}{L}}-\frac{\pi^3 L_1 L_2 (L_1+L_2)}{L^3 \sin^2\frac{\pi d}{L}\tan\frac{\pi d}{L}}+\cdots
\end{align}
This approximation breaks down if $d\approx 0,L$ because of the tangent in the denominator of the third term in the series expansion of $\frac{L_i}{L}$. While the series expansion is valid, the late time saturation value of the mutual information \eqref{SSDQuenchLateTimeValue} is approximately 
\begin{equation}
	\lim_{t\rightarrow\infty}I_{A\cup B,\text{univ.}}^{(n)} \approx \frac{n+1}{6n} \frac{\pi^2 L_1 L_2}{L^2 \sin^2 \frac{\pi d}{L}}.
\end{equation}
Thus,
the late time value of the mutual information 
decreases as a function of distance, which explains why the saturation value was smaller in case 2 than in cases 1 and 3, and why the saturation values appeared to be equal in cases 1 and 3 despite the intervals being located at different parts of the system.

\begin{figure}
	\centering
	\textbf{Case 1}\par\medskip
	\includegraphics[width=0.32\textwidth]{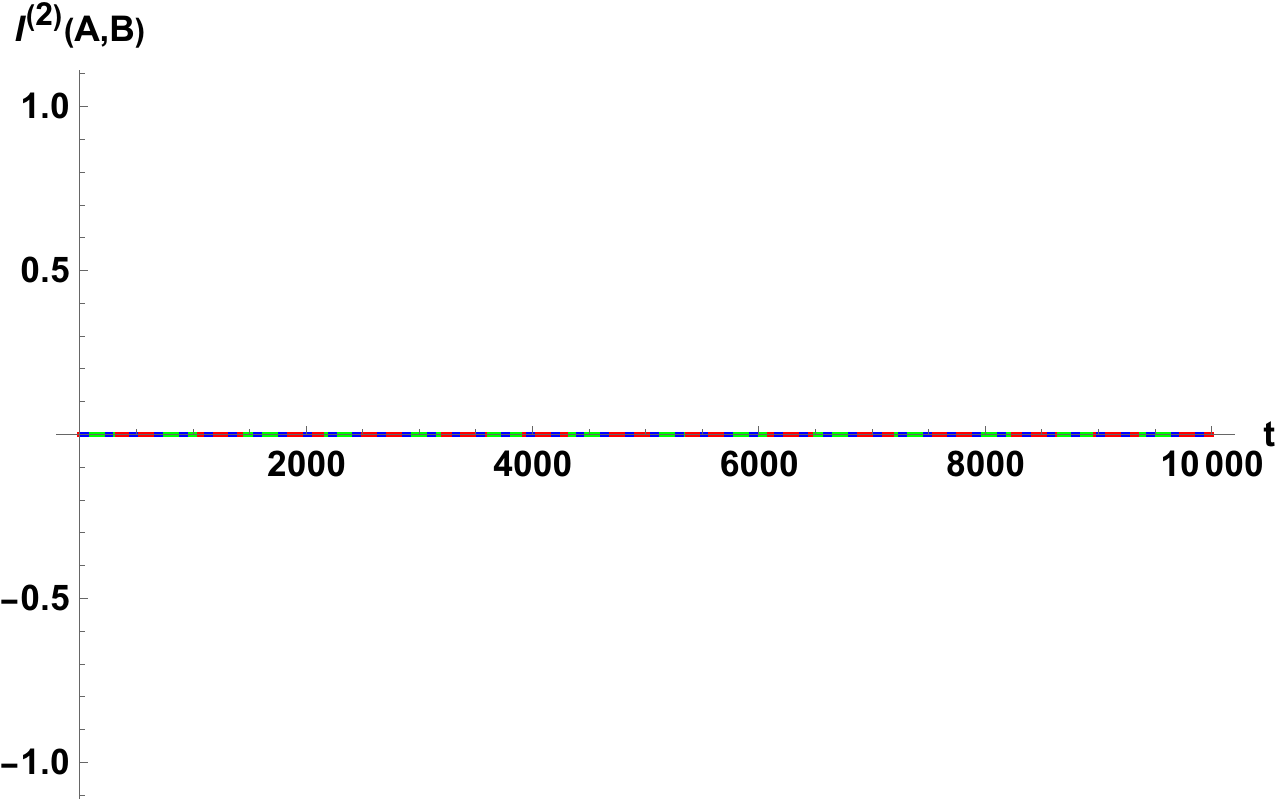}
	\includegraphics[width=0.32\textwidth]{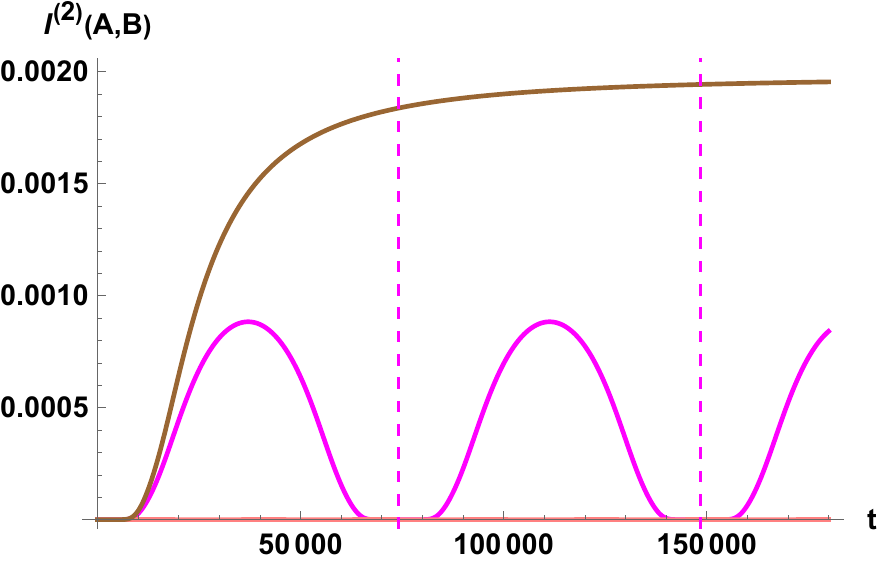}
	\includegraphics[width=0.32\textwidth]{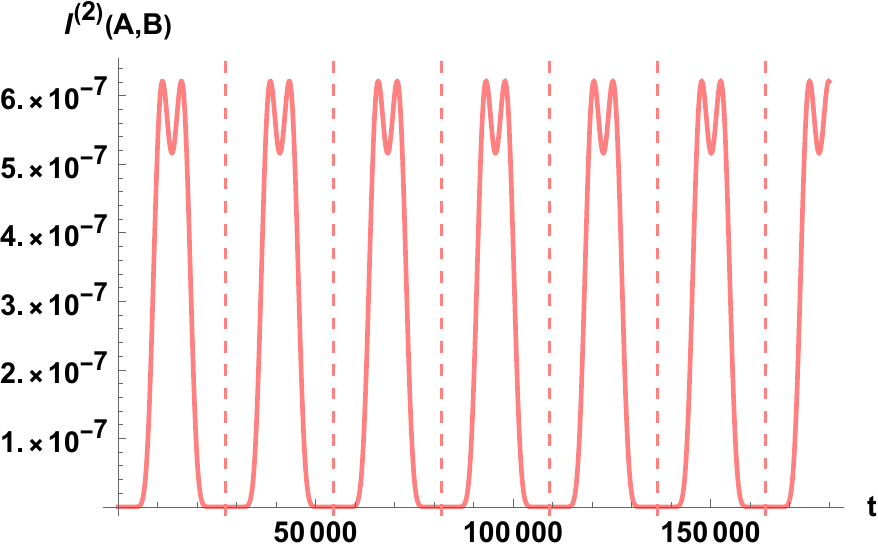}
		\textbf{Case 2}\par\medskip
	\includegraphics[width=0.32\textwidth]{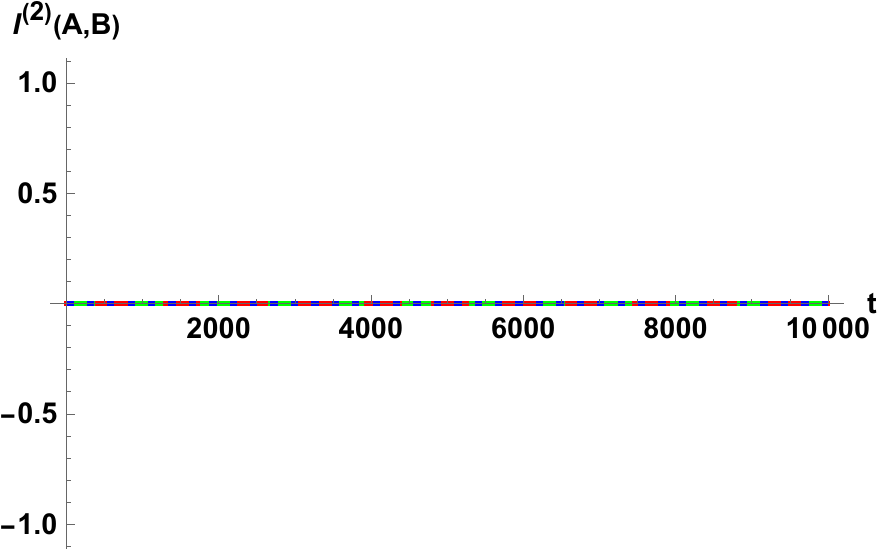}
\includegraphics[width=0.32\textwidth]{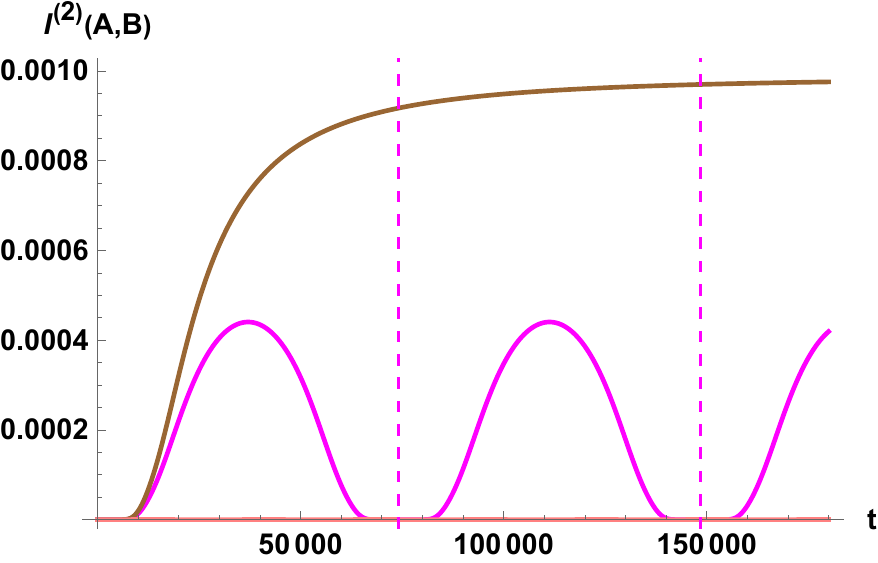}
\includegraphics[width=0.32\textwidth]{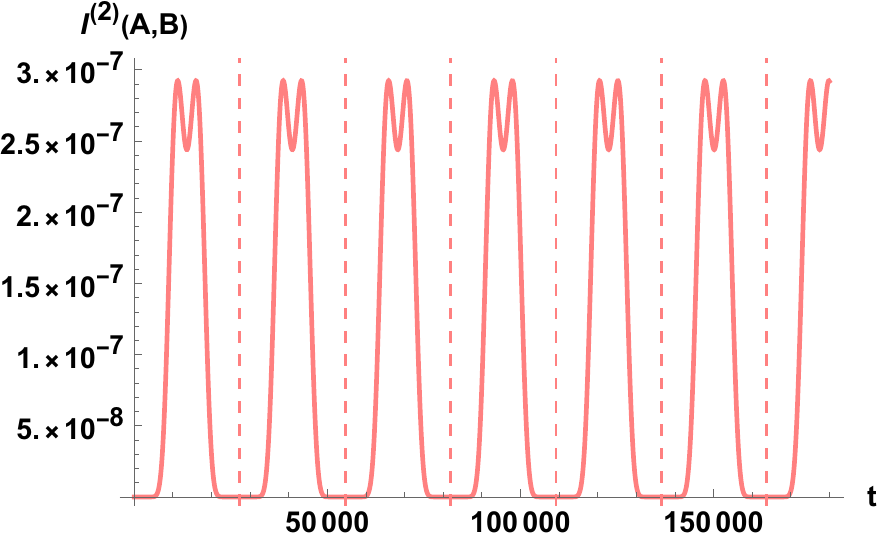}
	\textbf{Case 3}\par\medskip
	\includegraphics[width=0.32\textwidth]{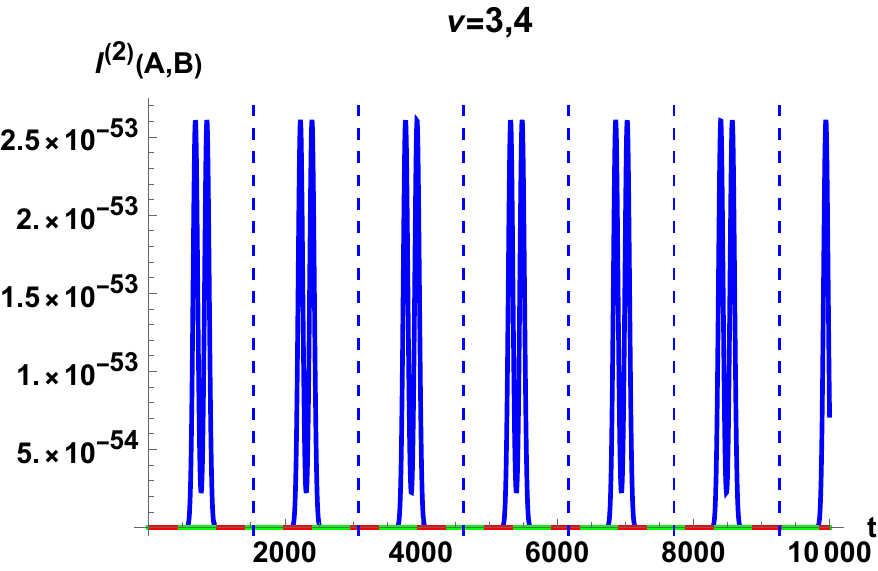}
\includegraphics[width=0.32\textwidth]{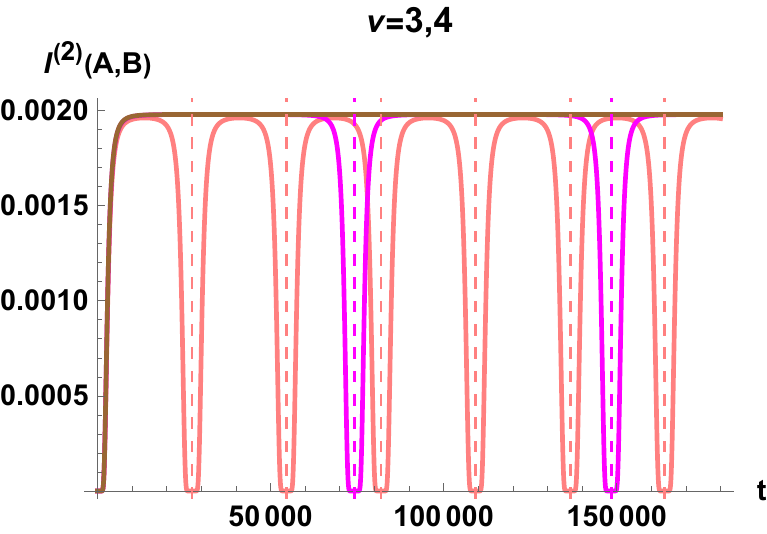}
\includegraphics[scale=0.4]{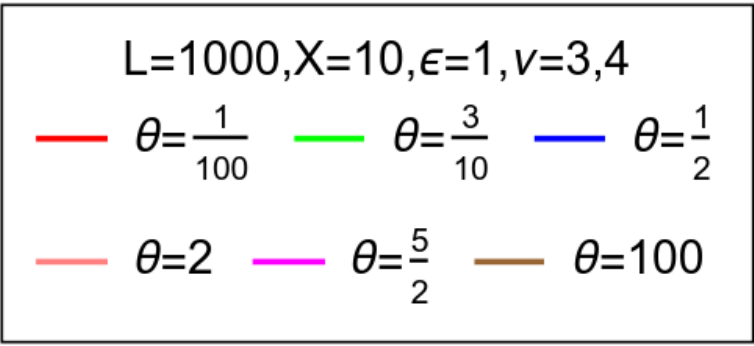}
	\caption{Plots of the second R\'{e}nyi mutual information of the thermal state after a M\"{o}bius quench for the three cases listed in \eqref{setup:MI} with $L=1000$, $X=10$ and $\epsilon=1$ for the two physical spin structures $\nu=3,4$ which turn out to be identical. The dotted vertical lines indicate the periods $L\cosh{2\theta}$.}
\label{MobiusQuenchThermalStateMI}
\end{figure}

Moving away from the SSD limit,
plots of the mutual information \eqref{FreeFermionTotalMI} after a general M\"{o}bius quench with finite deformation parameter $\theta$ are shown in Fig. \ref{MobiusQuenchThermalStateMI}. Just as in the SSD quench, the spin structure terms are negligible and the mutual information is the same for both $\nu=3$ and $\nu=4$. When $\theta$ is small, the deformed Hamiltonian is almost the uniform one, so the quench does nothing to the thermal state. Thus, the mutual information vanishes for small values of the deformation parameter $\theta$. As $\theta$ is increased, the mutual information starts to become non-zero and bumps with two peaks can be observed (c.f. $\theta=2$ for cases 1 and 2 and $\theta=\frac{1}{2}$ for case 3). As $\theta$ is increased further, the amplitude of the mutual information grows. Eventually, the bumps in the mutual information show only a single peak. In all cases, the period of oscillation is given by $L\cosh{2\theta}$. As $\theta$ becomes larger, the period keeps growing until the mutual information approaches that of the SSD quench as in Fig.\ \ref{SSD_ThermalQuench_BOMI}. Therefore, the mutual information in the SSD quench can be thought of as the limit of the M\"{o}bius quench with an infinite period.

Comparing the various cases also yield interesting insights into the dynamics of the M\"{o}bius quench. Since the SSD quench is a limit of the M\"{o}bius quench, the mutual information after the M\"{o}bius quench is upper bounded by the late time saturation value of the mutual information after the SSD quench \eqref{SSDQuenchLateTimeValue} which is a decreasing function of the separation between the two intervals. This explains why the mutual information in cases 1 and 3 are larger than the mutual information in case 2. However, the mutual information in case 3 is also larger than that in case 1. For instance, the mutual information when $\theta=\frac{1}{2}$ is non-negligible only in case 3 and the mutual information for $\theta=2,\frac{5}{2}$ only attains the upper bound in case 3. Furthermore, the mutual information grows much faster in case $3$ than in case $1$. This assortment of observations can likely be attributed to the fact that in case $3$, 
both intervals are located away from the SSD fixed point 
while one of the intervals contains the SSD fixed point in case 1.

\end{widetext}

\clearpage
\bibliography{reference}

\end{document}